\documentclass[aps,footinbib,superscriptaddress,showpacs,twocolumn,prb,longbibliography,bibnotes,showkeys]{revtex4-1}

\usepackage{changepage}
\usepackage{fancyhdr}
\usepackage{float}
\usepackage{floatflt}
\usepackage{soul}
\usepackage[colorlinks=true,citecolor=blue,linkcolor=magenta]{hyperref}
\usepackage[normalem]{ulem}
\usepackage{epstopdf}
\usepackage{amsmath,amssymb}
\usepackage{braket}
\usepackage{graphicx,graphics,subfigure}
\usepackage{color}
\usepackage{stackengine}
\usepackage{verbatim}

\newcommand{\Nf}{\mathcal{N}_F}

\def \titlename{Many-body critical phase in a quasiperiodic chain and dynamical Widom lines in Fock space properties}
\def \authornames{Nilanjan Roy$^{1,2}$, Subroto Mukerjee$^{1}$, and Sumilan Banerjee}
\def \affiliations{Centre for Condensed Matter Theory, Department of Physics,
Indian Institute of Science, Bangalore 560012 India}
\def \affiliationss {$^2$Division of Physics and Applied Physics,
Nanyang Technological University, Singapore 637371}

\begin{document}
\title{\titlename}
\author{\authornames}
\affiliation{\affiliations\\ \affiliationss}
\date{\today}
\begin{abstract}
We study a quasiperiodic model in one dimension, namely the extended Aubry-Andr\'e-Harper (EAAH) chain, that realizes a critical phase comprising entirely single-particle critical states in the non-interacting limit. In the presence of short-range interactions, the non-interacting critical phase transforms to a many-body critical (MBC) phase, separated by lines of MBC-ergodic, MBC-many-body localized (MBL) and ergodic-MBL phase transitions that meet at a triple point. We elucidate the unusual characteristics of the MBC phase compared to the ergodic and MBL phases through the localization properties of the excitations in real space and Fock space (FS), and eigenstate inverse participation ratio (IPR). We show that the MBC phase, like the MBL phase, is well described by a multifractal scaling of the IPR and a linear finite-size scaling ansatz near the transition to the ergodic and MBL phases. However, the MBC phase, at the same time, exhibits delocalization of all single-particle excitations and a system-size dependent Fock-space localization length, analogous to the ergodic phase. Remarkably, we find evidence of unusual Widom lines on the phase diagram in the form of lines of pronounced peaks or dips in the FS localization properties inside the MBC and MBL phases. These Widom lines either emerge as a continuation of the precursor phase transition line, terminating at the triple point, or originate from a phase boundary.
\end{abstract}

\maketitle

\section{Introduction}
Following Anderson's seminal work~\cite{anderson1958absence}, single-particle eigenstates of a non-interacting quantum particle are known to be either localized or delocalized in systems with quenched randomness, e.g., systems with random disorder potentials, due to the phenomenon of Anderson localization (AL). The AL phenomenon can even occur in systems with quasiperiodic potentials, as exemplified by the the well-known Aubry-Andr\'e-Harper (AAH) model~\cite{aubry1980analyticity}. However, a third category of states, namely critical states, intermediate to localized and delocalized states, are known to exist right at the critical point corresponding to the Anderson delocalization-localization transition. The Anderson transition can be realized in random systems in dimensions greater than one, like in 2D or 3D depending on symmetries \cite{abrahams1979scaling,evers2008anderson}, and even in 1D for quasiperiodic systems~\cite{aubry1980analyticity}. At the Anderson transition, all the eigenstates turn critical showing remarkable multifractal properties \cite{evers2008anderson}. An outstanding question is whether one can realize a critical phase, hosting entirely critical states over an extended region of parameter space, instead of at a single critical point. Moreover, the existence of such a critical phase leads to a further natural question of the stability of the phase in the presence of many-body interactions. 

Indeed there are non-interacting models, albeit very limited in number, with extended critical phases. All these models have quasiperiodic potentials. However, in addition, one needs to introduce either quasiperiodic hopping ~\cite{han1994critical,chang1997multifractal,liu2016phase}, a Fibonacci-sequenced potential~\cite{kohmoto1987critical,lin2023general}, spin-orbit coupling~\cite{wang2022quantum} or $p$-wave pairing~\cite{wang2016phase,wang2016spectral} alongside a quasiperiodic potential. Some of these models have also been experimentally realized recently using superconducting circuits~\cite{li2023observation} and ultra-cold atoms~\cite{xiao2021observation,wang2020realization}. 
There also exist a few one-dimensional models that show a phase with a fraction of critical states in the eigenspectrum~\cite{zhou2022exact,deng2019one,roy2021fraction,lee2023critical,ahmed2022flat}.
Once interactions are incorporated in the above systems, one expects the critical phase to be much more fragile compared to the Anderson localized phase with all single-particle states localized. In the last two decades, the stability of the Anderson localized phase in the presence of interactions, leading to the phenomenon of many-body localization (MBL) \cite{abanin2017recent,alet2018many,abanin2019colloquium}, has been indicated by many theoretical works~\cite{baa,gornyi2005interacting,oganesyan2007localization,pal2010many,luitz2015many,imbrie2016many,de2017stability}, at least, in 1D. Evidence of MBL have also been found in  experiments~\cite{schreiber2015observation,rispoli2019quantum,lukin2019probing}, although the regime of stability \cite{sels2021dynamical,morningstar2022avalanches,crowley2022constructive,long2022phenomenology,sels2022bath,Suntajs2020quantum,sierant2022challenges} of the MBL phase still remains under active debate. The MBL phase exhibits many fascinating properties, such as a lack of thermalization or ergodicity \cite{alet2018many,abanin2019colloquium,sierant2403many} and a violation of the eigenstate thermalization hypothesis (ETH) \cite{deutsch1991quantum,srednicki1994chaos} in isolated quantum systems. Universal properties of the putative thermal-MBL critical point have been studied in numerous works \cite{luitz2015many,zhang2018universal,dumitrescu2019kosterlitz,kiefer2020evidence,abanin2021distinguishing,Sierant2020,Laflorencie2020,kudo2018finite,bertrand2016anomalous,setiawan2017transport,sutradhar2022scaling,vosk2013many,goremykina2019analytically,morningstar2020many,de2017stability,morningstar2020many}.

Intriguingly, a recent numerical work~\cite{wang2021many} has shown the emergence of a many-body critical phase (MBC), as opposed to a thermal-MBL critical point, out of the non-interacting critical phase, when interactions are introduced in a fermionic extended AAH (EAAH) model with quasiperiodic hopping~\cite{han1994critical}. Ref.[\onlinecite{wang2021many}] classifies the eigenstates of the MBC phase as nonergodic extended (NEE) based on sub-thermal volume-law scaling of entanglement entropy of the states and the critical level statistics in the phase. However, when viewed from the perspective of a Fock space (FS) lattice or graph \cite{welsh2018simple,logan2019many,roy2020fock,roy2021fock,roy2022hilbert,roy2023diagnostics,ghosh2019many,ghosh2024scaling}, many-body eigenstates in the MBL phase, or even in the non-interacting Anderson localized phase, can be characterized as \emph{non-ergodic extended} based on the \emph{fractal} or \emph{multi-fractal} scaling of the inverse participation ratio (IPR) \cite{roy2020fock,roy2021fock} or the local FS propagator \cite{sutradhar2022scaling,roy2023diagnostics,ghosh2024scaling} with the number of FS basis states $\mathcal{N}_F$.  This situation is different from random regular graphs with uncorrelated disorder where the existence of NEE states has been highly debated \cite{altshuler2016nonergodic,tikhonov2016anderson,kravtsov2018non,herre2023ergodicity,vanoni2024renormalization}. Yet, another type of NEE state \cite{li2015many,modak2015many,deng2017many,Modak2018,ghosh2020transport} has been found in systems with a particular type of quasiperiodic disorder, namely the generalized Aubry-Andr\'e-Harper (GAAH) model ~\cite{ganeshan2015nearest,modak2015many}. These NEE states in the interacting GAAH model presumably arise due to the existence of a single-particle (SP) mobility edge in the non-interacting limit of the GAAH model \cite{modak2015many,ghosh2019many,roy2023diagnostics}, and the interaction-induced coupling between localized and delocalized single-particle states across the SP mobility edge. 

In this work, we ask how the NEE states in the MBC and non-interacting critical phases of the EAAH model are different in terms of their Fock and real-space properties from non-interacting delocalized and localized, ergodic, MBL and other previously studied NEE states \cite{modak2015many,ghosh2020transport,deng2017many,roy2023diagnostics}. We also study the nature of the transitions from delocalized and localized phases to the critical phase in the non-interacting model, and the transitions from the ergodic and MBL phases to the MBC phase in the interacting model. Intriguingly, we find the evidence of unusual Widom-like lines \cite{Xu2005,Franzese2007,Simeoni2010,Luo2014,Sordi2024} on the dynamical phase diagram of the EAAH model, namely the existence of loci of pronounced peaks or dips in the FS localization properties deep inside the MBC (critical) and MBL (localized) phases. These Widom-like lines either appear as a continuation of the antecedent phase transition line or emanate from a phase boundary.



To this end, we compute the local density of states (LDOS) for single-particle excitations, to characterize the many-body eigenstates in the \emph{delocalized, critical and localized} phases of the non-interacting EAAH model, and \emph{ergodic, MBC and MBL phases} of the interacting EAAH model. The LDOS captures localization properties of the single-particle excitations in real space. To obtain a complementary perspective, the localization properties in these phases are quantified in FS via IPR, and the local and non-local FS propagators \cite{sutradhar2022scaling,roy2023diagnostics,ghosh2024scaling}. 
A FS lattice or graph is constructed using the many-body basis of occupations of the real-space sites. Any non-interacting or interacting model on a real-space lattice can be viewed as a non-interacting hopping problem on the FS lattice~\cite{altland2017field,welsh2018simple,logan2019many,roy2020fock,roy2021fock,ghosh2019many,sutradhar2022scaling}. This approach has been explored in many recent works to emphasize the role of correlated disorder in the FS to give rise to MBL phenomena~\cite{pietracaprina2017total,logan2019many,altland2017field,welsh2018simple,roy2020fock,roy2021fock,ghosh2019many,scoquart2024role} and the inevitable multifractal nature of the MBL eigenstates~\cite{de2013ergodicity,luitz2015many,mace2019multifractal,roy2021fock,de2021rare}. Subsequently, numerical scaling theories of the MBL transition in terms of the FS inverse participation ratio (IPR) and FS propagator ~\cite{mace2019multifractal,roy2021fock,sutradhar2022scaling} have been developed based on the FS perspective. The latter has been useful \cite{roy2023diagnostics} to distinguish  thermal and MBL states from the NEE states in the GAAH model. 

In FS, we look into the scaling of IPR with FS dimension or the number of sites on the FS lattice $\mathcal{N}_F$, i.e., $IPR\simeq A_I\mathcal{N}_F^{-D_2}$, and the scaling of the typical value of the imaginary part of the local FS self energy, $\Delta_t\simeq A_s\mathcal{N}_F^{-(1-D_s)}$. Here $D_2$ and $D_s$ are fractal and spectral fractal dimensions, respectively, and $A_I$ and $A_s$ are the amplitudes of the power laws. One expects, $D_2=D_s=1$, $D_2=D_s=0$ and $0<D_s<1$, respectively, for the delocalized, localized and non-ergodic extended states in FS. Our main results for the EAAH model, summarized in Table~\ref{table_results}, are the following.\\
\begin{table*}
\centering 
\begin{tabular}{| p{0.2\textwidth} | p{0.2\textwidth} | p{0.2\textwidth} | p{0.2\textwidth} | p{0.2\textwidth} |}   
\hline 
 \centering {\bf  \vspace{0.1cm} PHASES} & \multicolumn{4}{c|}{\bf DIAGNOSTICS}  \\ 
\cline{2-5} 
\centering  & \centering \bf Single particle (SP) excitations in real-space & \centering \bf FS Inverse participation ratio $(IPR= A_I\mathcal{N}_F^{-D_2})$ & \centering \bf Typical value of FS self-energy $(\Delta_t= A_s\mathcal{N}_F^{-(1-D_s)})$ &  \bf FS localization length $(\xi_F)$ \\
 \hline
\centering  \bf Delocalized & \centering All SP excitations delocalized & \centering $A_I>1$, $D_2=1$ & \centering $A_s<1$, $D_s=1$ & $\xi_F$ increases with $L$ (slower than ergodic) \\ 
\hline
\centering  \bf Critical & \centering All SP excitations delocalized & \centering $A_I>1$, $0<D_2<1$ & \centering Anomalous $L$-dependence & $\xi_F$ increases with $L$ (slower than MBC) \\
\hline
\centering  \bf Localized &\centering All SP excitations localized & \centering $A_I<1$, $0<D_2<1$ & \centering  $A_s>1$, $0<D_s<1$ (deep within the phase) & $\xi_F$ independent of $L$ (more robust than MBL)\\  
\hline 
\centering  \bf Ergodic & \centering All SP excitations delocalized & \centering $A_I>1$, $D_2=1$  & \centering $A_s<1$, $D_s=1$ &  $\xi_F$ increases with $L$ \\ 
\hline 
\centering  \bf MBC & \centering All SP excitations tend to delocalize for $L\rightarrow\infty$ & \centering $A_I>1$, $0<D_2<1$ & \centering  Anomalous $L$-dependence &  $\xi_F$ increases with $L$ \\ 
\hline 
\centering  \bf MBL & \centering All SP excitations localized & \centering $A_I<1$, $0<D_2<1$ & \centering $A_s>1$, $0<D_s<1$ & $\xi_F$ independent of $L$ \\ 
\hline 
\end{tabular}
\caption{{\bf Classification of many-body phases of the noninteracting (delocalized, critical and localized) and interacting (ergodic, MBC and MBL) systems}: based on various real-space and
Fock-space diagnostics. Here $L$ and  $\mathcal{N}_F$ are the number of sites on the real-space and Fock-space lattices, respectively} 
\label{table_results} 
\end{table*}
(1) By calculating the typical value $\rho_t(\omega)$ of the LDOS at a SP excitation energy $\omega$, we find that all SP excitations in real-space are delocalized in the non-interacting critical and MBC phases. This feature of the critical phases is like that in the delocalized and ergodic phases, but unlike in the localized and MBL phases. The delocalized nature of all SP excitations in the critical phases is in contrast to that of the NEE phase in the GAAH model \cite{roy2023diagnostics}, where there exists a mobility edge between localized and delocalized SP excitations.\\
(2) From the scaling of IPR, we find that critical and MBC states are indeed extended but non-ergodic, having non-zero amplitude over an infinite number, $\mathcal{N}_F^{D_2}\to\infty$ ($0<D_2<1$), albeit zero fraction of sites in the thermodynamic limit $\mathcal{N}_F\to\infty$. Thus, in this regard, the critical phases are similar to the multifractal states with $0<D_2<1$ in the localized and MBL phases, as well as the NEE phase \cite{roy2023diagnostics} in the GAAH model, and dissimilar to the delocalized and ergodic states, where $D_2=1$. \\
(3) We find the existence of Widom-like lines in the form of a line of peaks or dips in FS quantities like $IPR$ and $\Delta_t$ inside the MBC (critical) and MBL (localized) phases. Such a line of non-monotonicity in physical quantities inside a phase is reminiscent of the well-known Widom or Fisher-Widom lines obseved in a supercritical fluid \cite{Xu2005,Franzese2007,Simeoni2010,Luo2014,Sordi2024} in the thermodynamic phase diagram. Here, however, we detect these lines on the dynamical phase diagram of the EAAH model. \\
(4) We find that the typical self energy $\Delta_t(\mathcal{N}_F)$ does not follow any systematic scaling with $\mathcal{N}_F$ in the non-interacting critical and MBC phases, often exhibiting a non-monotonic dependence with $\mathcal{N}_F$, as well as with the parameters of the EAAH model. This is in contrast to the well-defined scaling behaviour $\Delta_t\sim \mathcal{N}_F^{-(1-D_s)}$, seen in the delocalized and ergodic phases with $D_s=1$, AL and MBL phases with $0<D_s<1$, and the NEE phase\cite{roy2023diagnostics} in the GAAH model with $0<D_s<1$. \\
(5) For the finite system sizes accessible by our ED calculations, we find that the ergodic-MBC transition is well described by the same asymmetric scaling ansatz~\cite{garcia2017scaling} that has been found~\cite{mace2019multifractal,roy2021fock,sutradhar2022scaling} to describe the ergodic-MBL transition for similar system sizes. In particular, we find a \emph{volumic} finite-size scaling form, governed by a nonergodic FS volume which diverges with a Kosterlitz-Thouless (KT) like essential singularity for both the ergodic-MBL and ergodic-MBC transitions in the EAAH model. On the other hand, a \emph{linear} scaling form with power-law divergent correlation length describes the transitions approaching from the MBC and MBL phases. \\
(6) We extract a FS length scale $\xi_F$ from the decay of non-local FS propagators on the FS lattice. $\xi_F$ increases with $L$ in the MBC (critical) phase as in the ergodic (delocalized) phase and NEE phase \cite{roy2023diagnostics} in the GAAH model. This is in contrast to MBL (AL) phase where $\xi_F$ is almost independent of $L$. We find that $\xi_F(L)$ increases much faster with $L$ in the interacting ergodic and MBC phases compared to the non-interacting delocalized and critical phases. 

As shown in Table~\ref{table_results}, the amplitude $A_I$ of the power-law scaling of IPR in ergodic (delocalized) and MBC (critical) phases is greater than 1, whereas in the MBL (AL) phase $A_I<1$. The ergodic, MBC and MBL phases on the FS qualitatively look similar to the corresponding non-interacting phases in terms of the typical LDOS, FS IPR and self-energy. However, the distribution of the IPR and self energy show quite different behaviors in the ergodic phase and non-interacting delocalized phase.

The rest of the paper is organized as follows. In Sec.~\ref{sec2}, we describe the model. In Sec.~\ref{sec3} we define the single-particle excitations in real space and then use them for characterization of the phases for noninteracting and interacting systems, in Secs.~\ref{sec3a} and~\ref{sec3b}, respectively. Then in Sec.~\ref{sec4}, we define the FS propagator and discuss the FS lattice structure. In Secs.~\ref{sec4a} and \ref{sec4b}, we calculate the imaginary part of the Feenberg self energy and inverse participation ratio in FS to characterize phases in noninteracting and interacting systems respectively followed by a finite size scaling of inverse participation ratio across different phase transitions in the interacting systems. 
In Sec.~\ref{sec5}, we define the FS localization length and distinguish different phases using it. Then in Sec.~\ref{sec6} we discuss possibility of getting the Widom-like line in presence of random potential. Finally, we conclude in Sec.~\ref{sec7}.

\section{Model}\label{sec2}
The (EAAH) model Hamiltonian of the interacting spinless fermions is given by~\cite{wang2021many},
\begin{eqnarray}
H=\sum_{i=1}^{L-1} (t_ic_{i}^{\dagger}c_{i+1} + h.c.) +  \sum_{i=1}^{L} h_i n_i + V\sum\limits_{i=1}^{L-1} n_i n_{i+1}
\label{ham}
\end{eqnarray}
\begin{figure}[h!]
\centering
\stackon{\includegraphics[width=0.494\columnwidth,height=3.4cm]{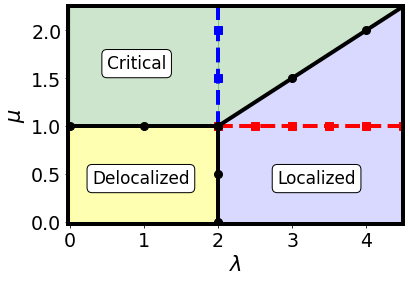}}{(a)}
\stackon{\includegraphics[width=0.494\columnwidth,height=3.4cm]{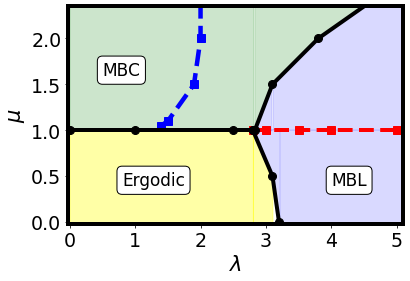}}{(b)}
\caption{ {\bf Phase diagrams:} (a) Phase diagram of the  noninteracting $(V=0)$ EAAH model. The phases are delocalized, localized and critical separated by the solid lines. The vertical dashed line ($\lambda=2$) in the critical phase and horizontal dashed line ($\mu=1$) in the localized phase indicate the non-monotonic behavior (`Widom line') in IPR and $\Delta_t$ which are also used to obtain the phase boundaries.  (b) Similar phase diagram for interacting $(V=1)$ EAAH model with ergodic, MBL and MBC phases separated by solid lines. The dashed lines in MBC and MBL phases denote similar Widom lines, implying non-monotonic behavior in IPR and $\Delta_t$ inside the phases, respectively. Here the strengths of the onsite quasi-periodic potential and quasi-periodic hoppings are denoted by $\lambda$ and $\mu$, respectively.}
\label{phase}
\end{figure}

Here $c_i$ is the fermion annihilation operator at site $i$ and $n_i=c^\dagger_ic_i$. We study the model at half filling.
The nearest-neighbor hopping strength $t_i=J+\mu\cos(2\pi \chi (i+1/2) + \phi)$ and the nearest-neighbor interaction strength $V=1$. The quasiperiodic potential $h_i=\lambda\cos(2\pi \chi i + \phi)$, where $\chi=(\sqrt{5}-1)/2$ and $\lambda$ is the strength of potential with a global phase $\phi\in (0,2\pi]$. The phase $\phi$ is used to generate different \emph{realizations of disorder} and perform disorder averaging.
By taking $\phi=0$ without loss of generality, followed by a special kind of Fourier transformation~\cite{aubry1980analyticity,biddle2010predicted} $c_j=\frac{1}{\sqrt{L}}\sum_k e^{-i2\pi jk\chi}c_k$, Eq.~\ref{ham} for $V=0$ in a closed chain with periodic boundary condition yields
\begin{eqnarray}
H=\sum_{k} (t^{'}_kc_{k}^{\dagger}c_{k+1} + h.c.) +  \sum_{k} h^{'}_k  c_{k}^{\dagger}c_{k},
\label{dual_ham}
\end{eqnarray}
where $t^{'}_k=\lambda/2 + \mu\cos(2\pi \chi (k+1/2) )$ and $h^{'}_k=2J\cos(2\pi \chi k)$ .
Hence the Hamiltonians in Eq.~\ref{dual_ham} and Eq.~\ref{ham} in the non-interacting limit ($V=0$) are related by a duality transformation and the model is self-dual at $\lambda=2J$ . 
For $\mu=0$, one obtains the Aubry-Andr\'e-Harper (AAH) model, with the self duality point $\lambda=2J$, which coincides with the delocalization-localization transition~\cite{aubry1980analyticity}. The above consideration suggests that the same AAH self-duality also exists for $\mu\neq0$, and $\lambda=2J$ remains the delocalization to localization phase transition. This is indeed the case as shown in Fig.\ref{phase}(a). Moreover, the model exhibits further intriguing features for $\mu\geq J$. In this case, a distribution of zeros appears in the quasiperiodic hopping $t_i$'s in the thermodynamic limit such that the system can no longer support a delocalized eigenstate. This mechanism has been put forward recently as a way to produce the critical states~\cite{zhou2022exact}.
 Only critical or localized eigenstates can exist in this parameter region. As a result, for small values of onsite potential strength $\lambda$ all the delocalized eigenstates turn out to be critical with $\mu=J$ being the delocalized-critical transition line for $\lambda<2J$.  For $\lambda>2J$, one obtains another critical line $\lambda=2 \mu$ separating the critical and localized phases, which was obtained numerically~\cite{han1994critical,chang1997multifractal,takada2004statistics,aditya2024family}. We set $J=1$ in our calculations. The schematic of the phase diagram of the non-interacting EAAH model is shown in Fig.~\ref{phase}(a). The phase diagram has been obtained numerically based on earlier works~\cite{han1994critical,chang1997multifractal,takada2004statistics,aditya2024family} and various diagnostics used in this work, as discussed later.
 
 Given the real-momentum space duality of the non-interacting model under $\lambda\rightleftarrows 2J $ for any $\mu$, 
 we find that $\lambda=2$ remains to be the self-dual line, even in the critical phase for $\mu>1$. 
 However, the phase boundaries $\lambda=2\mu$ for $\lambda>2$ and $\mu=1$ for $\lambda<2$ in Fig.\ref{phase}(a) are dual to one another. 
 As we discuss later, Widom like lines $\lambda=2, \mu>1$ and $\mu=1, \lambda>2$ (dual to $\lambda=2\mu$, $\mu<1$ in delocalized phase) appear in the phase diagram for the non-interacting model, both in real-space and Fock-space quantities. Remarkably, the vestiges of these lines seem to persist as Widom-like lines even in the interacting model ($V\neq 0$) where the self-duality is broken. In principle, one may expect to find a $\lambda=2\mu$ line for $\mu<1$ (dual to $\mu=1$ line for $\lambda>2$ in localized phase) in the non-interacting delocalized phase. Nevertheless, no strong signature of this line is seen within the delocalized phase for the quantities studied in this work.

 In the presence of repulsive many-body nearest-neighbor interactions, the self-duality is broken. As a result, all the critical phase boundaries shift in parameter space except the ergodic-MBC transition that still appears at $\mu=1$, as shown in the schematic Fig.\ref{phase}(b). This is because the argument based on the (near) zeros of the quasiperiodic hopping, as explained previously, still holds in the interacting limit where the interaction strength $V$ presumably acts as an irrelevant parameter. The many-body phases and critical lines have been obtained in Ref.~[\onlinecite{wang2021many}] by calculating mainly the energy-level statistics and the half-chain entanglement entropy of the half-filled system. The level spacing ratio $\langle r \rangle=0.53$ and $0.386$ in the ergodic and MBL phases, respectively whereas  $\langle r \rangle$ shows an intermediate value in the MBC phase. A further analysis of the many-body energy level spacings and their correlations can be found in Appendix~\ref{appA}. In the ergodic phase, the half-chain eigenstate entanglement entropy $S_A$ shows the thermal volume-law scaling ($\propto L$), with a coefficient of proportionality consistent with the thermal entropy density in the thermodynamic limit \cite{wang2021many}, and $S_A=S_A^{th}$. In the MBL phase, $S_A$ follows area-law scaling, whereas in the MBC phase, the entanglement entropy is sub-thermal, $S_A<S_A^{Th}$, although $S_A$ still satisfies a volume law. 
A schematic of the phase diagram the of the interacting system is shown in Fig.~\ref{phase}(b) based on earlier works~\cite{han1994critical,chang1997multifractal,takada2004statistics,aditya2024family} and our results. 

\section{Real-space single-particle excitations}\label{sec3}
Here we use local one-particle Green's function to probe the localization properties of the single particle excitations in the real space. 
The single-particle Green's function in the $n$-th many-body eigenstate $\ket{\Psi_n}$ can be defined as $\mathcal{G}_n(i,j,t)=-i\Theta(t)\bra{\Psi_n}\{c_i(t),c^\dagger_j(0)\}\ket{\Psi_n}$ for sites $i$ and $j$. The Fourier transform of the local or onsite term $\mathcal{G}_n(i,i,t)$ can then be written as         
\begin{eqnarray}
\mathcal{G}_n(i,\omega)=\sum\limits_{m} \left[\frac{|\bra{\Psi_{m}^+}c_i^\dagger \ket{\Psi_n}|^2}{\omega^+ - E_{m} + E_n} +  \frac{|\bra{\Psi_{m}^-}c_i \ket{\Psi_n}|^2}{\omega^+ + E_{m} - E_n}\right],\nonumber\\
\label{green}
\end{eqnarray}
where $\ket{\Psi_{m}^+}$ and  $\ket{\Psi_{m}^+}$ are the $m$-th eigenstate with energy $E_m$ of the system with $N+1$ and $N-1$ particles, respectively, and $\omega^+=\omega+\mathrm{i}\eta$, with $\omega$ the excitation energy. The broadening parameter $\eta$ is chosen to be the typical value or the geometric mean of the many-body energy level spacing $(\sim e^{-L})$ at energy $E_n$ for the interacting system ($V\neq 0$) and the average (arithmetic mean) single-particle energy level spacings $(\sim1/L)$ for the non-interacting case $(V=0)$. The particular scaling of $\eta $ with system size, depending on non-interacting or interacting system, allows to distinguish localized and delocalized states/phases while approaching the thermodynamic limit ($L\to \infty$), e.g., as discussed in Sec.\ref{sec3a} below for the non-interacting EAAH model. For the interacting case, the geometric mean of many-body level spacing has been computed over an $\mathcal{O}(1)$ energy range around $E_n$. In this case, instead of the geometric mean, arithmetic mean can also be chosen. We have not studied the dependence of our results for various choices of $\eta$ in this work. However, the choice of $\eta$ has been discussed in detail in Ref.\onlinecite{roy2023diagnostics} in the context of the quasiperiodic GAAH model. For the non-interacting EAAH model, the arithmetic mean of the single-particle level spacing has been calculated for ranges of $\omega$ with finite average local density of states, i.e., excluding the gapped regions (see Sec.\ref{sec3a}).

In particular, we calculate the spectral function characterized by local density of states (LDOS) $\rho_n(i,\omega) = -(1/\pi)Im[\mathcal{G}_n(i,\omega)]$ for an excitation energy $\omega$ at the middle of the many-body energy spectrum ($E_n\simeq 0$) for the interacting EAAH model. The LDOS does not depend on many-body energy for the non-interacting case ($V=0$).
We obtain the typical LDOS defined as $\ln\rho_t(\omega)=\langle \ln \rho_n(i,\omega)\rangle$ and average LDOS defined as $\rho_a(\omega)=\langle \rho_n(i,\omega)\rangle$, where $\langle... \rangle$ denotes an arithmetic average over the lattice sites and $\phi$. 
The single particle excitations originate from the poles of the Green's function $\mathcal{G}_n(i,\omega)$. For a localized state, a finite number of discrete poles, corresponding to the localization volume (length), gives rise to discrete peaks of $\mathcal{G}_n(i,\omega)$ in $\omega$ having zero measure in the  LDOS even in the thermodynamic limit. In contrast, for a delocalized state the poles of $\mathcal{G}_n(i,\omega)$ form a continuum for $\omega$ lying within the energy band. Hence, $\rho_t(\omega)\rightarrow0$ and $\mathcal{O}(1)$ in the thermodynamic limit for localized and delocalized phases, respectively~\cite{ganeshan2015nearest}. On the other hand, the arithmetic mean $\rho_a(\omega)$ approaches a non-zero value with increasing system size both in the localized and delocalized phases. Thus $\rho_t(\omega)$ or the ratio $\rho_t(\omega)/\rho_a(\omega)$ can be used as an order parameter to detect the localization of an excitation with energy $\omega$~\cite{dobrosavljevic2003typical,dobrosavljevic1997mean,jana2021local,roy2023diagnostics}.

\subsection{Noninteracting systems}\label{sec3a}
In the noninteracting limit $(V=0)$, Eq.~$\ref{green}$ can be simplified to write the LDOS as,
\begin{eqnarray}
\rho(i,\omega) =\frac{1}{\pi} \sum\limits_{k} |\psi_k(i)|^2 \frac{\eta}{(\omega-\epsilon_k)^2 + \eta^2},
\label{ldos}
\end{eqnarray}
\begin{figure}[h!]
\centering
\stackon{\includegraphics[width=0.4925\columnwidth,height=3.5cm]{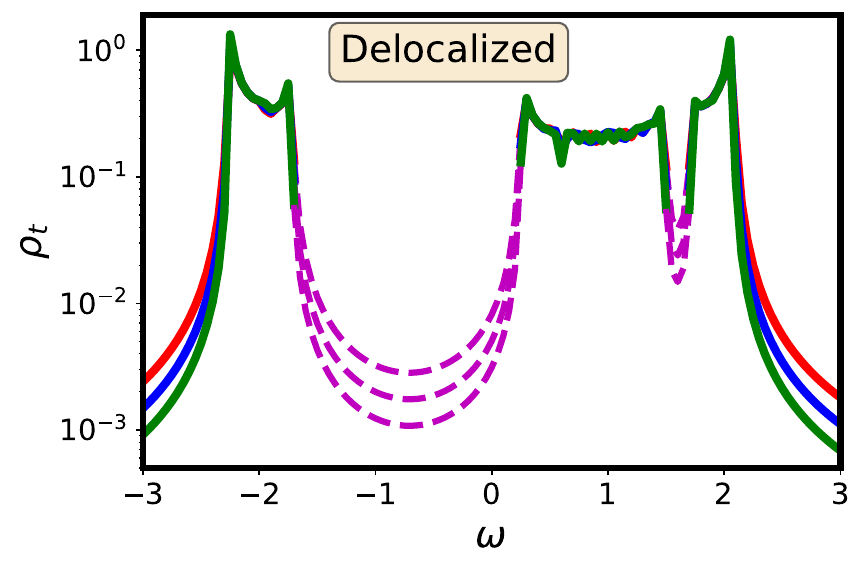}}{(a)}
\stackon{\includegraphics[width=0.4925\columnwidth,height=3.5cm]{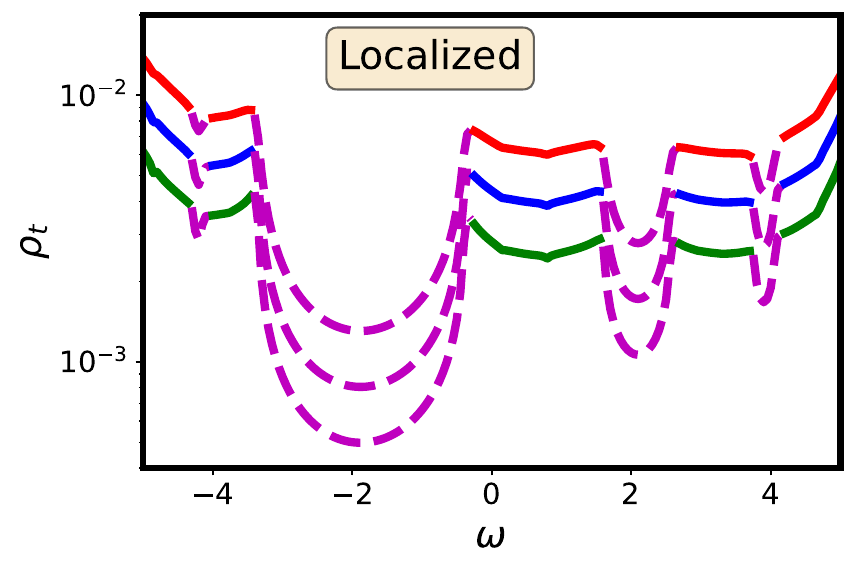}}{(b)} 
\stackon{\includegraphics[width=0.4925\columnwidth,height=3.5cm] {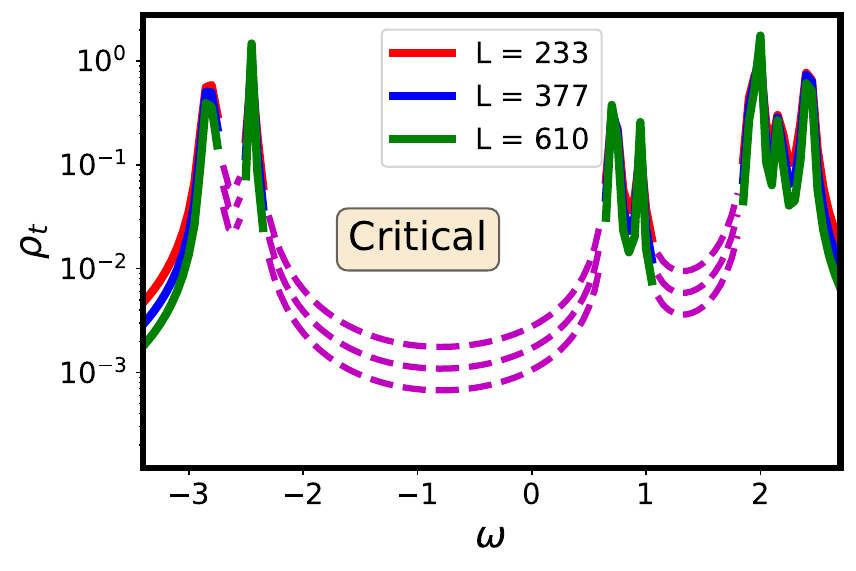}}{(c)}
\stackon{\includegraphics[width=0.4925\columnwidth,height=3.5cm] {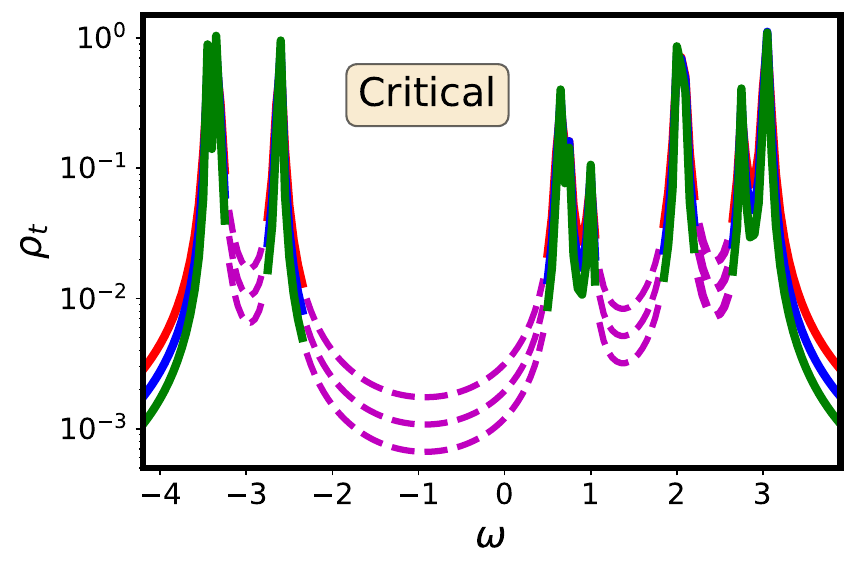}}{(d)}

\caption{{\bf Real-space single-particle excitations in noninteracting system}: (a-b) Typical values of LDOS $\rho_t(\omega)$ for increasing $L$ in the delocalized and localized phases, for $\mu=0.5,~\lambda=1.0$ and $\mu=0.5,~\lambda=5.0$, respectively.  The dashed curves in the figure represent the {\it gapped} excitations, where both $\rho_t(\omega)$ and $\rho_a(\omega)$ (not shown) decrease with $L$. (c-d) $\rho_t(\omega)$ for increasing $L$ at two different points, $(\mu=1.5,\lambda=1.0)$ and $(\mu=2.0,\lambda=1.0)$, in the critical phase on the phase diagram of Fig.\ref{phase}(a).}
\label{ldos_nonint}
\end{figure}
where $\psi_k(i)$ and $\epsilon_k$ are the single particle eigenfunction and energy, respectively.
The $L$-dependence of both $\rho_t(\omega)$ and $\rho_a(\omega)$ can be useful to detect different types of SP excitations, delocalized, localized and gapped. Gaps are $\mathcal{O}(1)$ interval of $\omega$ in the single particle energy spectra without any spectral weight in the thermodynamic limit $L\to\infty$. Eq.~\ref{ldos} can be further written as $\rho(i,\omega) \approx  (1/\pi\eta)\sum_{|\omega-\epsilon_k|\lesssim \eta}  |\psi_k(i)|^2 +  (\eta/\pi)\sum_{|\omega-\epsilon_k|\gtrsim \eta}  |\psi_k(i)|^2/(\omega-\epsilon_k)^2$ . For an excitation energy $\omega$ that falls within a gap, the second term contributes and $\rho(i,\omega)\sim |\psi_k(i)|^2 /L$, where $\psi_k$ corresponds to state(s) at the gap edge withing the band. This leads to both $\rho_t(\omega)$ and $\rho_a(\omega)$ decreasing at least as fast as $1/L$. For excitation energy $\omega$ within an energy band, $\rho(i,\omega) \sim L |\psi_k(i)|^2 + \mathcal{O}(1/L)$. Now, for delocalized and localized states $|\psi_k(i)|^2 \sim 1/L$ and $|\psi_k(i)|^2 \sim e^{-|i-i_k|/\xi}$, respectively, where $\xi$ is the single particle localization length and $i_k$ is localization center. As a result, for delocalized SP excitations, $\rho_t(\omega)$ and $\rho_a(\omega)$ are similar and remains nonzero in the thermodynamic limit. In contrast, the typical value of $\rho(i,\omega)$, $\rho_t(\omega)$, vanishes for the localized SP excitations for $L\to\infty$. However, the arithmetic mean $\rho_a(\omega)$ remains nonzero in the thermodynamic limit due to averaging over $i$ and/or disorder realizations, since the discrete peaks of $\rho(i,\omega)$ form a continuum when sampled over all sites and/or disorder realizations. 

The results for $\rho_t(\omega)$ are shown in Fig.~\ref{ldos_nonint}(a) and Fig.~\ref{ldos_nonint}(b) for delocalized and localized phases of the EAAH model, where all the SP excitations are delocalized and localized, respectively. The dashed curves in the figure represent the {\it gapped} excitations, where both $\rho_t(\omega)$ and $\rho_a(\omega)$ (not shown) decrease with $L$. 
Interestingly, in the critical phase $\rho_t(\omega)$ saturates to $L$-independent nonzero values, as shown in Figs.~\ref{ldos_nonint}(c) and \ref{ldos_nonint}(d) for two different points in the critical phase, qualitatively similar to that of the delocalized phase.

\subsection{Interacting systems}\label{sec3b}
\begin{figure}[h]
\centering
\stackon{\includegraphics[width=0.49\columnwidth,height=3.5cm]{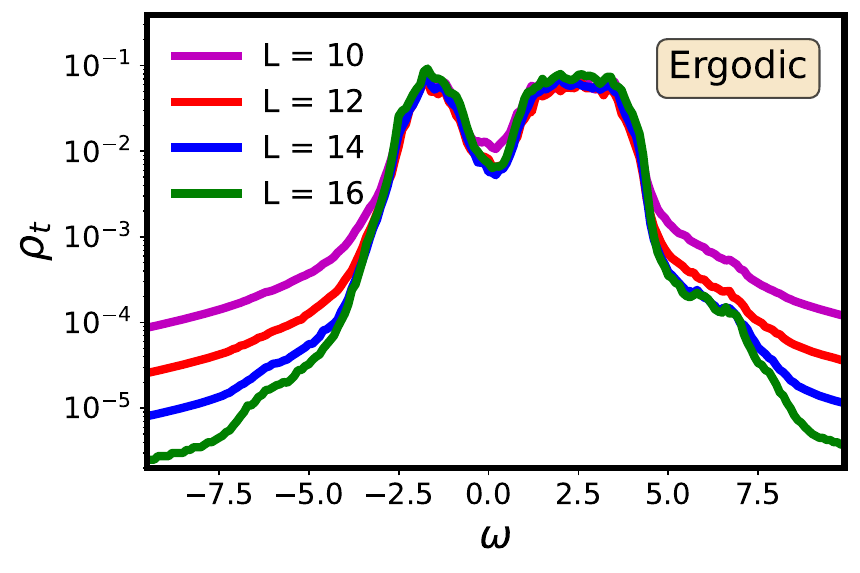}}{(a)}
\stackon{\includegraphics[width=0.49\columnwidth,height=3.5cm]{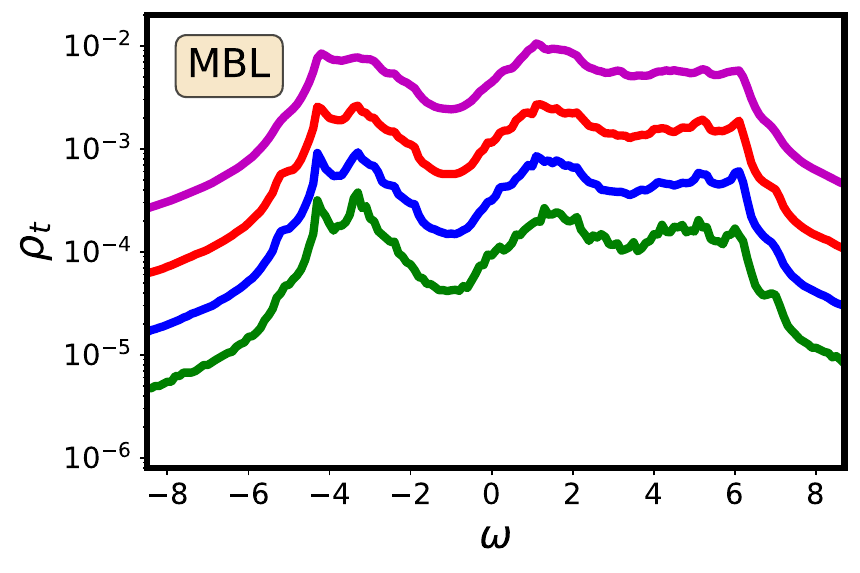}}{(b)}
\stackon{\includegraphics[width=0.49\columnwidth,height=3.5cm]{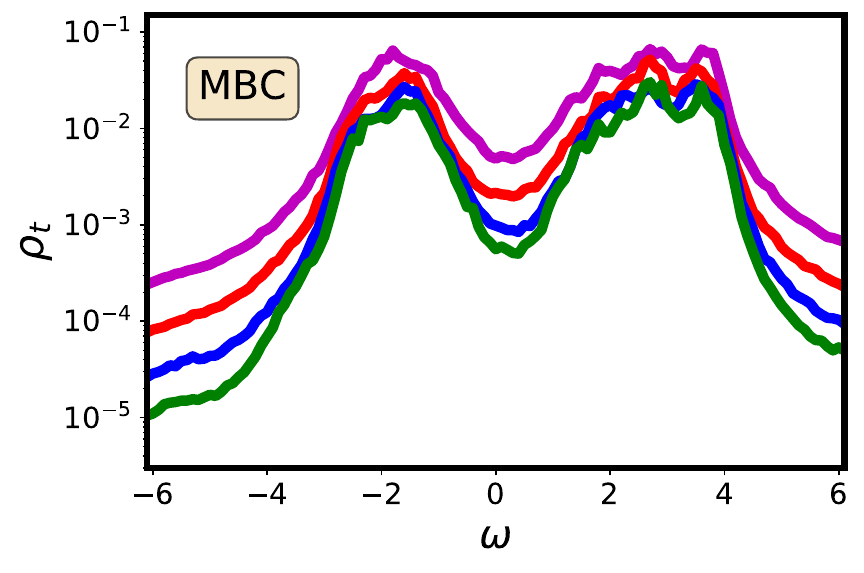}}{(c)}
\stackon{\includegraphics[width=0.49\columnwidth,height=3.5cm]{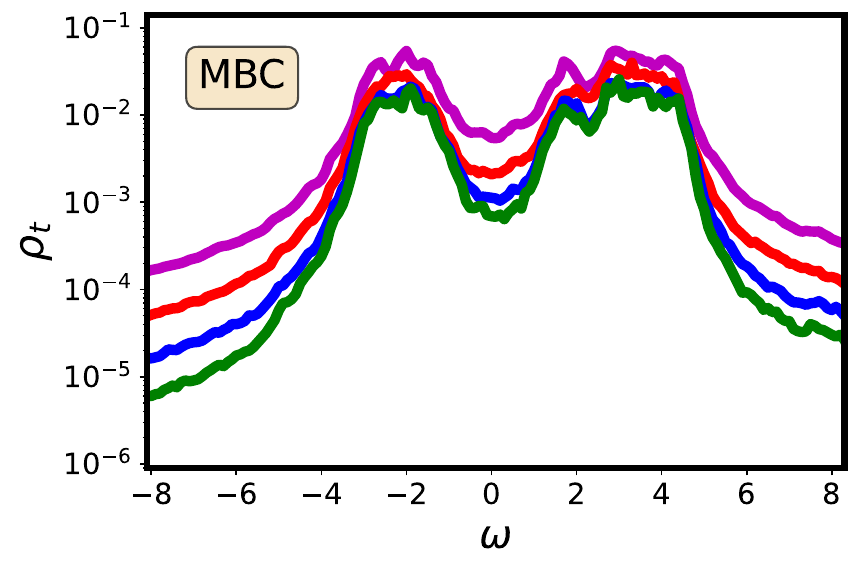}}{(d)}
\caption{{\bf Real-space single-particle excitations in interacting system}: (a-b) Typical values of LDOS $\rho_t(\omega)$ as a function of system size $L$ in ergodic and MBL phases, for $\mu=0.5,~\lambda=2.0$ and $\mu=0.5,~\lambda=5.0$, respectively. 
(c-d) $\rho_t(\omega)$ for increasing $L$ at two different points, $(\mu=1.5,\lambda=1.0)$ and $(\mu=2.0,\lambda=1.0)$, in the MBC phase on the phase diagram of Fig.\ref{phase}(b).}
\label{ldos_int}
\end{figure}

In this subsection, we characterize the phases of the interacting EAAH model through the spectral function, namely, the typical value $\rho_t(\omega)$ of LDOS for many-body eigenstates chosen from a sufficiently small energy window at the middle of the many-body energy spectrum. In particular, we ask how the SP excitations in the delocalized, localized and critical phases of the non-interacting EAAH model, discussed in the preceding section, are impacted by the effects of interaction in the ergodic, MBL and MBC phases of the interacting model. To obtain $\rho_t(\omega)$, We  perform an average over $2000,500,250$ and $100$ samples of global phase $\phi$ for system of sizes $L=10,12,14$ and $16$, respectively. Typical LDOS $\rho_t(\omega)$ for excitation of energy $\omega$ in the ergodic and MBL phases are shown in Fig.~\ref{ldos_int}(a) and Fig.~\ref{ldos_int}(b), respectively. In the ergodic phase, $\rho_t(\omega)$ saturates to a finite value with increasing $L$, indicating delocalization of all excitations within the bandwidth $|\omega|\lesssim 4$. In contrast, in the MBL phase $\rho_t(\omega)$ decreases rapidly with $L$. For example, we have verified that $\rho_t(\omega)/\rho_a(\omega)\sim e^{-L}$. This implies localization of SP excitations at all $\omega$, as also found for the GAAH model~\cite{roy2023diagnostics}. In Figs.~\ref{ldos_int}(c-d), we show $\rho_t(\omega)$ for two different choices of parameters in the MBC phase. Qualitatively, in both the figures, $\rho_t(\omega)$  tends to saturate to a finite value as $L$ increases, albeit with more pronounced $L$ dependence in compared to that for the ergodic phase in Fig.~\ref{ldos_int}(a). The saturation of $\rho_t(\omega)$ to a finite value indicates delocalization of the single particle exitations at all energies $\omega$. However, the saturation value of $\rho_t(\omega)$ in the MBC phase is order of magnitude smaller than that in the ergodic phase. This is also seen by comparing (not shown) the typical LDOS $\rho_t(\omega)$ with mean LDOS $\rho_a(\omega)$ in these phases. In the the ergodic phase, $\rho_t(\omega)\lesssim \rho_a(\omega)$, whereas $\rho_t(\omega)\ll \rho_a(\omega)$ in the MBC phase. In contrast, in the non-interacting EAAH model, $\rho_t(\omega)\lesssim \rho_a(\omega)$ for both delocalized and critical phases.


$\rho_t(\omega)$ in the MBC phase of interacting systems shows qualitatively similar behavior as that in the corresponding critical phase of the noninteracting systems,  although $L$-dependence for the non-interacting critical phase in \textcolor{red}{Fig.\ref{ldos_nonint}(c)} cannot be detected due to much larger system sizes accessed there for the non-interacting EAAH. These results are in contrast with that in an earlier work~\cite{roy2023diagnostics} for GAAH model which hosts a single particle mobility edge in the non-interacting limit over a range of parameters. The mobility edge persists in the SP excitations even in the presence of interaction in the nonergodic extended (NEE) phase that exists over a broad range of many-body energy in the finite-sized systems accessed by ED. The NEE phase in the GAAH model emerges as a result of the interaction-induced mixing of delocalized and localized single-particle excitations separated by a mobility edge in the noninteracting systems \cite{modak2015many,ghosh2020transport,ganeshan2015nearest,deng2017many,roy2023diagnostics}. 
However, in the EAAH model, the MBC phase is obtained due to interaction-induced mixing of the single-particle critical states, in the absence of any SP mobility edge in the noninteracting limit. Thus the feature of the non-interacting critical phase, namely the delocalized nature of all SP excitations, tends to persist even in the MBC phase. Moreover, the MBL phase in the GAAH model has been argued~\cite{modak2015many,ghosh2020transport,roy2023diagnostics} to be an outcome of the mechanism called as the ``MBL proximity effect"~\cite{nandkishore2015many}. On the contrary, the MBC phase in the EAAH model arises due to the stability of the critical phase to interaction, at least in the finite systems accessed by ED. 


\section{Dynamical transitions and Widom lines in Fock-space self energy and inverse participation ratio}\label{sec4}
In this section, we study the localization properties of the different phases of the non-interacting and interacting EAAH model through the local Fock-space propagator \cite{sutradhar2022scaling,roy2023diagnostics}, or the associated local Feenberg self energy \cite{logan2019many,sutradhar2022scaling,roy2023diagnostics}, and the inverse participation ratio (IPR) of the many-body eigenstates. 
The Hamiltonian in Eq.~\ref{ham} can be rewritten as a tight-binding model in terms of the occupation number basis $\{\ket{I}\}$ as~\cite{welsh2018simple,logan2019many,ghosh2019many}
\begin{eqnarray}
H = \sum_{IJ} {T_{IJ}} \ket{I}\bra{J} + E_I \ket{I}\bra{I} ,
\end{eqnarray}
where $\ket{I}=\ket{n_{I1}n_{I2}...n_{IL}}$ with onsite real-space occupation $n_i\in0$ or $1$. The \emph{FS hopping} $T_{IJ}=t_i$ when $\ket{I}$ and $\ket{J}$ are connected by a single nearest neighbor hop and $T_{IJ}=0$ otherwise. The onsite potential at the \emph{FS site} $I$, given by, $E_I=\sum_i h_i n_{Ii} + V\sum_i n_{Ii}n_{Ii+1}$, generates a correlated disorder for a \emph{fictitious particle} on the \emph{FS lattice or network} ~\cite{welsh2018simple,logan2019many,ghosh2019many,roy2020fock,altland2017field,roy2020localization}. The disorder-averaged distribution of the many-body density of states of the half-filled system is a Gaussian as a function of $E$, with the mean $\propto L$ and variance $\sigma_E^2=(t^2+\mu^2/2+\lambda^2/6+V^2/8)L/2$~\cite{welsh2018simple}.
In order to have a well-defined thermodynamic limit, we divide all the parameters by $\sqrt{L}$ and work with the rescaled hamiltonian $\tilde{H}=H/\sqrt{L}$~\cite{welsh2018simple,roy2020fock,logan2019many}. Since we consider here the `infinite-temperature state' corresponding to the middle of the many-body energy spectrum, we set the mean energy manually to zero by implementing the transformation $\tilde{H}\rightarrow (\tilde{H}-Tr\tilde{H}/\Nf)$ with $\Nf$ being the dimension of the Fock space, for each disorder realization. This is done to compensate for the fluctuations of the the middle of the spectrum with disorder realizations~\cite{sutradhar2022scaling} for finite systems.
\begin{figure}[h!]
\centering
\includegraphics[width=0.85\columnwidth,height=5.6cm]{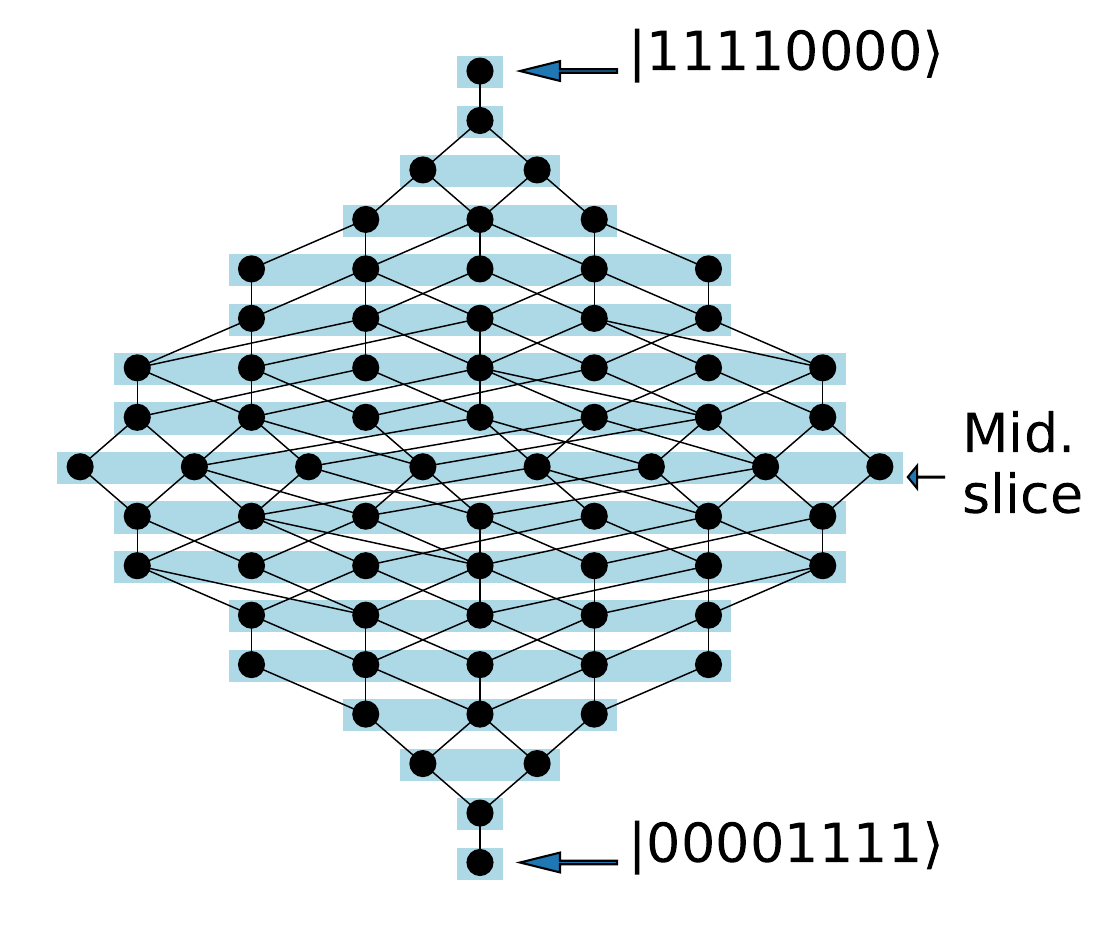}
\caption{{\bf Fock-space (FS) lattice}: FS lattice constructed out of real-space occupation-number basis states (black circles), illustrated for $L = 8$ at half filling, starting at the top with $\ket{11110000}$, i.e., all particles on the left side, and ending at the bottom
with all particles on the right. The hoppings (black lines) and the slices (in lightblue) are indicated.}
\label{FSlattice}
\end{figure}
The FS sites are arranged in slices such that any site in a particular slice is connected to the sites of the nearest slices by a single FS hopping, which is shown in Fig.~\ref{FSlattice}. The local structure of the FS lattice allows one to implement a recursive Green's function method~\cite{mackinnon1980conductivity,mackinnon1983scaling,lee1981anderson,verges1999computational}, which has been recently applied to the FS lattice of similar models with random~\cite{sutradhar2022scaling} and quasiperiodic disorders~\cite{roy2023diagnostics}. We compute the elements of the retarded FS propagator $G(E)=[E + i\eta - \tilde{H}]^{-1}$ at the middle of the energy spectrum $E=0$, corresponding to the infinite-temperature state. Here we choose a broadening parameter $\eta(\mu,\lambda)=\sqrt{2\pi}\sigma_E/(\sqrt{L}\Nf)$~\cite{welsh2018simple,logan2019many}, i.e., the mean energy-level spacing, which is a function of parameters $\mu$ and $\lambda$.

{\it Fock-space self energy:} We calculate $G_{IJ}(E)=\bra{I}G(E)\ket{J}$ for $I,J\in$ the middle (largest) slice of the FS lattice. From the diagonal elements $G_{II}(E)$, or the local FS propagator, we extract the imaginary part $\Delta_I(E)$ of the local self energy, or the Feenberg self energy \cite{,logan2019many}, $\Sigma_I(E)=X_I(E)-i\Delta_I(E)$, which is related to diagonal elements as $G_{II}(E)=[E+i\eta - E_I - \Sigma_I(E)]^{-1}$. The imaginary part of the local self-energy $\Delta_I(E)=Im[G_{II}^{-1}(E)] - \eta$ quantifies the inverse lifetime of a \emph{fictitious excitation} created at FS site $I$ with energy $E$. Such a self energy in real space has been used earlier in the development of theory of Anderson localization~\cite{anderson1958absence}.
We compute the typical value $\Delta_t$ of $\Delta_I(E)$, defined as $\ln \Delta_t=\langle\ln\Delta_I\rangle$, where $\langle...\rangle$ represents the average over disorder realizations and FS sites $I\in$ the middle slice. In the delocalized phase $\Delta_t\sim\mathcal{O}(1)$,  whereas in the localized or non-ergodic phases, one may expect $\Delta_t\rightarrow0$ as $\Nf\rightarrow\infty$ in the thermodynamic limit. Thus $\Delta_t$ serves as an order parameter for the nonergodic-ergodic phase transition as found in the earlier studies~\cite{roy2023diagnostics,sutradhar2022scaling}.
We expect $\Delta_t$ to decay as a power law, i.e., $\Delta_t \simeq A_s\mathcal{N}_F^{-1+D_s}$ with the spectral fractal dimension $D_s$, which is a fraction in the nonergodic phase and $1$ in the ergodic phase. As discussed later, we extract the proportionality constant $A_s$ and spectral dimension $D_s$ in the phases of the noninteracting and interacting systems. We use $16000,8000,4000,1000,500,200$ disorder realizations for system sizes $L=10,12,14,16,18,20$ respectively to generate the distributions of $\Delta_I$. 

{\it Fock-space inverse participation ratio:} We also calculate many-body IPR in Fock-space, which has been another important quantifier for distinguishing the many-body ergodic phase from the nonergodic phases of interacting systems~\cite{mace2019multifractal,roy2023diagnostics}.
For a normalized eigenstate $\ket{\Psi}=\sum_{I=1}^{\mathcal{N}_F} \Psi_I \ket{I}$, IPR is defined as $IPR=\sum_{I} |\Psi_I|^4$ $\sim A_I\mathcal{N}_F^{-D_2}$. The fractal dimension $D_2=1$ in the ergodic phase and $0<D_2<1$ in the MBL phase~\cite{mace2019multifractal}. Moreover, the NEE phase in the interacting GAAH model has been reported to show $0<D_2<1$, similar to the behavior of $D_s$ in the same phase~\cite{roy2023diagnostics}. The coefficient $A_I$ has also been shown to be a potential indicator of the MBL transition in an earlier study~\cite{mace2019multifractal} and also related to the measure of nonergodic volume in ergodic phase~\cite{mace2019multifractal,roy2021fock}. In our work, we extract the coefficient $A_I$ and fractal dimension $D_2$ from averaged IPR, calculated over disorder realizations and an energy window centered around $E=0$, for various phases in the noninteracting and interacting systems, as discussed below.

\begin{figure}
\centering
\stackon{\includegraphics[width=0.493\columnwidth,height=3.5cm]{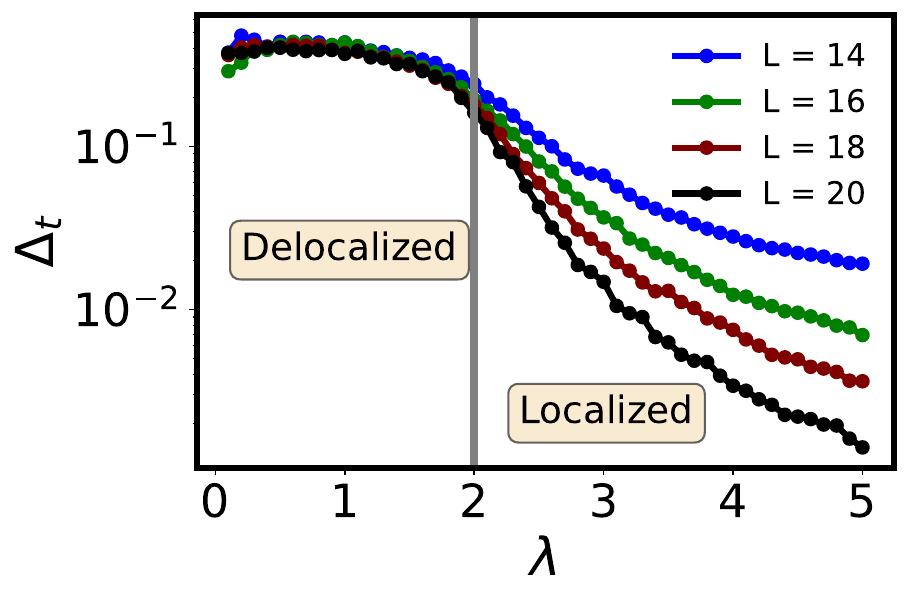}}{(a)}
\stackon{\includegraphics[width=0.493\columnwidth,height=3.4cm]{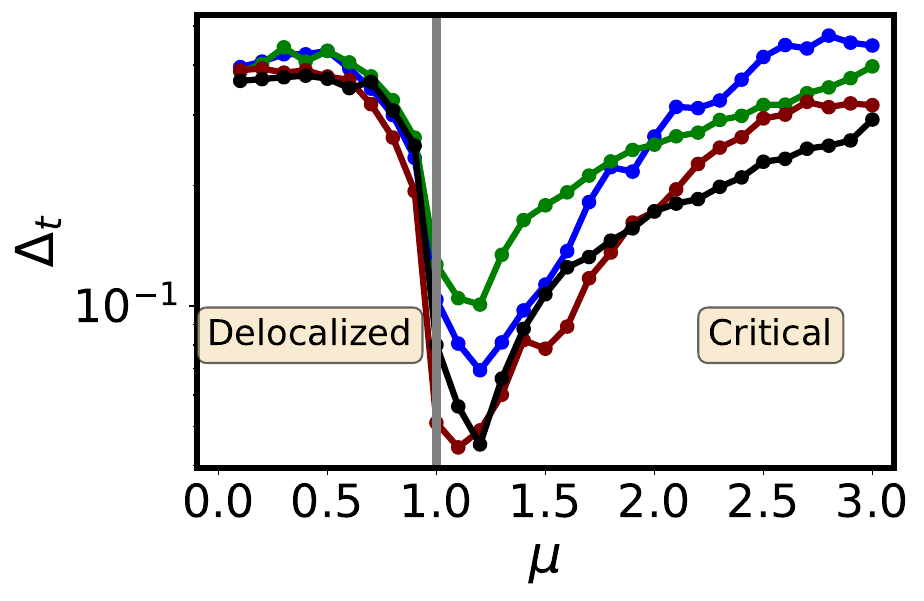}}{(b)}
\stackon{\includegraphics[width=0.493\columnwidth,height=3.5cm]{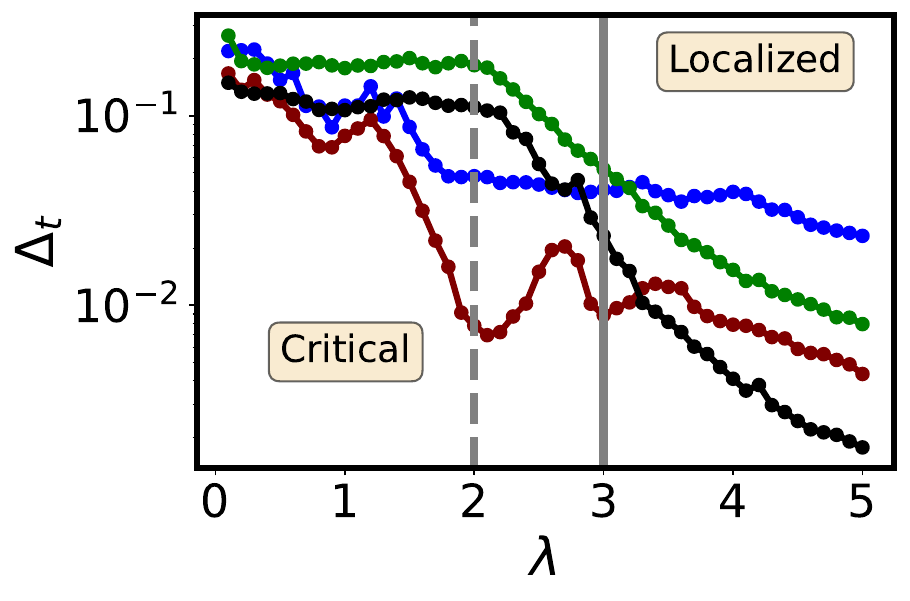}}{(c)}
\stackon{\includegraphics[width=0.493\columnwidth,height=3.5cm]{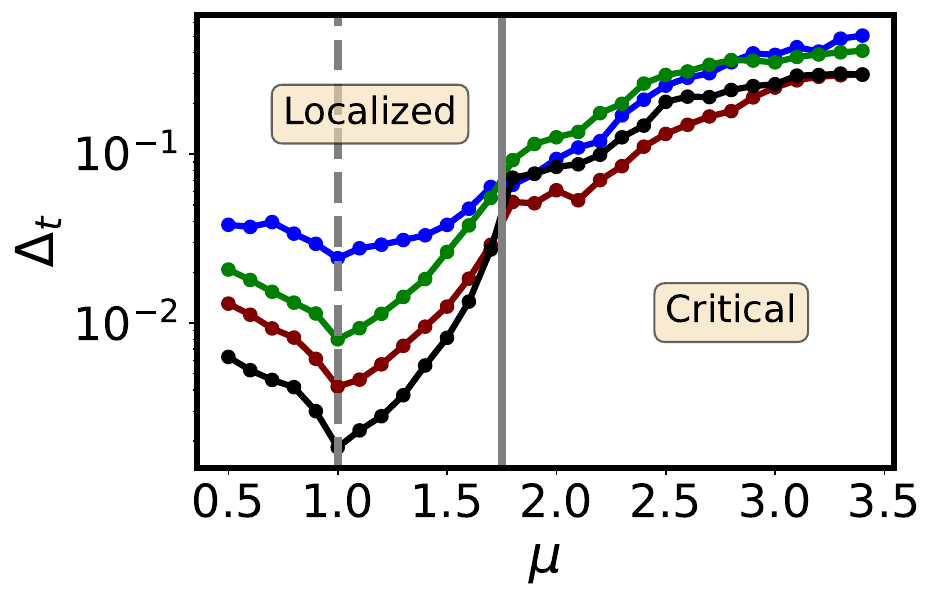}}{(d)}
\caption{{\bf Fock-space self-energy for noninteracting systems}: Typical values of self-energy $\Delta_t$ across (a) the delocalization-localization transition as a function of $\lambda$ for $\mu=0.5$, (b) delocalization-critical 
transition as a function of $\mu$ for $\lambda=1.0$, (c) critical-localization transitions as a function of $\lambda$ for $\mu=1.5$, and (d) localization-critical transition as a function of $\mu$ for $\lambda=3.5$. The phase transition points are denoted by the vertical solid lines. In figures (c) and (d), the vertical dashed lines denote the peak/dip in $\Delta_t$ across the Widom-like lines in Fig.\ref{phase}(a) inside the critical and localized phases, respectively. }
\label{selfenergy_noint}
\end{figure} 
\begin{figure}
\centering
\stackon{\includegraphics[width=0.493\columnwidth,height=3.5cm]{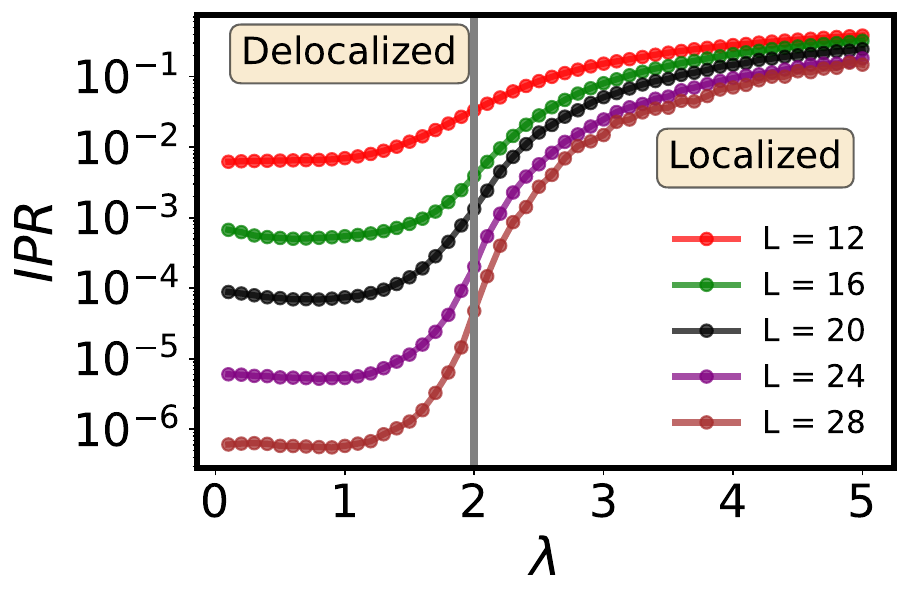}}{(a)}
\stackon{\includegraphics[width=0.493\columnwidth,height=3.5cm]{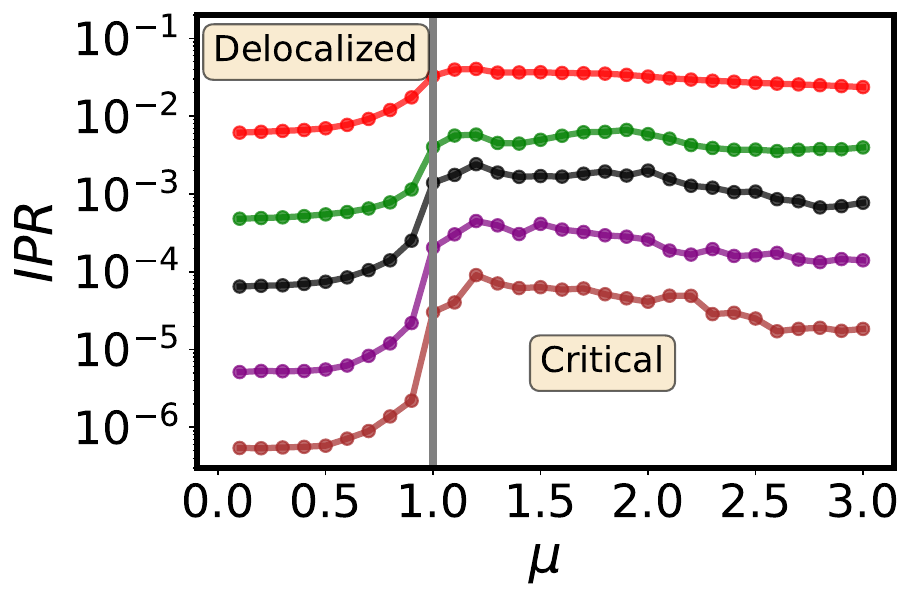}}{(b)}
\stackon{\includegraphics[width=0.493\columnwidth,height=3.5cm]{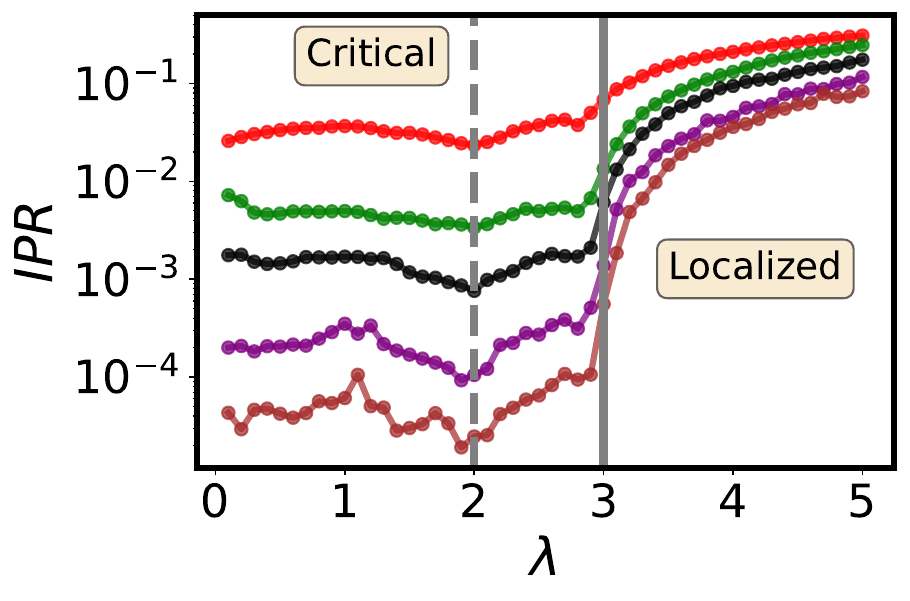}}{(c)}
\stackon{\includegraphics[width=0.493\columnwidth,height=3.5cm]{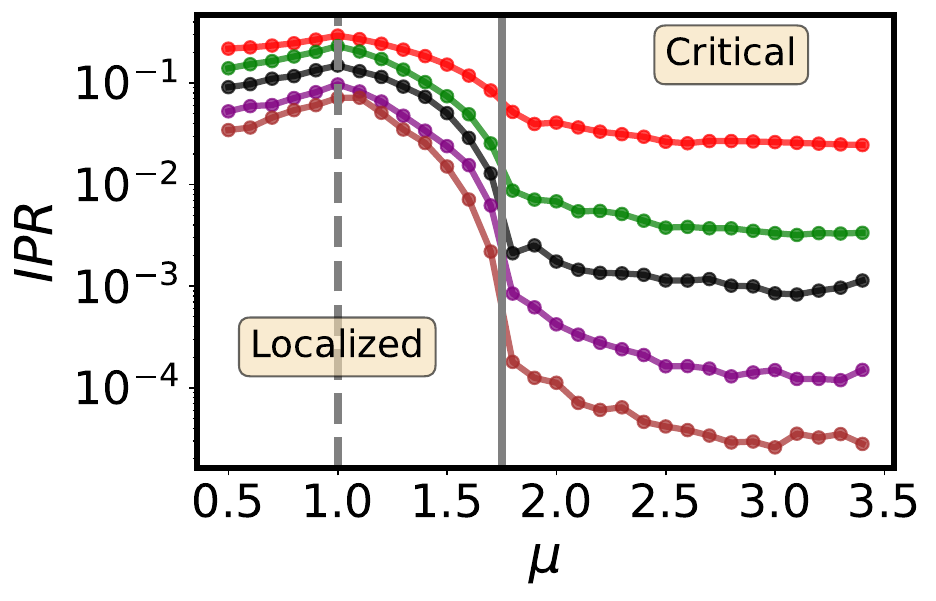}}{(d)}
\caption{{\bf Fock-space $IPR$ for noninteracting systems}: IPR across (a) the delocalization-localization transition as a function of $\lambda$ for $\mu=0.5$, (b) delocalization-critical 
transition as a function of $\mu$ for $\lambda=1.0$, (c) critical-localization transitions as a function of $\lambda$ for $\mu=1.5$, and (d) localization-critical transition as a function of $\mu$ for $\lambda=3.5$. Vertical solid lines denote the points of phase transitions.
In figures (c) and (d), the vertical dashed lines denote the peak/dip in IPR across the Widom-like lines in Fig.\ref{phase}(a) inside the critical and localized phases, respectively.
}
\label{mipr_noint}
\end{figure} 

\subsection{Noninteracting systems}\label{sec4a}
We begin with the discussion of $\Delta_t$ in the noninteracting system. In Fig.~\ref{selfenergy_noint}(a-d), we show the variation of $\Delta_t$ across the phase transitions along four chosen cuts on the phase diagram of Fig.~\ref{phase}(a) to study $\Delta_t$ across delocalized to localized, delocalized to critical, and critical to localized phase transitions. In Fig.~\ref{selfenergy_noint}(a), $\Delta_t$ is shown across the delocalization-localization phase transition (at $\lambda=2$). As evident in Figs.~\ref{selfenergy_noint}(a) and (b), $\Delta_t\sim \mathcal{O}(1)$ and does not vary much with system size $L$ in the delocalized phase. On the contrary, $\Delta_t$ rapidly decreases with $L$ in the localized phase. The delocalization-critical phase transition at $\mu=1$ for $\lambda=1$ is shown in \ref{selfenergy_noint}(b). $\Delta_t$ is non-monotonic in the critical phase, exhibiting a sudden drop near the transition. In the critical phase, $\Delta_t$ shows an anomalous non-monotonic dependence, and thus no systematic scaling with $\mathcal{N}_F$, at least up to the largest system sizes accessed here ($L=20$). 
Fig.~\ref{selfenergy_noint}(c) and Fig.~\ref{selfenergy_noint}(d) show variation of $\Delta_t$ across the critical-localization and localization-critical phase transitions as a function of $\lambda$ ($\mu=1.5$), and as a function of $\mu$ ($\lambda=3.5$), respectively. 
$\Delta_t$ in the critical phase again shows a non-monotonic system size and/or parameter dependence. 

Remarkably, there is a distinct dip, indicating stronger localizing tendency, at $\mu=1$ in the localized phase in Fig.~\ref{selfenergy_noint}(d). We have verified that the dip is present along the $\mu=1$ line as a function of $\lambda$ throughout the localized phase. The $\mu=1$ line, emanating from the delocalized-localized-critical \emph{triple} point, is a continuation of the dip in $\Delta_t$ [Fig.~\ref{selfenergy_noint}(d)] near the delocalized-critical phase boundary, as elucidated in Fig.\ref{phase}(a). This marked non-monotonic feature also appears as a peak in IPR as we discuss below. 
The locus of distinct dips (peaks) in $\Delta_t$ (IPR) along the $\mu=1$ line also persists in the MBL phase in the interacting EAAH model, as shown in Fig.\ref{phase}(b), and discussed later. 
Generally, such a line of dips or peaks in physical quantities within a phase as an extension of a phase transition line or critical point is reminiscent of the well-known Widom or Fisher-Widom lines seen in supercritical fluid \cite{Xu2005,Franzese2007,Simeoni2010,Luo2014,Sordi2024} or near the metal-insulator critical point \cite{Sordi2012,Sordi2013,Vucicevic2013} in the equilibrium thermodynamic phase diagram. There, the Widom line, manifests as a peak in the correlation length or peaks/dips in various thermodynamic quantities like the specific heat, compressibility etc., may indicate a crossover or weak singularity or even in the presence of some hidden critical point \cite{Xu2005}.

Curiously, our work reveals the presence of Widom-like lines even in the dynamical phase diagram, presumably indicating a crossover within the localized phase due to the antecedent delocalized-critical phase transition line along $\mu=1$. Moreover, a weaker non-monotonic feature, in the form of a dip or peak in $\Delta_t$ [Fig.~\ref{selfenergy_noint}(c)] and dip in IPR [Fig.\ref{mipr_noint}(c)], is also seen as a function of $\lambda$ within the critical phase along $\lambda=2$ line, as shown in Fig.\ref{phase}(a). The $\lambda=2$ line is in the critical phase is an extension the delocalized-localized phase boundary as well as the self-dual line. This Widom-like line, shown in Fig.\ref{phase}(b), inside the critical phase becomes even more prominent in the interacting case, as we discuss in the next section. 



In addition to the self-energy, we also calculate the averaged FS IPR as shown in Fig.~\ref{mipr_noint}(a-d) for the same transitions shown in Fig.~\ref{selfenergy_noint}(a-d). Since, here we deal with noninteracting many-body eigenstate which can be written using single particle eigenfunctions as $\ket{\Psi}=\prod_{\nu=1}^{N} c_\nu^{\dagger}\ket{0}$ where $c_\nu=\sum_i \psi_\nu(i)c_i$ are the diagonalizing operator of the single particle Hamiltonian $H_{sp}$, the IPR can be written as $IPR=\sum_I |\braket{I|{\Psi}}|^4/(\sum_I |\braket{I|{\Psi}}|^2)^2$, where $\braket{I|{\Psi}}$'s are given by Slater determinants. This way we can avoid the exact diagonalization of the many body Hamiltonian and study the noninteracting system in FS for larger system sizes up to $L=28$. An alternative compact expression of IPR is given in Ref.~\onlinecite{turkeshi2024hilbert} that uses replica of the state concerned and also implementable using tensor network.  We choose the energy window and number of disorder realizations such that we average over at least a total of $5000$ eigenstates. The Widom-like line $\lambda=2$ is manifested as weak dip in IPR in the critical phase [Fig.~\ref{mipr_noint}(c)] and the Widom line $\mu=1$ appears as a broad peak in IPR in the localized phase [Fig.~\ref{mipr_noint}(d)]. 

\begin{figure}
\centering
\stackon{\includegraphics[width=0.49\columnwidth,height=3.5cm]{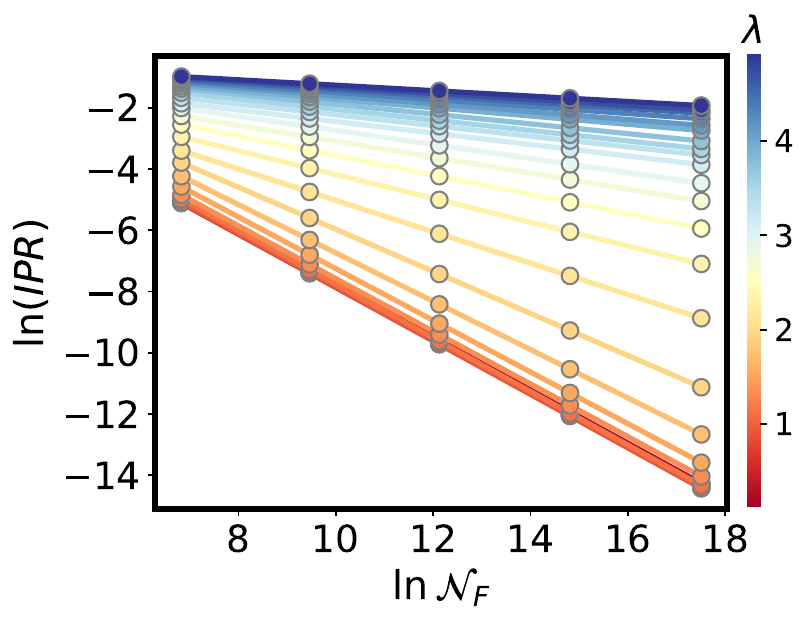}}{(a)}
\stackon{\includegraphics[width=0.49\columnwidth,height=3.5cm]{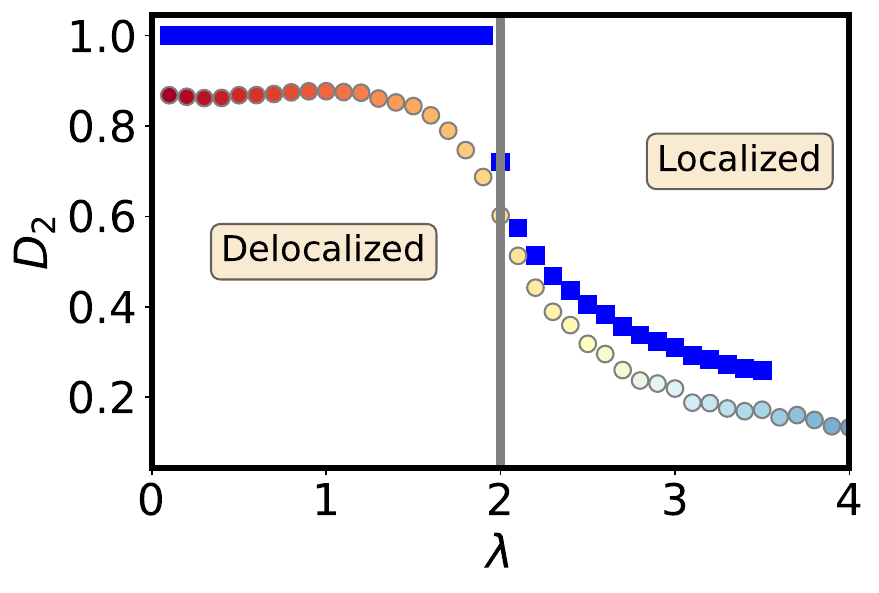}}{(b)}
\stackon{\includegraphics[width=0.49\columnwidth,height=3.5cm]{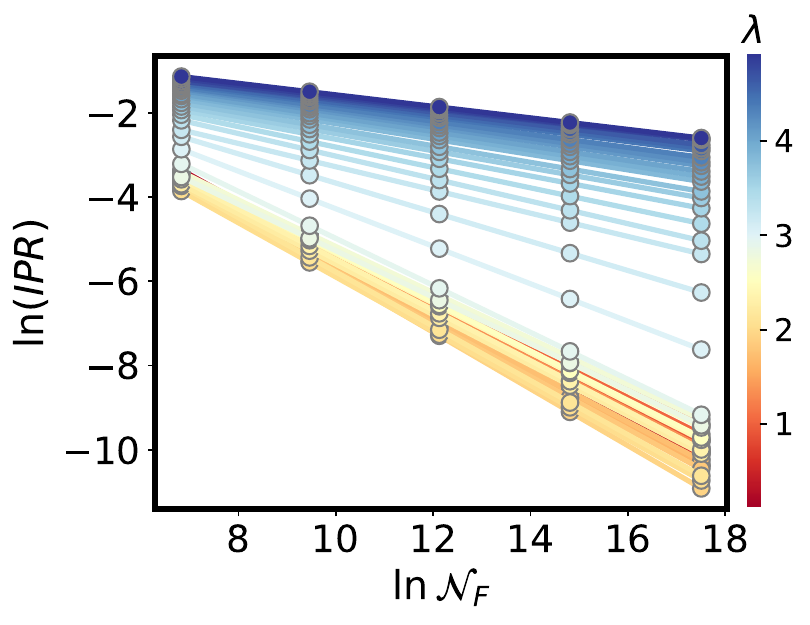}}{(c)}
\stackon{\includegraphics[width=0.49\columnwidth,height=3.5cm]{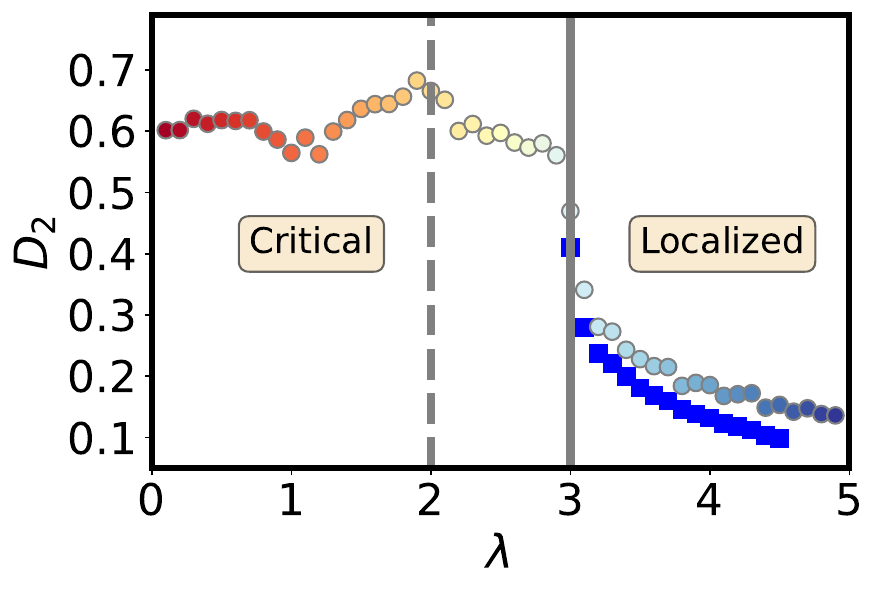}}{(d)}
\stackon{\includegraphics[width=0.49\columnwidth,height=3.5cm]{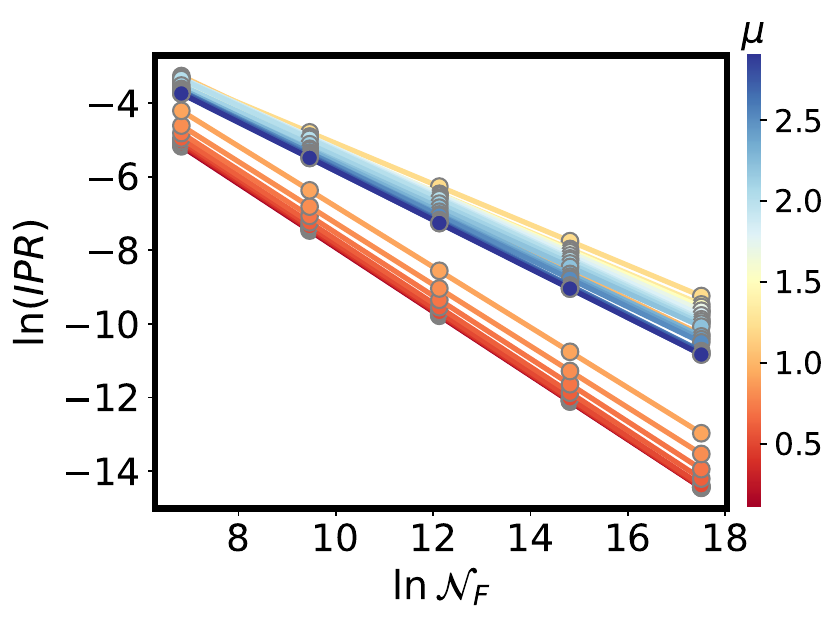}}{(e)}
\stackon{\includegraphics[width=0.49\columnwidth,height=3.5cm]{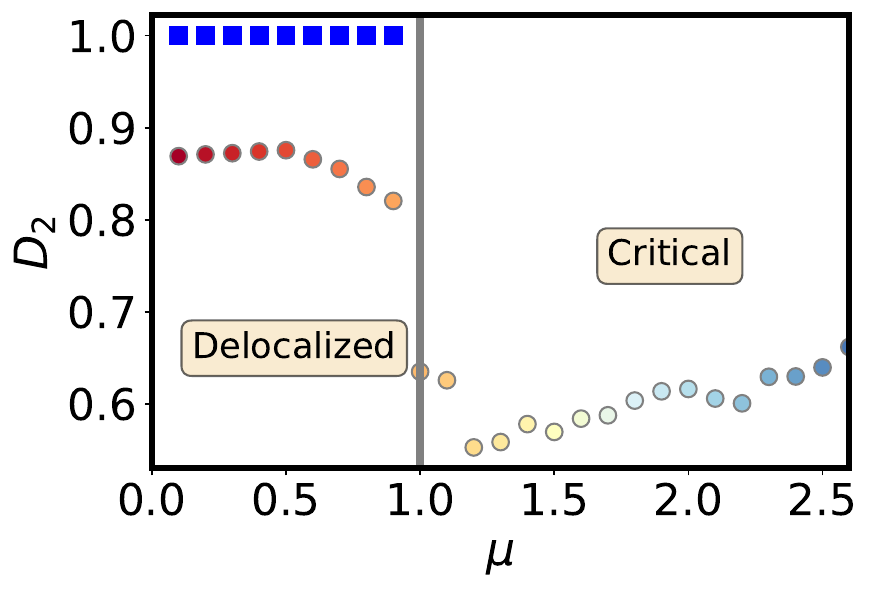}}{(f)}
\caption{{\bf Finite-size dependence of $IPR$ for noninteracting systems}: 
(a) Log-log plots of $IPR$, after linear fitting, as a function of FS dimension $\mathcal{N}_F$ across the delocalization-localization transition as a function of $\lambda$ for $\mu=0.5$.
(b) Spectral dimension $D_2$ extracted from $IPR=A_I \mathcal{N}_F^{-D_2}$ across the delocalized-localized transition in (a). 
(c,d) and (e,f) Similar figures as (a,b) but for the critical-localized transition as a function of $\lambda$ for $\mu=1.5$ and the delocalized-critical transition as a function of $\mu$ for $\lambda=1.0$, respectively.
Blue squares in figures (b,d,f) indicate $D_2$ obtained from the finite-size scaling analysis (not shown here) in the delocalized and localized phases. 
In figure (d) the vertical dashed line corresponds to the Widom line along $\lambda=2$ in the critical phase. 
Vertical solid lines denote phase transitions in all the plots.}
\label{mipr_D2_noint}
\end{figure} 
We extract the coefficient $A_I$ and the fractal dimension $D_2$ from the $IPR\simeq A_I\mathcal{N}_F^{-D_2}$ through a linear fitting of the data shown in Fig.~\ref{mipr_D2_noint}(a,c,e). 
From this analysis we conclude that $D_2\approx 1$, $A_I<1$ in the delocalized phase, whereas $0<D_2<1$, $A_I>1$ in the localized phase. The critical phase exhibits IPR in between the delocalized and localized phases, namely $0<D_2<1$ like the localized phase, but $A_I>1$ as in the delocalized phase. We also extract similar coefficients and exponents, $A_s$ and $D_s$, respectively, from $\Delta_t$, as summarized in Table \ref{table_results} (see also Appendix \ref{appB}). However, $\Delta_t$ is rather irregular as a function of $\mathcal{N}_F$ in the critical phase, unlike the IPR [Fig.~\ref{mipr_D2_noint}(c)] and, as a result, we could not extract $A_s$ and $D_s$ reliably in the critical phase.

In Figs.\ref{mipr_D2_noint}(b,d,f), we also plot $D_2$ obtained from the finite-size scaling analysis of the IPR in the delocalized and localized phases approaching the delocalized-localized, localized-critical and delocalized-critical transitions. The finite-size scaling ansatz~\cite{mace2019multifractal,garcia2017scaling,roy2021fock,sutradhar2022scaling,roy2023diagnostics,ghosh2024scaling}, discussed in detail in the next section for the interacting EAAH model, assumes a \emph{volumic} scaling for the delocalized phase and a linear scaling for the localized phase. As evident, $D_2$ estimated from finite-size scaling in the localized phase [Figs.\ref{mipr_D2_noint}(b,d)] agrees reasonably with the $D_2$ extracted by directly the fitting the data for $IPR$ as a function of $\mathcal{N}_F$ in the localized phase in Figs.\ref{mipr_D2_noint}(a,c). The volumic finite-size scaling in the delocalized phase assumes $D_2=1$ by default. This value is only expected in the asymptotic thermodynamic limit $\mathcal{N}_F\to \infty$. In Figs.\ref{mipr_D2_noint}(b,f), we see that $D_2$ extracted directly from the data in Figs.\ref{mipr_D2_noint}(a,e) for the delocalized phase deviates by $\sim 10-20\%$ from the asymptotic $D_2=1$ even deep inside the delocalized phase for the system sizes accessed in this work. Due to the non-monotonic, and somewhat irregular, behavior of the $IPR$ as a function of $\mathcal{N}_F$ and the parameters $\lambda$ or $\mu$ in the critical phase, $D_2$ could not be extracted from finite-size scaling in the critical phase, using either volumic or linear scalings.

Note that within the sizes that we are able to study $(L=28)$, the many-body noninteracting localized phase is multifractal in the FS, unlike in the case of Anderson localization in real-space which has a fractal dimension zero. Interestingly, we find that the change of $A_I$  (not shown) and $D_2$ [Figs.\ref{mipr_D2_noint}(d,f)] across the critical-localized and the delocalized-critical phase transitions are much more abrupt than those across the delocalized-localized transition, e.g., for $D_2$ in Fig.\ref{mipr_D2_noint}(b). The abrupt jumps of values of $D_2$ in Figs.\ref{mipr_D2_noint}(d,f) coincide very well with the critical-localized and delocalized-critical transitions estimated from other diagnostics~\cite{wang2021many}.  This provides us with confidence to study these quantities further the interacting systems, as discussed in the next section.

\subsection{Interacting systems}\label{sec4b}
\begin{figure}
\centering
\stackon{\includegraphics[width=0.493\columnwidth,height=3.5cm]{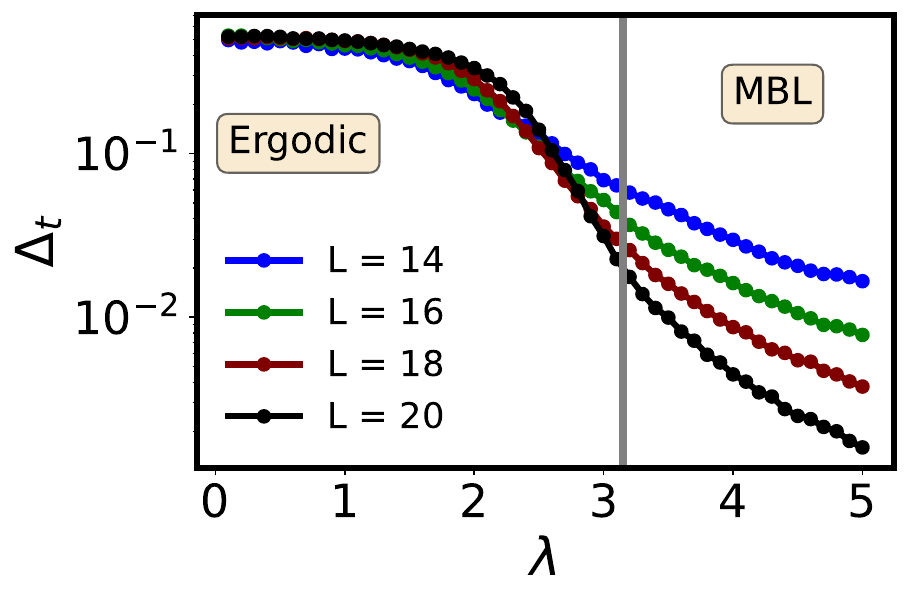}}{(a)}
\stackon{\includegraphics[width=0.493\columnwidth,height=3.5cm]{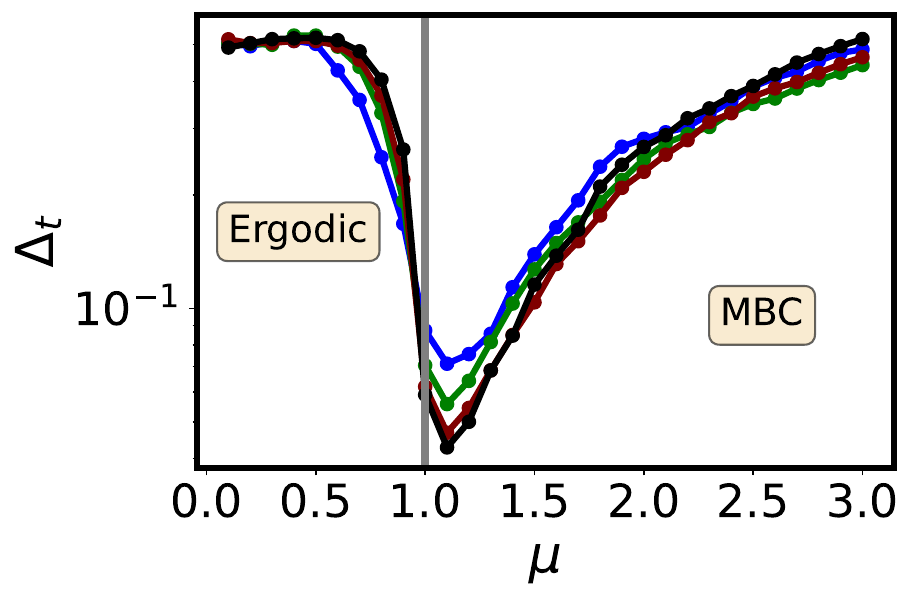}}{(b)}
\stackon{\includegraphics[width=0.493\columnwidth,height=3.5cm]{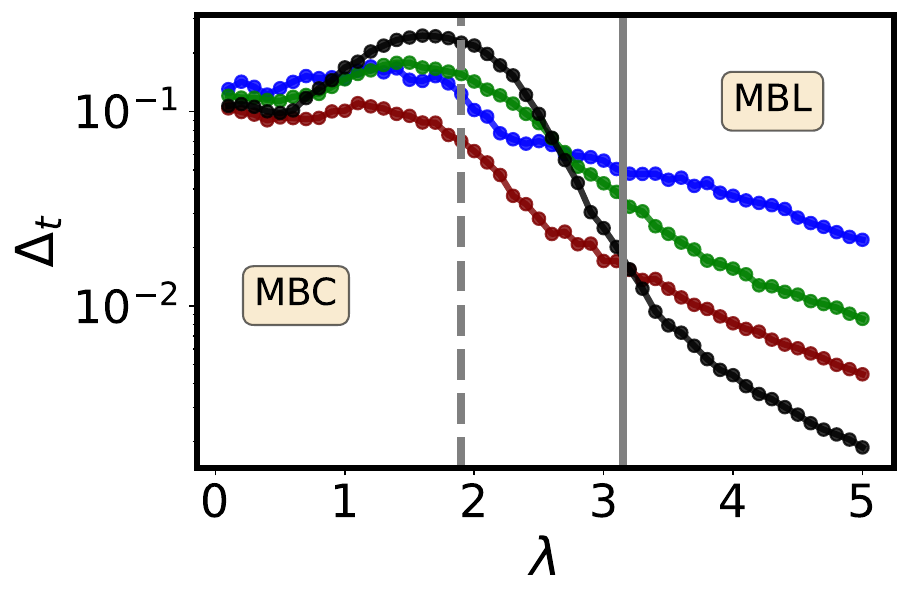}}{(c)}
\stackon{\includegraphics[width=0.493\columnwidth,height=3.5cm]{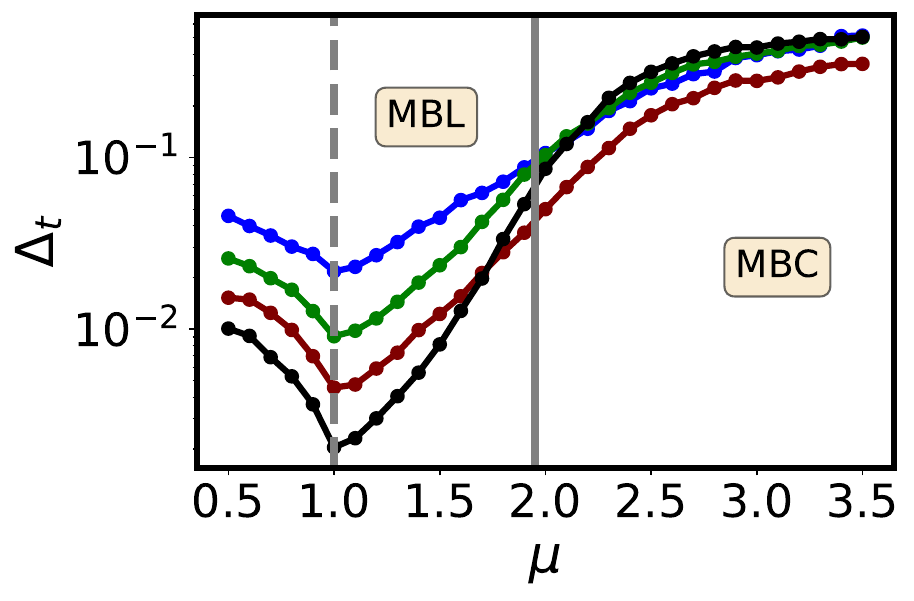}}{(d)}
\caption{{\bf Fock-space self-energy for interacting systems}: 
Typical values of self-energy $\Delta_t$ across
(a) the ergodic-MBL transition as a function of $\lambda$ for $\mu=0.5$, (b) ergodic-MBC 
transition as a function of $\mu$ for $\lambda=1.0$, (c) MBC-MBL transitions as a function of $\lambda$ for $\mu=1.5$, and (d) MBL-MBC transition as a function of $\mu$ for $\lambda=3.5$, for increasing system sizes $L$ . In figures (c) and (d), the vertical dashed lines denote the non-monotonic dependence of $\Delta_t$ and IPR on the parameters in MBC and MBL phases, respectively. The solid vertical lines denote the points of the phase transition. }
\label{selfenergy_int}
\end{figure}

Here we discuss our results for FS self-energy and IPR in the interacting EAAH model with $V=1$. In Fig.~\ref{selfenergy_int}(a-d), we show the variation of $\Delta_t$ of interacting systems for the same choices of parameters as in Fig.~\ref{selfenergy_noint}(a-d), respectively, involving all three phases [see Fig.~\ref{phase}(b)]. In Fig.~\ref{selfenergy_int}(a), $\Delta_t$ is shown across the ergodic-MBL phase transition, where $\Delta_t\sim \mathcal{O}(1)$ in the ergodic phase and rapidly decreases with $L$ in the MBL phase. The ergodic-MBC phase transition at $\mu=1$ is shown in \ref{selfenergy_int}(b), where $\Delta_t$ abruptly drops near the transition, like in the non-interacting system, and then increases again deeper in the MBC phase. Fig.~\ref{selfenergy_int}(c) and Fig.~\ref{selfenergy_int}(d) show variation of $\Delta_t$ across the MBC-MBL and MBL-MBC phase transitions, respectively. $\Delta_t$ in the MBC phase shows an anomalous $L$-dependence. In Fig.~\ref{selfenergy_int}(c), we see an enhanced delocalization tendency around $\lambda=2$ as a broad peak in the MBC phase. The peak in $\Delta_t$ appears as a dip in IPR, as we discuss below. The peak is indicative of the Widom line in the MBC phase, as shown in Fig.\ref{phase}(b). However, unlike in the non-interacting model where the Widom line comes out from the triple point $\lambda=2,\mu=1$, the locus of the peak in the MBC phase emanates from the ergodic-MBC phase boundary around $\lambda\approx 1.5,\mu=1$, away from the triple point ($\lambda\approx 3, \mu=1$) in the interacting EAAH model. However, deeper inside the MBC phase, the Widom line eventually merges with the $\lambda=2$ line, which is the Widom line in the critical phase of the non-interacting EAAH. Similarly, in the MBL phase, $\Delta_t$ has a minimum along the $\mu=1$ Widom line, implying stronger localization, as shown in Fig.~\ref{selfenergy_int}(d). A recent numerical work, based on the supervised machine learning approach to detect various phases of our many-body quasi-periodic system~\cite{ahmed2025supervised}, has also indicated the possibility of the  existence of crossovers around the lines $\lambda=2$ and $\mu=1$ within the MBC and MBL phases, respectively .
\begin{figure}
\centering
\stackon{\includegraphics[width=0.493\columnwidth,height=3.5cm]{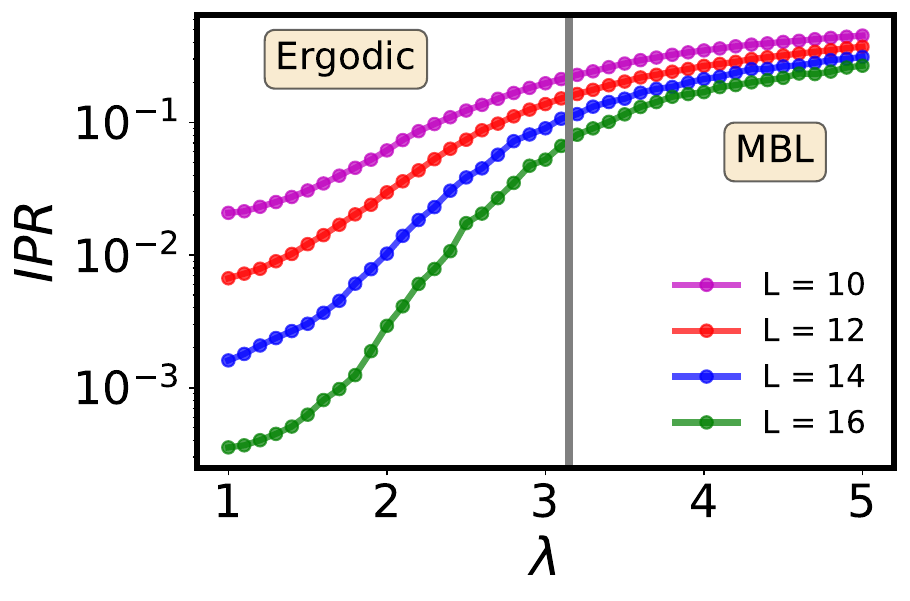}}{(a)}
\stackon{\includegraphics[width=0.493\columnwidth,height=3.5cm]{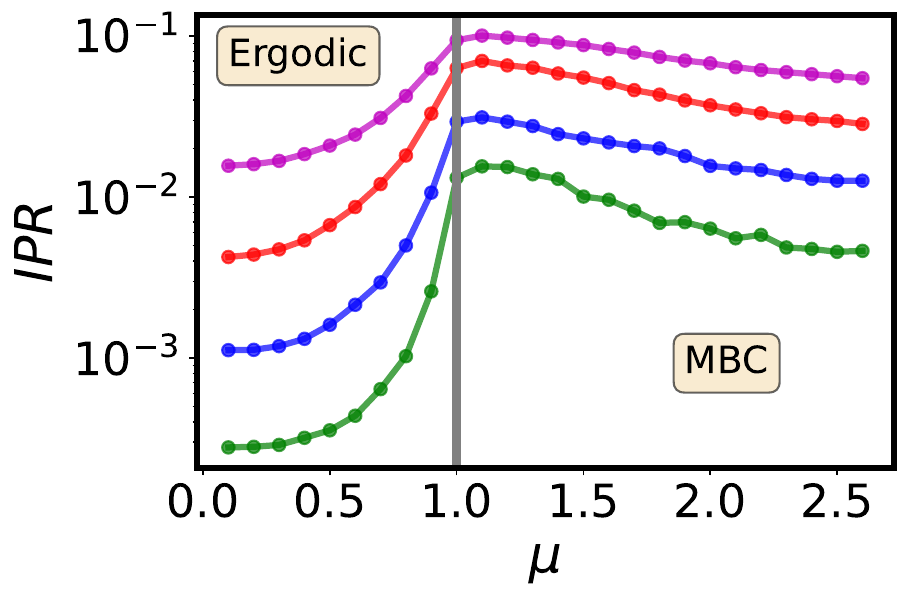}}{(b)}
\stackon{\includegraphics[width=0.493\columnwidth,height=3.5cm]{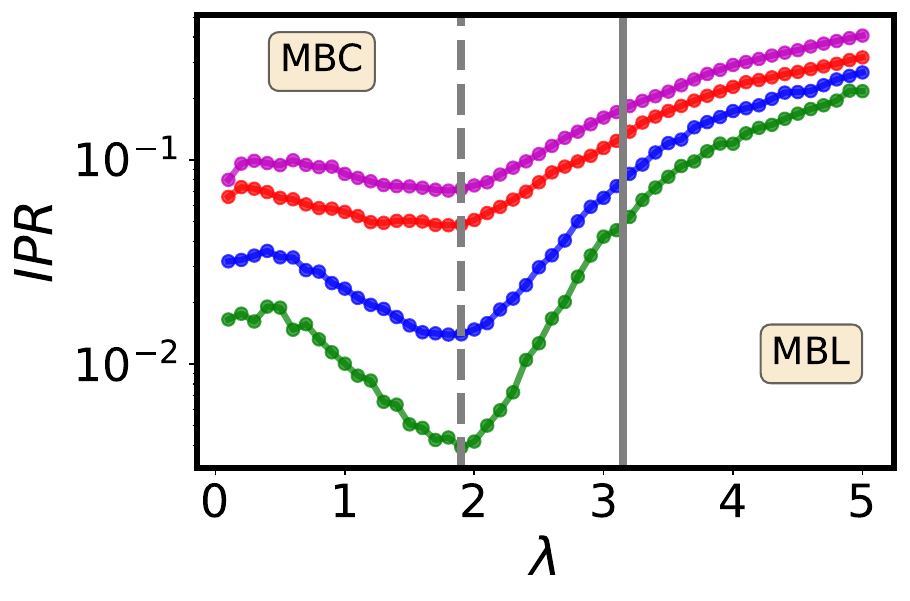}}{(c)}
\stackon{\includegraphics[width=0.493\columnwidth,height=3.5cm]{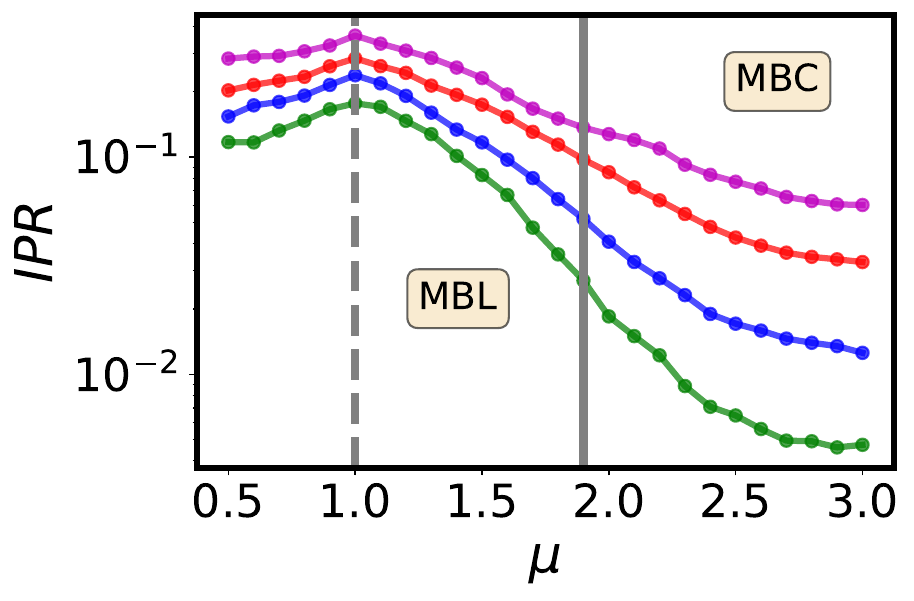}}{(d)}
\caption{{\bf Fock-space $IPR$ for interacting systems}: $IPR$ across (a) the ergodic-MBL transition as a function of $\lambda$ for $\mu=0.5$, (b) ergodic-MBC 
transition as a function of $\mu$ for $\lambda=1.0$, (c) MBC-MBL transitions as a function of $\lambda$ for $\mu=1.5$, and (d) MBL-MBC transition as a function of $\mu$ for $\lambda=3.5$, for increasing system sizes $L$.
In figures (c) and (d), the vertical dashed lines denote the peak/dip of $\Delta_t$ across the Widom lines in Fig.\ref{phase}(b). The solid vertical lines denote the phase transitions. 
}
\label{mipr_int}
\end{figure} 
\begin{figure}
\centering
\stackon{\includegraphics[width=0.49\columnwidth,height=3.5cm]{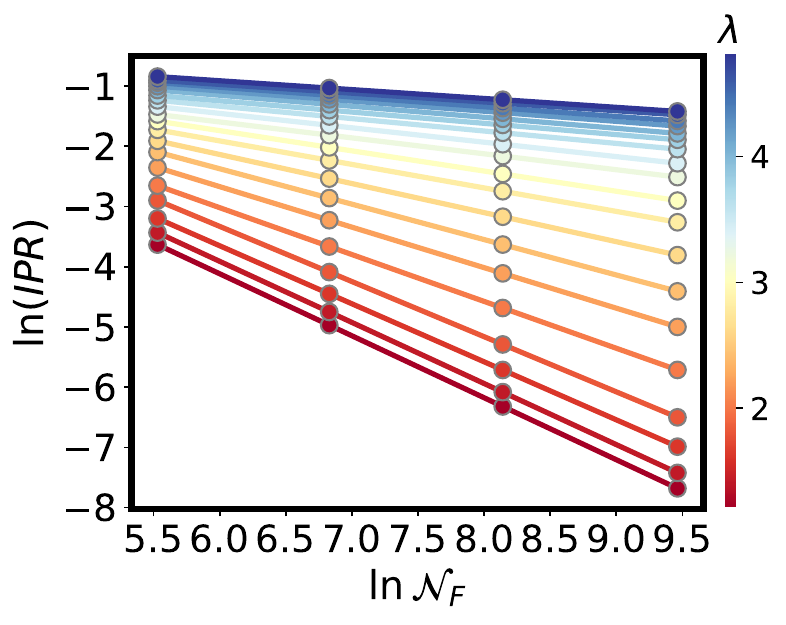}}{(a)}
\stackon{\includegraphics[width=0.49\columnwidth,height=3.5cm]{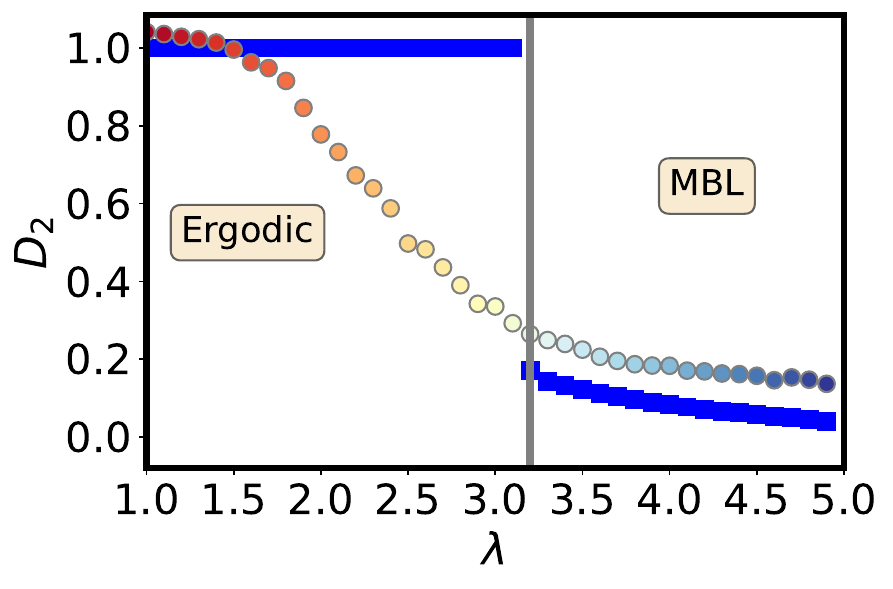}}{(b)}
\stackon{\includegraphics[width=0.49\columnwidth,height=3.5cm]{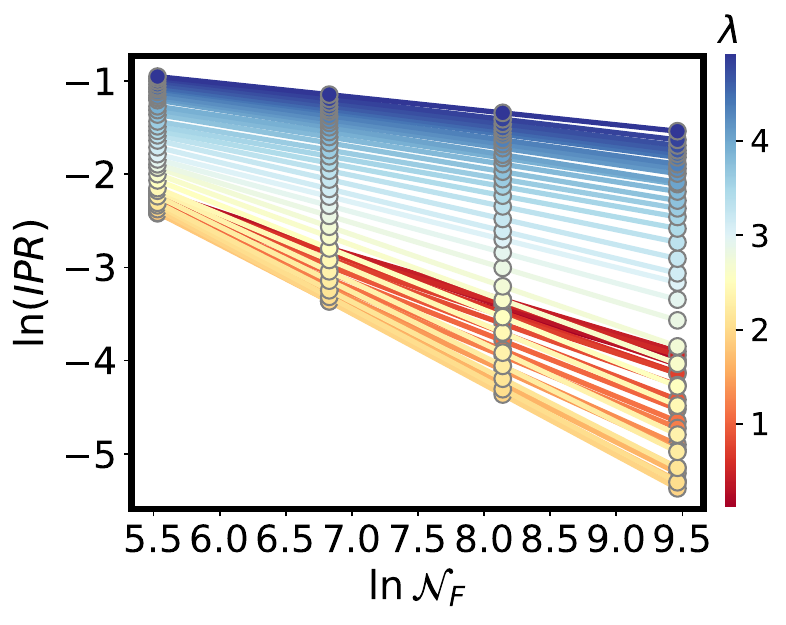}}{(c)}
\stackon{\includegraphics[width=0.49\columnwidth,height=3.5cm]{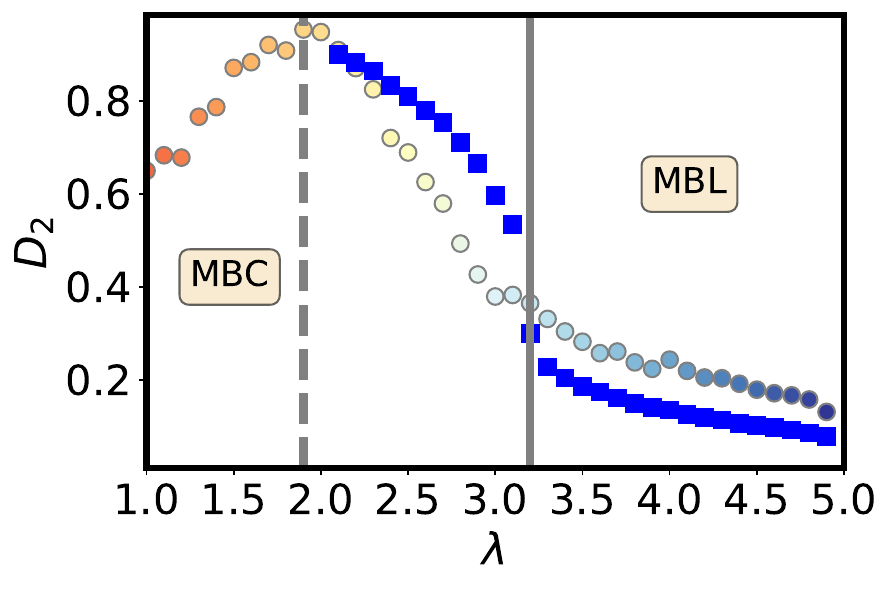}}{(d)}
\stackon{\includegraphics[width=0.49\columnwidth,height=3.5cm]{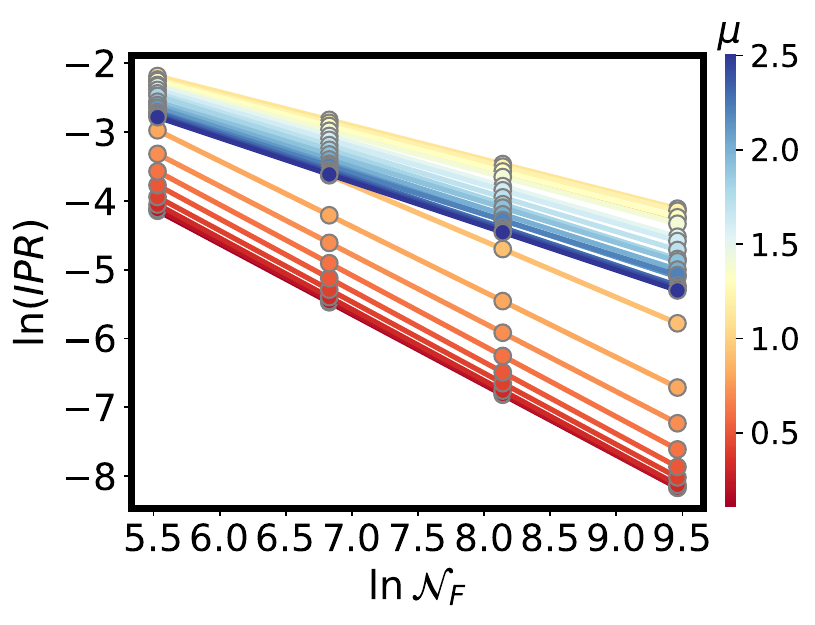}}{(e)}
\stackon{\includegraphics[width=0.49\columnwidth,height=3.5cm]{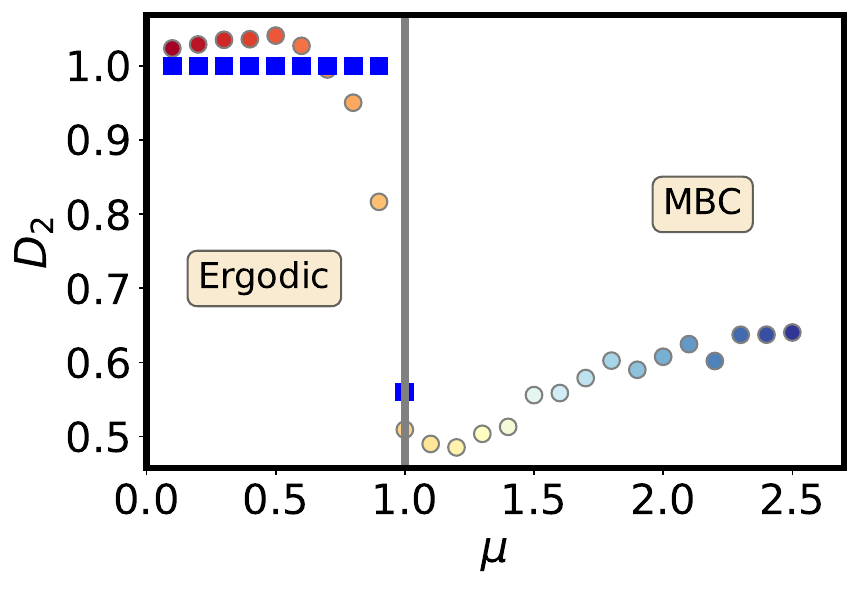}}{(f)}
\caption{{\bf Finite-size dependence of $IPR$ for interacting systems}: (a) Log-log plots of $IPR$, after linear fitting, as a function of FS dimension $\mathcal{N}_F$ across the ergodic-MBL transition with $\lambda$ at $\mu=0.5$.
(b) Spectral dimension $D_2$ extracted from $IPR=A_I \mathcal{N}_F^{-D_2}$ for the ergodic-MBL transition in (a). 
(c,d) and (e,f) Similar figures as (a,c) for MBC-MBL transition as a function of $\lambda$ at $\mu=1.5$ and ergodic-MBC transition as function of $\mu$ at $\lambda=1.0$. The vertical solid lines denote the phase transitions in all plots. $D_2$ obtained from the finite-size scaling analysis of Fig.~\ref{mipr_scaling} is shown with blue squares in figures (b,d,f). In figure (d) the vertical dashed line corresponds to the Widom line along $\lambda=2$ in the critical phase.
}
\label{mipr_D2_int}
\end{figure} 

We now discuss our results for FS IPR, which is shown in Fig.~\ref{mipr_int}(a-d) for the same transitions shown in Fig.~\ref{selfenergy_int}(a-d), respectively. Here we employ exact diagonalization of the interacting Hamiltonian. As a result, we are able to study the system only up to a system size $L=16$. In all the three phases, i.e., ergodic, MBL and MBC, the $IPR$ decreases with increasing $L$. The non-monotonicity across the Widom lines [Fig.\ref{phase}(b)] are again visible in IPR in Figs.\ref{mipr_int}(c) and (d). 
The $IPR$ as a function of FS dimension $\Nf$ is shown in Fig.~\ref{mipr_D2_int}(a,c,e) for the same parameters as in Fig.~\ref{mipr_int}(a,b,c), respectively. 
The coefficient $A_I$ and $D_2$ [Figs.\ref{mipr_int}(b,d,f)] are obtained through a linear fitting of the data in Fig.~\ref{mipr_D2_int}(a,c,e). We find that $(D_2\approx1,A_I>1)$, $(0<D_2<1,A_I<1)$ and $(0<D_2<1,A_I>1)$ in the ergodic, MBL and MBC phases, respectively. Similar results are obtained for $A_s$ and $D_s$ extracted from $\Delta_t$ (Table \ref{table_results}, see also Appendix \ref{appB}), except in the MBC phase where we could not extract $A_s$ and $D_s$ to the non-monotonic dependence of $\Delta_t$ on $\mathcal{N}_F$. The signature of the Widom line along $\lambda=2$ is visible as a broad peak in $D_2$ as a function of $\lambda$ in Fig.~\ref{mipr_D2_int}(d). The solid vertical lines in Fig.~\ref{mipr_D2_int}(b,d,f) denote the phase transitions at the critical values $\lambda_c\approx3.2$ ($\mu=0.5$), $\lambda_c\approx3.2$ ($\mu=1.5$) and $\mu_c\approx1.0$ ($\lambda=1$) obtained from the finite-size scaling analysis, as we discuss next. 


For the finite-size scaling analysis of $IPR$ across the different transitions of the interacting system, we consider the volumic and linear scaling ansatzs~\cite{mace2019multifractal,garcia2017scaling,roy2021fock} which are given by,
\begin{eqnarray}
-\ln\frac{I_t}{I_c}=
\begin{cases}
      \mathcal{F}_{vol}\big(\frac{\mathcal{N}_F}{\Lambda}\big)  & ~~:~ \text{Ergodic}
      \\
      \mathcal{F}_{lin}\big(\frac{\ln\mathcal{N}_F}{\xi}\big)  & ~~:~ \text{MBC, MBL}.
    \end{cases} 
\label{scaling_ansatz}    
\end{eqnarray}
\begin{figure}
\centering
\includegraphics[width=0.8\columnwidth,height=6.9cm]{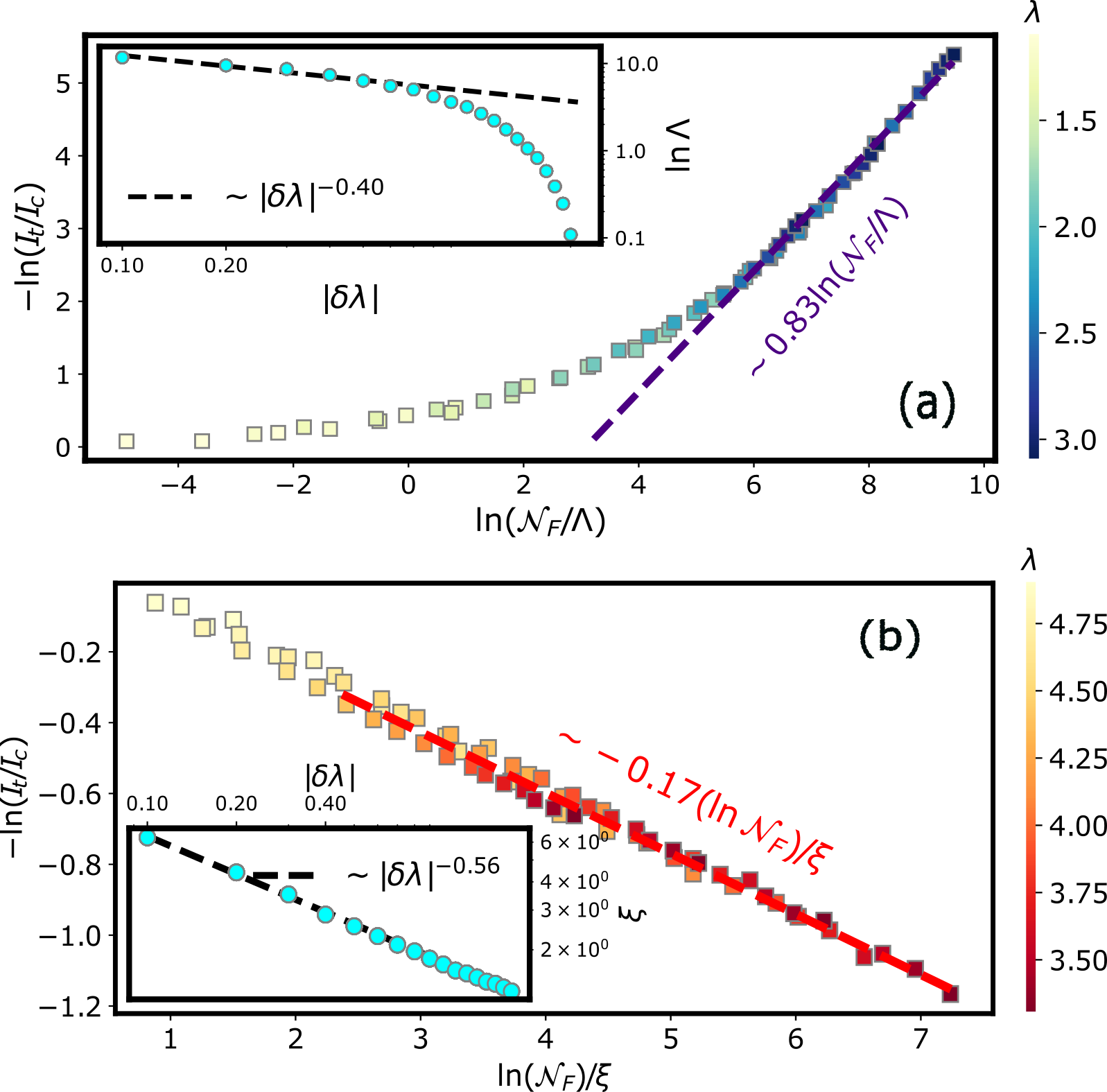} \\
\vspace{0.3cm}
\includegraphics[width=0.8\columnwidth,height=6.9cm]{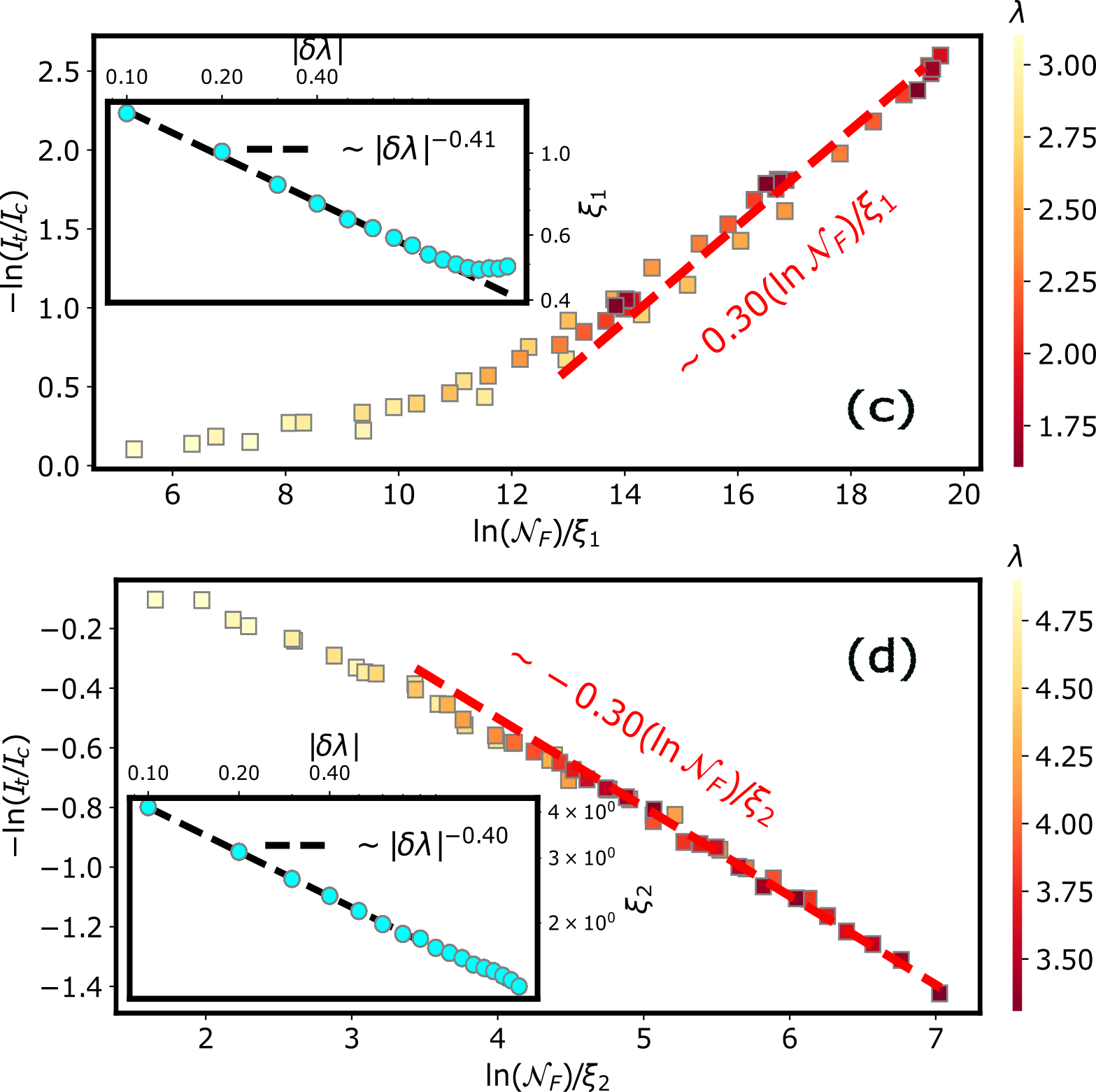} \\
\vspace{0.3cm}
\includegraphics[width=0.8\columnwidth,height=3.45cm]{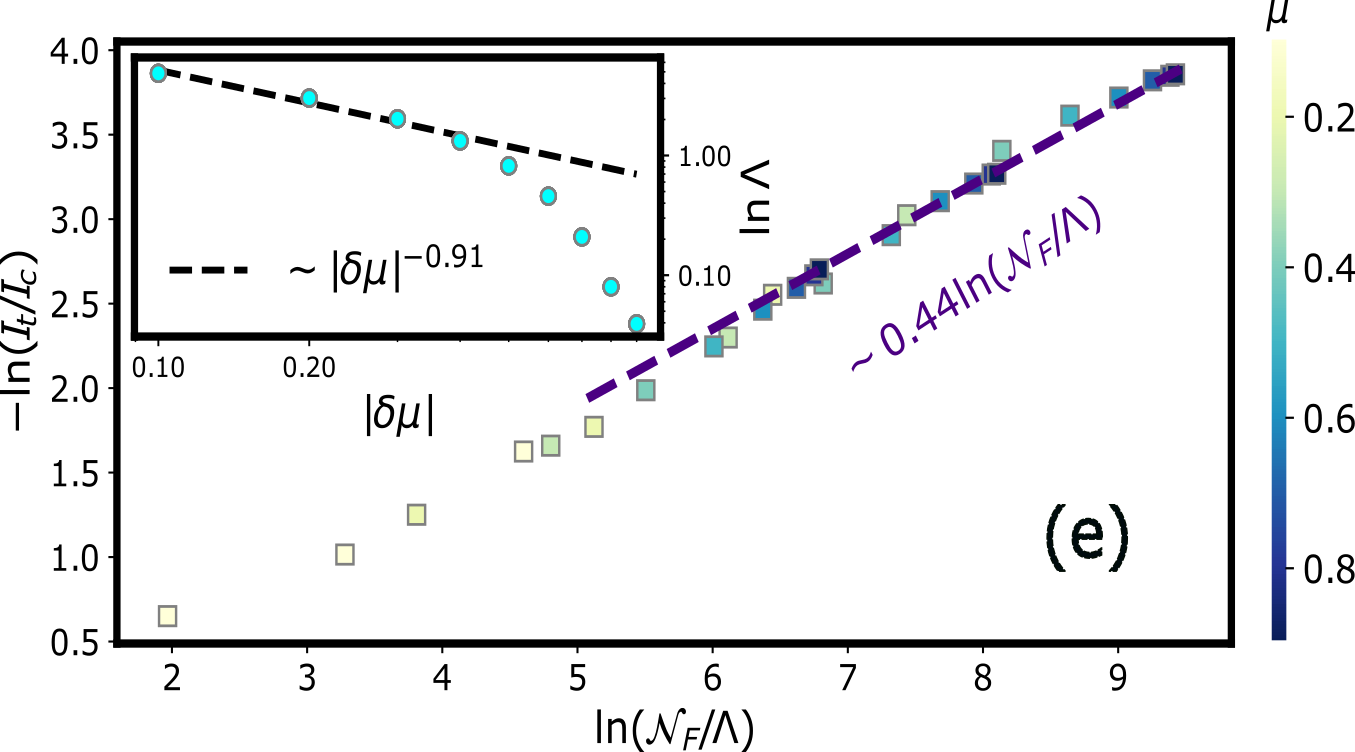}
\caption{{\bf Scaling collapse of $IPR$ for interacting systems}: (a,b) {\bf Ergodic-MBL transition:} Volumic scaling and linear scaling of $IPR$ in the ergodic and MBL phases, respectively, for the ergodic-MBL transition ($\lambda_c=3.2$) as a function of $\lambda$ for $\mu=0.5$ corresponding to Fig.~\ref{mipr_D2_int}(a,b). (c,d) {\bf MBL-MBC transition:} Linear scaling of $IPR$ in the MBC and MBL phases, respectively for the MBC-MBL transition $(\lambda_c=3.2)$ as a function of $\lambda$ for $\mu=1.5$, corresponding to Fig.~\ref{mipr_D2_int}(c,d). (e) {\bf Ergodic-MBC transition:} Volumic scaling of $IPR$ in ergodic phase for the ergodic-MBC phase transition $(\mu_c=1.0)$ as a function of $\mu$ for $\lambda=1.0$, corresponding to Fig.~\ref{mipr_D2_int}(e,f). Insets show the KT-like divergence of the non-ergodic volume $\Lambda$ and the power-law divergence of correlation length $\xi$ near phase transitions in case of volumic and linear scaling, respectively. }
\label{mipr_scaling}
\end{figure} 
Here $I_t$ is the typical value of IPR that depends on $\mathcal{N}_F$ and other parameters like $\lambda,~\mu$; $I_t=I_c$ at the critical point. In the thermodynamic limit $\mathcal{N}_F\to \infty$, we expect $I_t\sim \Nf^{-D_2}$ ($I_c\sim \Nf^{-D_{2c}}$) with an exponent $D_2$ ($D_{2c}$). According to the scaling ansatz, the scaling form inside the ergodic phase is volumic controlled by a `nonergodic volume' $\Lambda$ that diverges when the ergodic system approaches a phase transition towards a nonergodic phase. On the other hand, in the nonergodic phases, like MBC and MBL phases, we assume a linear scaling form controlled by a length scale $\xi$ that diverges as the system approaches a non-ergodic to ergodic phase transition. The scaling ansatzs of Eq.\eqref{scaling_ansatz} can also be used to study the FS IPR in the non-interacting EAAH model across the delocalized-localized, critical-localized, and delocalized-critical transitions. For example, we have used the volumic scaling in the delocalized phase and the linear scaling in the localized and critical phases to extract the fractal dimension $D_2$ (see the discussion below), as shown in Figs.\ref{mipr_D2_noint}(b,d,f).

The asymptotic behavior \cite{roy2021fock,sutradhar2022scaling} of volumic scaling function for $IPR$ [Eq.\eqref{scaling_ansatz}] in the ergodic phase for $\mathcal{N}_F\gg \Lambda$ is given by $\mathcal{F}_{vol}(\mathcal{N}_F/\Lambda)\sim (1-D_{2c}) \ln{(\mathcal{N}_F/\Lambda)}$ where we assume $D_2=1$ for the ergodic phase. The asymptotic form of the linear scaling function in the limit $\ln{N_F}\gg \xi$ in both the nonergodic phases is given by $\mathcal{F}_{lin}(\ln\mathcal{N}_F/\xi)\sim -D_{2c} \ln{(\mathcal{N}_F)}/\xi$ where $\xi=D_{2c}/(D_{2c}-D_2)$. Thus, the values of $\xi$ extracted as a function of $\lambda$ or $\mu$ from the finite-size scaling collapse in the MBL and MBC phases, can be used to obtain the fractal dimension $D_2$ in these non-ergodic phases. We also use the above asymptotic scaling functions to obtain estimates of $D_{2c}$ by approaching the transitions from either sides. For example, for the ergodic-MBL transition we extract $D_{2c}$ using the volumic scaling from the ergodic side, and using the linear scaling from the MBL side. Similarly, we use two different linear scaling collapses for the MBC-MBL transition, one from the MBC, and the other from the MBL side. In all the cases, where reliable scaling collapses can be obtained and $D_{2c}$ can be estimated independently in two different sides of the transition, $D_{2c}$ values match. This gives confidence in our finite-size scaling procedure.

{\bf Ergodic-MBL transition}.-- In \textcolor{green}{Figs.~\ref{mipr_scaling}(a,b)} we show the volumic and linear finite-size scaling collapses in the ergodic and MBL phases, respectively, for the ergodic-MBL transition as a function of $\lambda$ at $\mu=0.5$, as shown in Fig.~\ref{mipr_D2_int}(a,b). The quality of the collapse is estimated from the $\chi^2$ difference between the data points and fitted scaling functions. The best quality of the volumic scaling collapse shown in Fig.~\ref{mipr_scaling}(a) is obtained for the critical value $\lambda_c=3.2$. From the extracted scaling function in the ergodic phase [Fig.~\ref{mipr_scaling}(a)], we find $\mathcal{F}_{vol}(\mathcal{N}_F/\Lambda\gg 1)\sim 0.83 \ln{(\mathcal{N}_F/\Lambda)}$, implying $D_{2c}\approx0.17$. The extracted non-ergodic volume follows $\Lambda\sim \exp{[b(\lambda_c-\lambda)^{-0.4}]}$ [Fig.~\ref{mipr_scaling}(a)(inset)] with $b\sim \mathcal{O}(1)$. In the MBL phase, from Fig.~\ref{mipr_scaling}(b), we find $\mathcal{F}_{lin}(\ln\mathcal{N}_F/\Lambda)\sim -0.17 \ln{(\mathcal{N}_F)}/\xi$, i.e., with the same value of $D_{2c}\simeq 0.17$ as obtained from the ergodic side. The extracted length scale $\xi$ [Fig.~\ref{mipr_scaling}(b)(inset)] diverges as $\xi\sim |\lambda-\lambda_c|^{-0.5}$. Similar exponents have been reported in a recent study on ergodic-MBL transition in quasiperiodic AAH chain using imaginary part of FS self-energy~\cite{ghosh2024scaling}. 

{\bf MBC-MBL transition}.-- \textcolor{green}{Figs.~\ref{mipr_scaling}(c,d)} demonstrate the finite-size scaling analysis for the MBC-MBL phase transition of Fig.~\ref{mipr_D2_int}(c,d) as a function of $\lambda$ at $\mu=1.5$. We find that the linear scalings are governed by correlation lengths $\xi_1$ and $\xi_2$ in the MBC and MBL phases, respectively. The analysis leads to $\lambda_c=3.2$, $D_{2c}\approx0.3$ and $\xi_1, \xi_2 \sim |\lambda-\lambda_c|^{-0.4}$. 

{\bf Ergodic-MBC transition}.--\textcolor{green}{Fig.~\ref{mipr_scaling}(e)} shows the volumic scaling in the ergodic phase corresponding to ergodic-MBC transition with $\mu$ at $\lambda=1.0$, as depicted in Fig.~\ref{mipr_D2_int}(e,f). The data collapse in the ergodic phase indicates that $\mathcal{F}_{vol}(\mathcal{N}_F/\Lambda)\sim 0.44 \ln{(\mathcal{N}_F/\Lambda)}$ with  $\mu_c=1$, $D_{2c}\approx0.56$ and $\Lambda\sim \exp{[b_2(\mu_c-\mu)^{-0.9}]}$ with $b_2\sim \mathcal{O}(1)$. Based on the $\chi^2$ analysis, a reasonable collapse of the data in the MBC phase was not possible to achieve. Our scaling analysis of $IPR$ shows that the asymptotic form of the volumic scaling, governed by $\Lambda$, in the ergodic phase can change substantially depending on whether the ergodic-to-non-ergodic transition is approached from ergodic to a MBL phase or a MBC phase.

Scaling ansatz \cite{sutradhar2022scaling,roy2023diagnostics,ghosh2024scaling} similar to Eq.\eqref{scaling_ansatz} can be written for the FS self-energy $\Delta_t$. Using these scaling ansatz for $\Delta_t$, we have also performed the finite-size scaling analysis for the data of self-energy across ergodic-MBL, ergodic-MBC and MBC-MBL transitions in Figs.\ref{selfenergy_int}(a,b,c,d). The analysis across the ergodic-MBL transition, as discussed in Appendix \ref{appB}, leads to qualitatively similar results as in the earlier studies on random and other quasiperiodic systems\cite{sutradhar2022scaling,roy2023diagnostics,ghosh2024scaling}. However, due to more irregularity and non-monotonicity of $\Delta_t(\mathcal{N}_F,\lambda,\mu)$ compared to that in $IPR$ in the MBC phase, we could not obtain reliable data collapse across the transitions to MBC phase. Hence, we have only discussed the finite-size scaling analysis of $IPR$ above. The distributions of IPR and $\Delta_I$ for different phases show distinct features which are discussed in Appendix~\ref{appC}.


\section{Fock-space localization length}\label{sec5}
\begin{figure}[h!]
\centering
\stackon{\includegraphics[width=0.49\columnwidth,height=3.5cm]{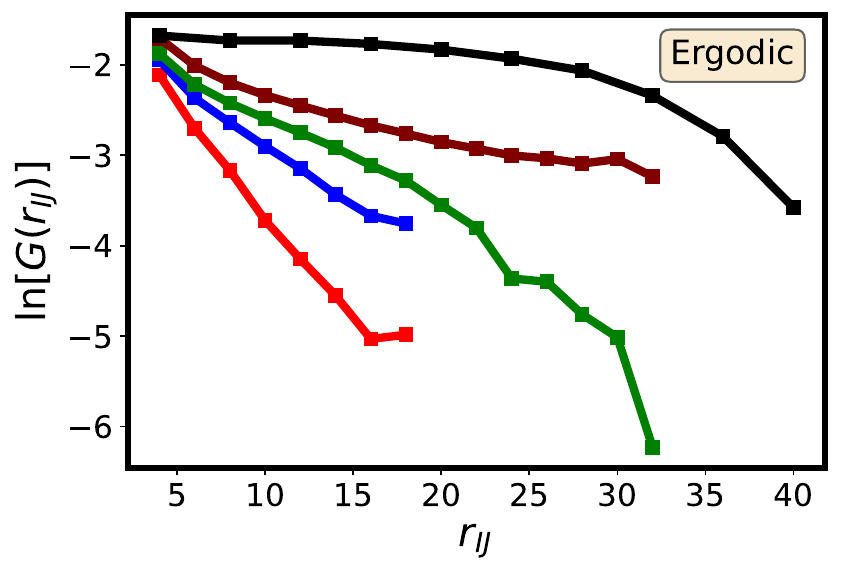}}{(a)}
\stackon{\includegraphics[width=0.49\columnwidth,height=3.5cm]{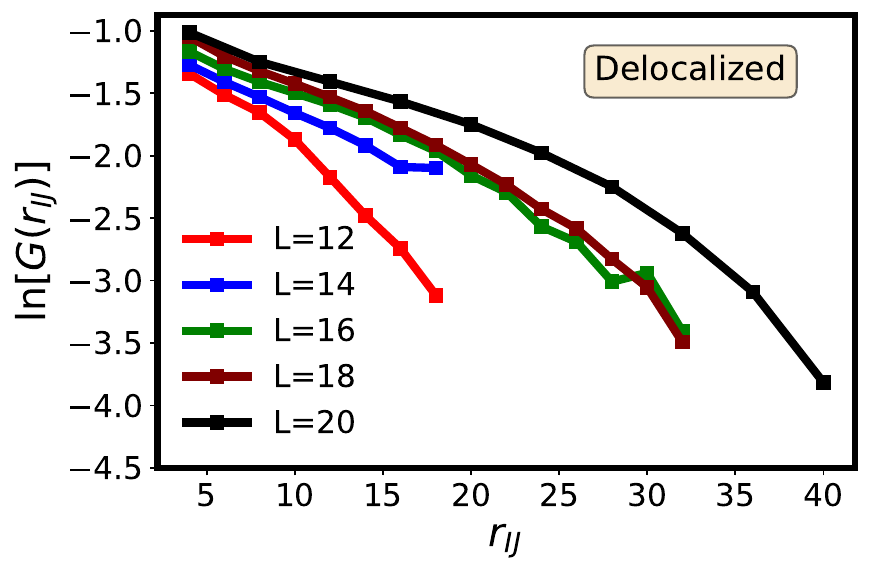}}{(d)}
\stackon{\includegraphics[width=0.49\columnwidth,height=3.5cm]{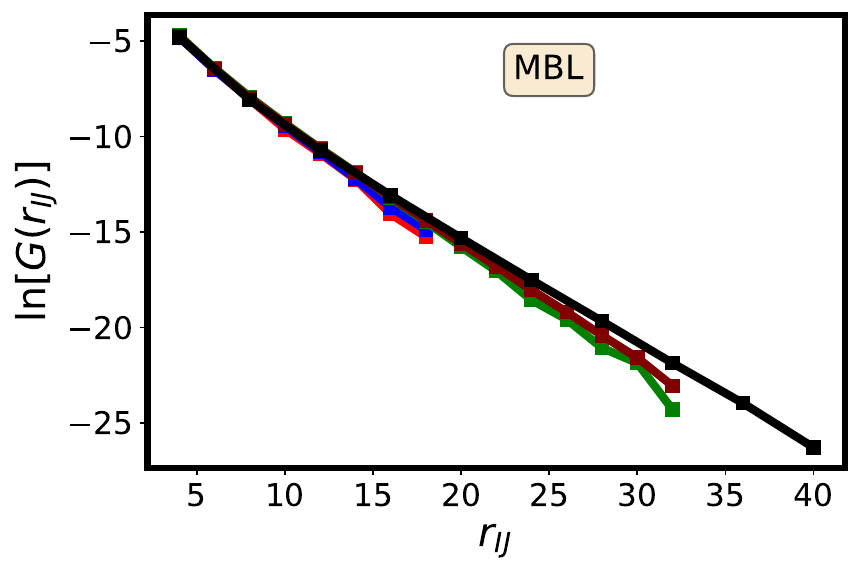}}{(b)}
\stackon{\includegraphics[width=0.49\columnwidth,height=3.5cm]{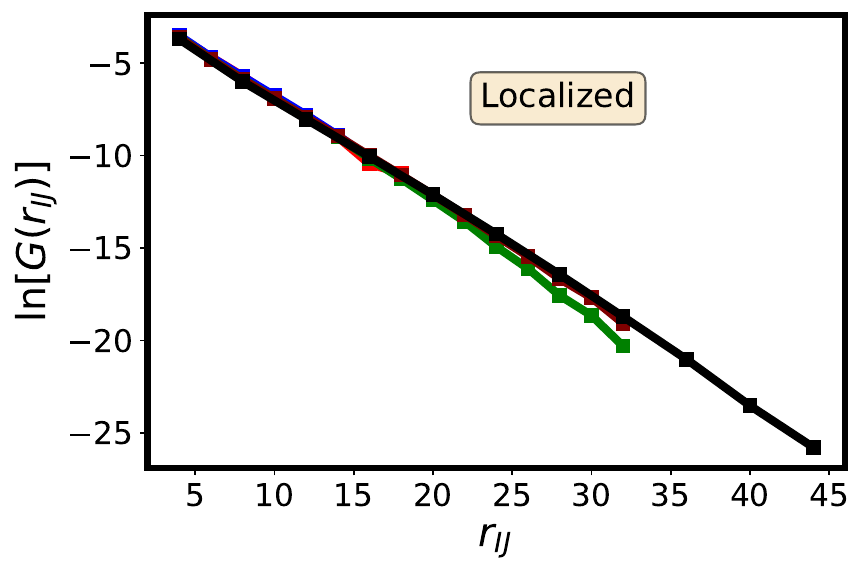}}{(e)}
\stackon{\includegraphics[width=0.49\columnwidth,height=3.5cm]{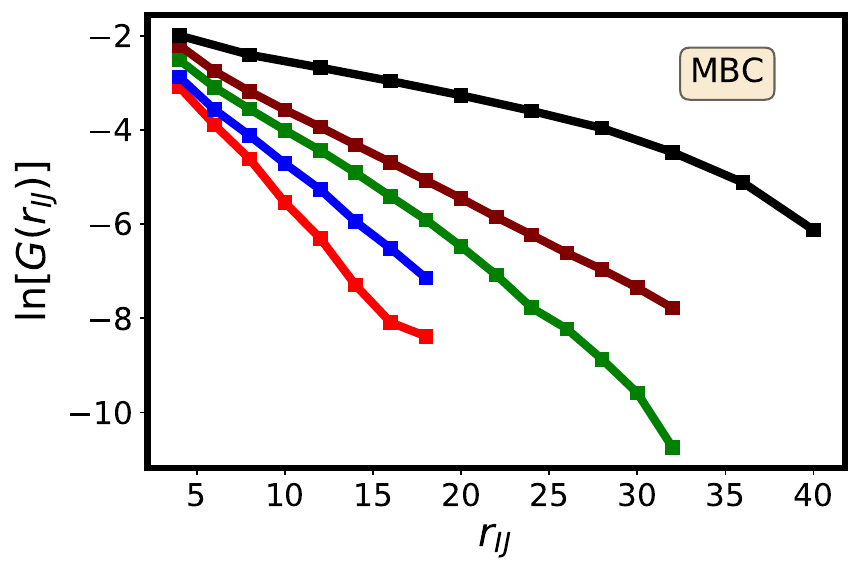}}{(c)}
\stackon{\includegraphics[width=0.49\columnwidth,height=3.5cm]{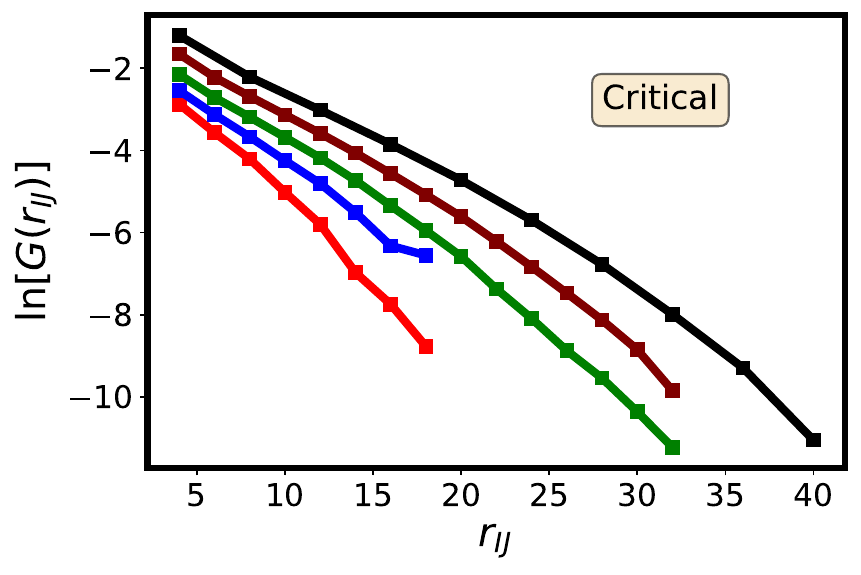}}{(f)}
\caption{{\bf Decay of two-point correlations in Fock-space}: (a-c) Decay of typical values $G(r_{IJ})$ of correlations with FS-hopping distance $r_{IJ}$ in the ergodic ($\lambda=1.0,~\mu=0.5$), MBL ($\lambda=5.0,~\mu=0.5$) and MBC ($\lambda=1.0,~\mu=1.5$) phases, respectively, of the interacting EAAH model. (d-f) Similar plots in the delocalized, localized and critical phases, respectively, of noninteracting model.}
\label{lnGij_loc}
\end{figure}

In this section, we analyze the off-diagonal elements $G_{IJ}(E=0)$ of the FS propagator~\cite{roy2023diagnostics,ghosh2024scaling} in the ergodic, MBL and MBC phases of the interacting EAAH model, and compare them with the phases of the non-interacting model. We calculate the geometric mean or the typical value, defined as $G(r_{IJ})=\text{exp}[\langle \ln G_{IJ}\rangle]$, where $\langle...\rangle$ denotes an average over
$\phi$ and all the off-diagonal elements $G_{IJ}$ for a pair of FS sites are at a hopping distance $r_{IJ}$. The latter is the minimum number of nearest-neighbor hops to go from
$I$ to $J$ in the middle slice. The number of realizations used for $\phi$ are $2000,1000,400,200,100$ for $L=12,14,16,18,20$, respectively.
Plots of $\ln[G(r_{IJ})]$ as a function of $r_{IJ}$ are shown in Figs.~\ref{lnGij_loc}(a-c) and Figs.~\ref{lnGij_loc}(d-f) for the phases of the interacting and non-interacting EAAH model.
In all the phases, the plots show a linear regime with negative slope, i.e., $\ln[G(r_{IJ})] \propto -r_{IJ}$, before it deviates from linearity depending on $L$. This deviation from the linear regime corresponds to rare hopping distances and associated multiple length scales in FS~\cite{roy2023diagnostics,ghosh2024scaling}. The linear regime implies the existence of a decay length $\xi_F(L)$, dubbed as the `FS localization length'~\cite{roy2023diagnostics,ghosh2024scaling}, such that $G(r_{IJ})\sim\text{exp}[-r_{IJ}/\xi_F(L)]$. 

In the ergodic phase, $\xi_F(L)$ rapidly increases with $L$ [Fig.~\ref{lnGij_loc}(a)]. In comparison, $\xi_F(L)$ increases much more slowly with $L$ in the non-interacting delocalized phase [Fig.~\ref{lnGij_loc}(d)]. Similarly, in the MBC phase, $\xi_F(L)$ shows more rapid increases with $L$ [Fig.~\ref{lnGij_loc}(c))] compared to $\xi_F(L)$ in the non-interacting critical phase [Fig.~\ref{lnGij_loc}(f)]. The faster increase of the FS localization length with $L$ in the interacting model presumably reflects the effect of interaction that induces more correlation in the FS lattice, leading to a greater  tendency to localize. In contrast to the ergodic (delocalized) and MBC (critical) phases, in the MBL (localized) phase, $\xi_F(L)$ for different values of $L$ overlap up to a certain $L$-dependent value of $r_{IJ}$ and then show a weak dependence on $L$ [Fig.~\ref{lnGij_loc}(b) (Fig.~\ref{lnGij_loc}(e))]. The non-interacting AL phase shows a more stable linear regime and weaker dependence on $L$. 
\begin{figure}[h!]
\centering
\stackon{\includegraphics[width=0.49\columnwidth,height=3.5cm]{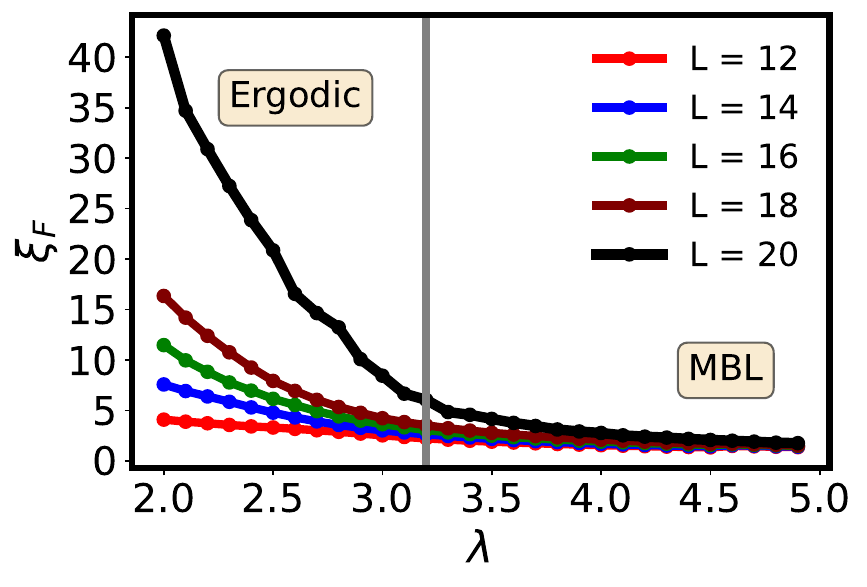}}{(a)}
\stackon{\includegraphics[width=0.49\columnwidth,height=3.5cm]{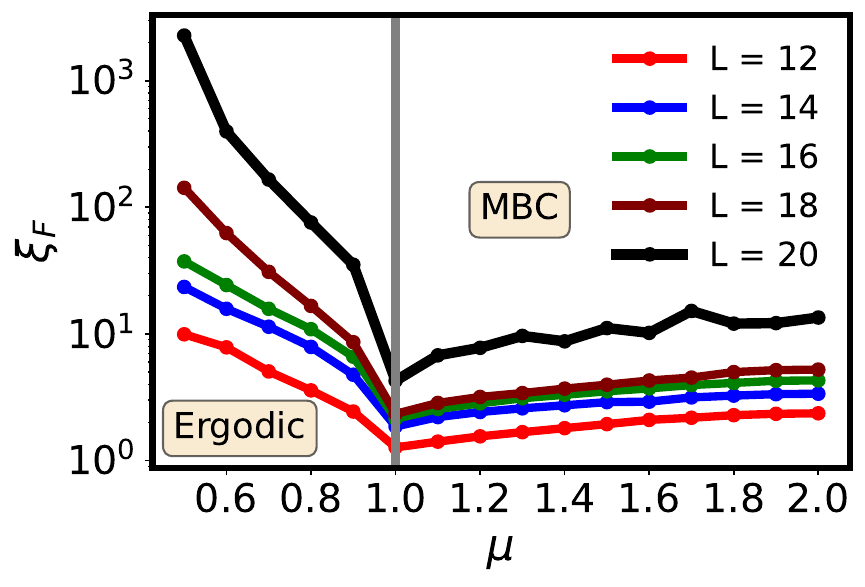}}{(b)}
\stackon{\includegraphics[width=0.49\columnwidth,height=3.5cm]{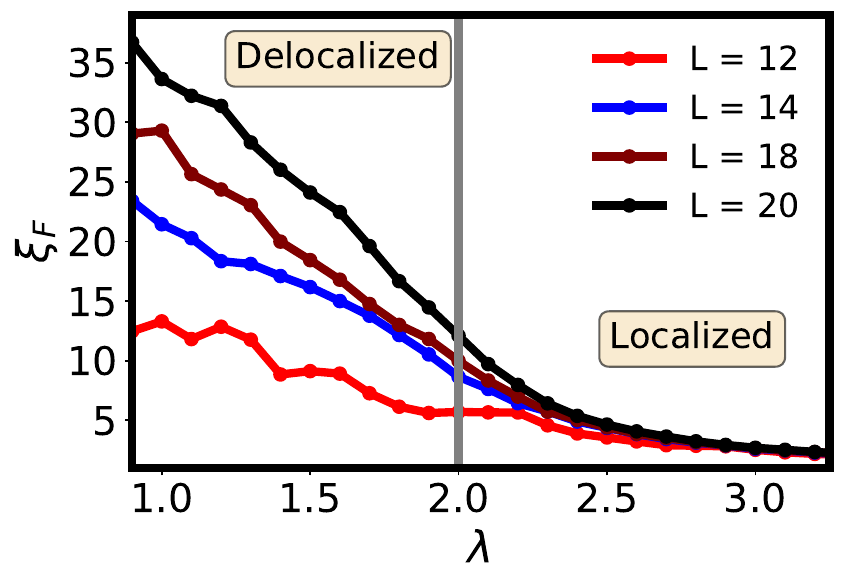}}{(c)}
\stackon{\includegraphics[width=0.49\columnwidth,height=3.5cm]{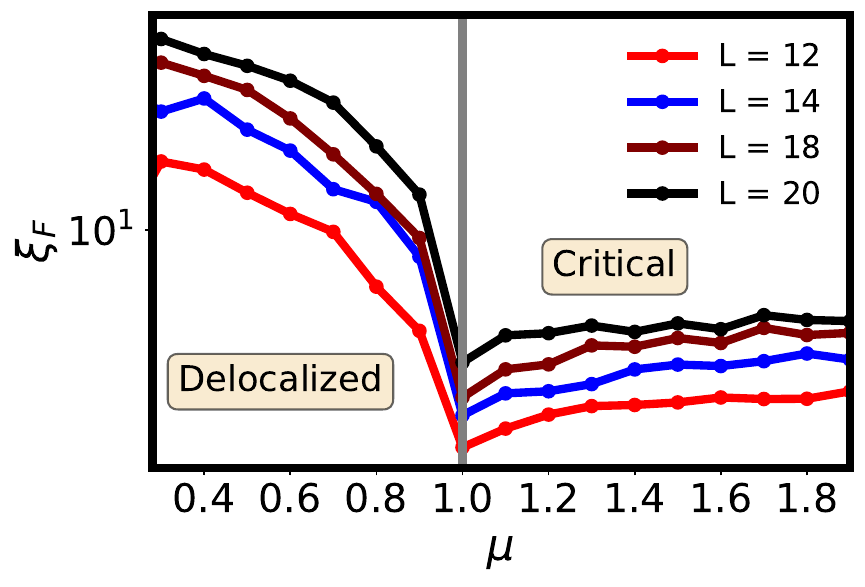}}{(d)}
\caption{{\bf Fock-space localization length}: (a-b) FS localization length $\xi_F$ extracted from the decay of $G_t(r_{IJ})$ across the ergodic-MBL transition as a function of $\lambda$ for $\mu=0.5$, and ergodic-MBC transition as function of $\mu$ for $\lambda=1.0$ for the  interacting EAAH model. (c-d) $\xi_F$ across the delocalization-localization transition as a function of $\lambda$ for $(\mu=0.5)$ and delocalization-critical transitio as a fucntion of $\mu$ for $\lambda=1.0$ for the non-interacting EAAH model. Vertical dashed lines denote the transitions in the non-interacting and interacting systems. }
\label{FS_loc}
\end{figure}  

Plots in Figs.~\ref{FS_loc}(a,b) show the variation of $\xi_F$ with $\lambda$ or $\mu$ for increasing $L$ across the ergodic-MBL and ergodic-MBC transitions, respectively. The FS localization length $\xi_F$ is almost independent of $L$ in the MBL phase, whereas it depends strongly on $L$ in the ergodic phase, and weakly in the MBC phase. Similarly, plots for the non-interacting system are shown in Figs.~\ref{FS_loc}(c,d) across the delocalization-localization and delocaliazation-critical phase transitions, respectively. Deep in the localized phase, $\xi_F$ is independent of $L$, whereas it increases with $L$ in the delocalized and critical phases, albeit slowly compared to $\xi_F(L)$ in the ergodic and MBC phases of the interacting EAAH model. The non-monotonic localization properties across the ergodic (delocalized) to MBC (critical) phase transition are nicely captured by $\xi_F$, similar to $\Delta_t$ and IPR [Figs.\ref{selfenergy_int},\ref{mipr_int}]. 


\section{Widom-like line in the presence of random  potential}\label{sec6}
\begin{figure}
\centering
\stackon{\includegraphics[width=0.49\columnwidth,height=3.5cm]{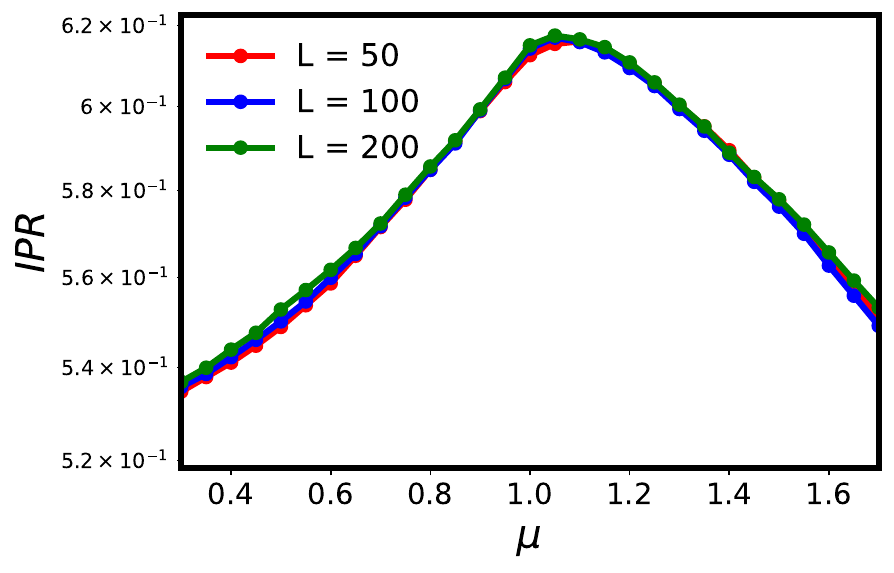}}{(a)}
\stackon{\includegraphics[width=0.49\columnwidth,height=3.6cm]{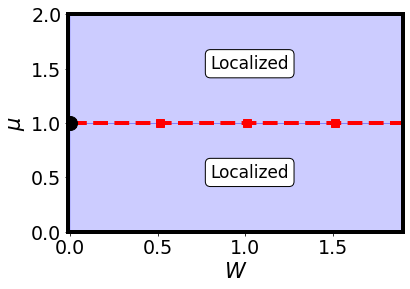}}{(b)}
\caption{{\bf Tight-binding model with quasiperiodic hopping and random onsite potential}: 
(a) For $W\neq 0$, $IPR$ as a function of $\mu$ shows a peak at around $\mu=1$ tracing the presence of Widom-like line.
(b) Phase diagram: In absence of onsite random potential $(W=0)$, the single particle system undergoes delocalized-to-critical phase transition at $\mu=1$ denoted by the black dot. In presence of random potential $(W\neq 0)$, all the single particle eigenstates become localized. However, a Widom-like line denoted by the red dot-dashed line emanates from the critical point $\mu=1$ showing a line of the strongest localization as demonstrated in figure~(a).}
\label{phase_random}
\end{figure}

In this short section, we make an attempt to find Widom-like line in presence of random disorder. It is not clear at the moment how to find a Widom-like line in a truly random system as appearance of Widom line would typically need presence of a special critical line or a triple point. However, one can think of a single particle model with both a regular hopping $(J=1)$ and quasiperidoc hopping (of strength $\mu$), like in EAAH model plus a random onsite potential (of strength $W$) instead of the quasiperiodic potential. In that case, for $W=0$ there will be a transition from delocalized to critical at $\mu=1$, according to Fig.~\ref{phase}(a) in the main text. For finite non-zero $W$ all single particle states become localized for all values of $\mu$. Interestingly, we find that the states show the strongest localization along the $\mu=1$ line for $W>0$. Hence, one can say that a Widom-like line emanates from $(\mu=1, W=0)$ point and continues for finite $W$ (see Fig.~\ref{phase_random}). Presumably this $\mu=1$ line will continue to persist even in the stipulated MBL phase in presence of interaction. 

\section{Discussion and Conclusions}\label{sec7}
To summarize, we provide a characterization of the MBC (critical) phase in the interacting (non-interacting) extended Aubry-Andr\'e-Harper (EAAH) model through localization properties of excitations in real and Fock space, and eigenstate IPR. We also analyze critical properties of the transitions between the MBC (critical) phase and other phases, like ergodic (delocalized) and MBL (localized).
We show that the MBC (critical) phase is described by multifractal scaling of FS IPR, as well as a linear finite-size scaling ansatz for IPR with a diverging length scale near the MBC (critical) to ergodic (delocalized) and MBL (localized) transition. However, the MBC (critical) phase hosts delocalized single-particle excitations in real-space and exhibits a system-size dependent FS localization length, like the ergodic (delocalized) phase. Most interestingly, we show that the ergodic (delocalized)- MBC (critical) phase transition line, even after terminating at triple point, continues as a Widom-like line in the MBL (localized) phase. Similarly, another Widom line, emanating from the ergodic-MBC phase boundary (triple point) persists in the MBC (critical) phase. These Widom lines are manifested in the FS localization properties, namely as peaks/dips (dips/peaks) in the FS self energy (FS IPR).

In the thermodynamic limit, the NEE phases, like the MBC phase, is expected to be more fragile than the MBL phase.
The stability of the MBL phase in the thermodynamic limit for systems with random disorder has been called to question due to the possibility of long-range resonances and avalanche instabilities from rare weak-disorder regions in larger system sizes, not accessible via exact diagonalization $(L\le24)$~\cite{sels2021dynamical,sels2022bath,crowley2022constructive,long2022phenomenology,morningstar2022avalanches,sierant2022challenges}. However, unlike the systems with random disorder, the quasiperiodic systems are not susceptible to the usual avalanche instability~\cite{de2017stability} due to the absence of rare weak-disorder regions~\cite{agarwal2017rare}. Hence the MBC phases, like the MBL phase, in the quasiperiodic EAAH model could be more robust. The mechanism of obtaining NEE states from the single-particle critical states is a new one and may produce more stable NEE states than found in previous works~\cite{li2015many,deng2017many} on the quasiperiodic GAAH model. Nevertheless, the issue with extrapolating the existence of MBC phase, based on the small finite-size numerics, to the thermodynamic limit is expected to be more complex than that for the MBL phase. In this regard, it might be useful to apply matrix-product state (MPS) or density-matrix renormalization group based approaches for the EAAH model to access larger system sizes, as has been attempted \cite{Pomata2020} for the GAAH model. However, the MBC states have volume-law entanglement and thus harder to describe through MPS-based approaches than the MBL phase.

The EAAH model and a related model have been realized experimentally with superconducting circuits~\cite{li2023observation} and ultra-cold atoms~\cite{xiao2021observation}, respectively. In future, with development and improvement of energy-resolved spectroscopies~\cite{kollath2007scanning,stewart2008using,perali2011evolution,jiang2011single} for the cold-atomic systems, it might be possible to probe LDOS that would be able to track the localization phenomena in real-space. It will also be interesting to look at the short-range and long-range resonances~\cite{crowley2022constructive,long2022phenomenology,morningstar2022avalanches,garratt2022resonant,garratt2021local} in the MBC phase, as well as entanglement growth and the related four-point correlations~\cite{roy2022hilbert} in Fock-space. A better understanding of the nature of the Widom lines and the mechanism behind their origin in the dynamical phase diagram of the EAAH model, and realizations of such lines in other models would be interesting research directions to pursue in future.

\section*{Acknowledgements}
SB acknowledges support from CRG, SERB (ANRF), DST, India (File No. CRG/2022/001062) and STARS, MoE, Govt. of India (File. No. MoE-STARS/STARS-2/2023-0716)




\appendix

\section{Energy level-spacings and spectral form factor}\label{appA}
\begin{figure}[h!]
\centering
\stackon{\includegraphics[width=0.49\columnwidth,height=3.4cm]{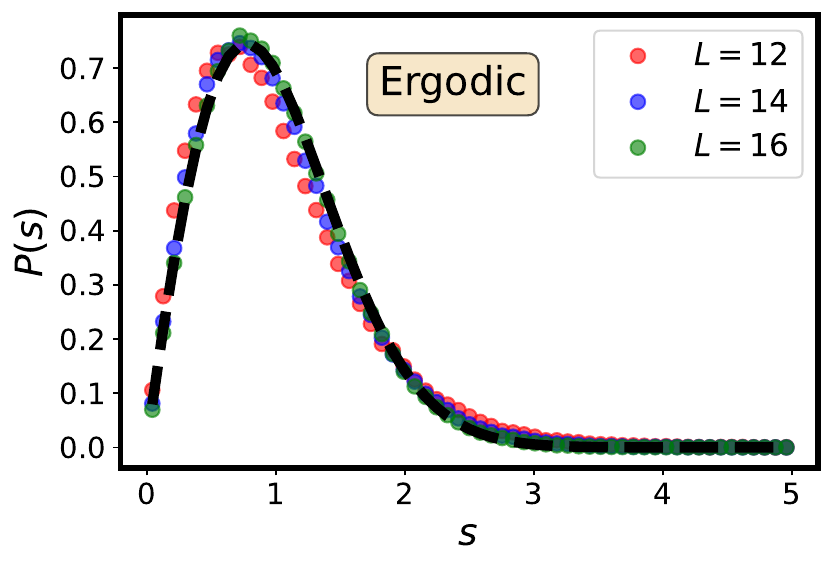}}{(a)}
\stackon{\includegraphics[width=0.49\columnwidth,height=3.4cm]{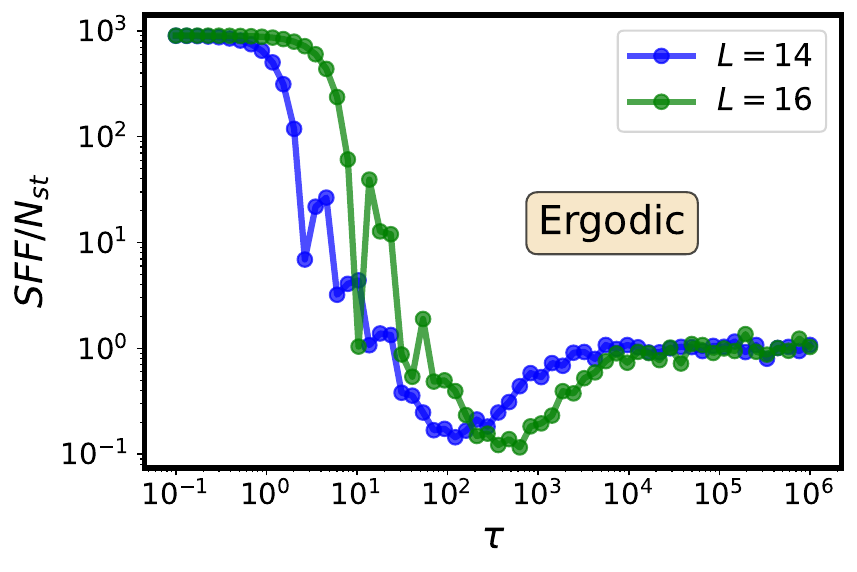}}{(d)}
\stackon{\includegraphics[width=0.49\columnwidth,height=3.4cm]{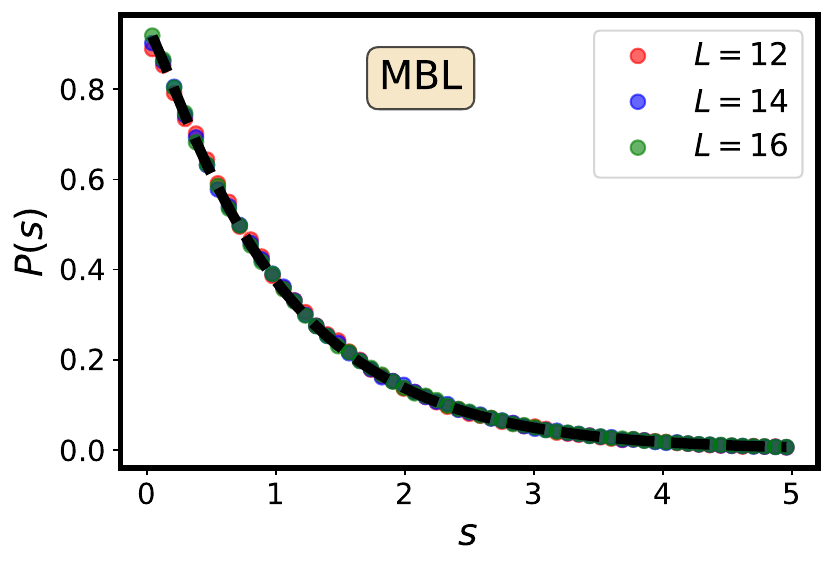}}{(b)}
\stackon{\includegraphics[width=0.49\columnwidth,height=3.4cm]{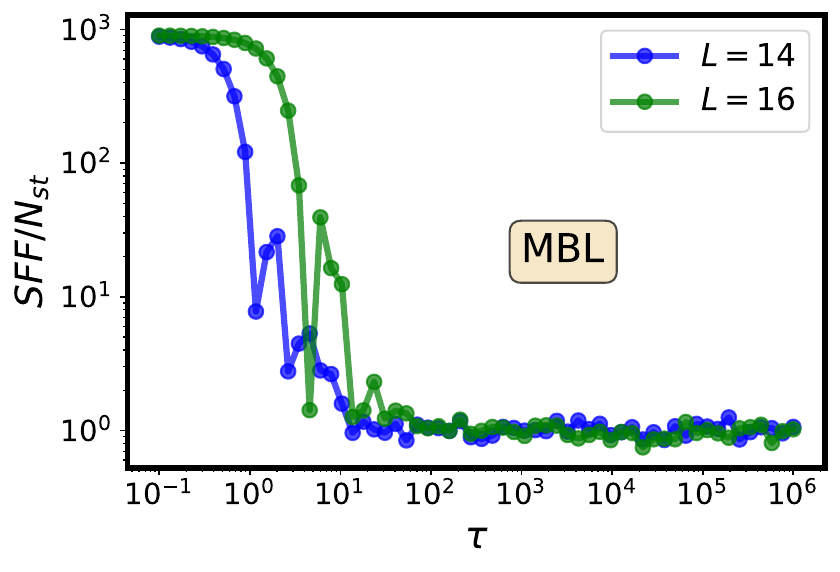}}{(e)}
\stackon{\includegraphics[width=0.49\columnwidth,height=3.4cm]{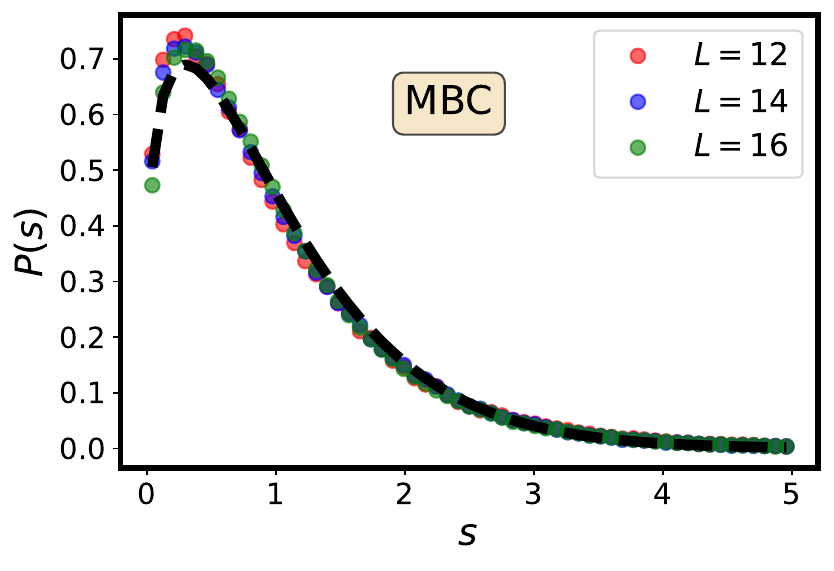}}{(c)}
\stackon{\includegraphics[width=0.49\columnwidth,height=3.4cm]{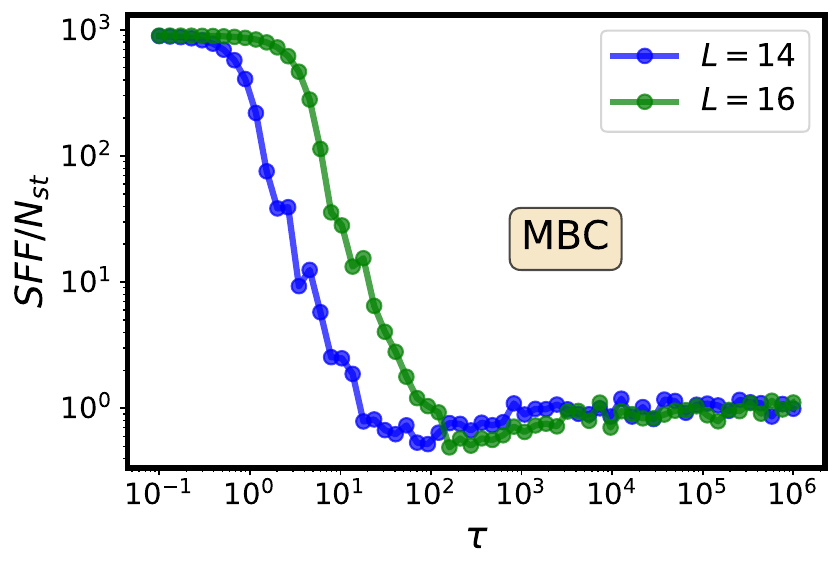}}{(f)}
\caption{ {\bf Energy level statistics and correlations in the interacting EAAH model}: (a-c) Level spacing distributions $P(s)$ in the ergodic $(\mu=0.5,\lambda=1.0)$, MBL $(\mu=0.5,\lambda=5.0)$, MBC $(\mu=1.5,\lambda=2.0)$ phases, respectively. The dashed lines are fit to the Brody distribution to the data for $L=16$ .
(d-f) Spectral form factor ($SFF$) of the interacting systems for the same parameters as in figures (a-c), respectyively. For calculation of $SFF$, the number of mid-spectrum states $N_{st}=1000$. }
\label{int_enspacing_sff}
\end{figure}

We discuss here the analysis of distributions $P(s)$ of consecutive many-body energy level-spacings $s$,
normalized by the (arithmetic) mean level spacing, in the ergodic, MBL and MBC phases of the interacting EAAH model.
The nearest neighbor level-spacing distributions for ergodic, MBL and MBC phases are shown in Fig.~\ref{int_enspacing_sff}(a-c), respectively, for increasing $L$. For the largest system size $L=16$, we fit the data to the Brody distribution given by $P(s)=As^a\exp(-As^{a+1}/(a+1))$ where $A=(a+1)\Gamma(\frac{a+2}{a+1})^{(a+1)}$. $a=1$ and $a=0$ correspond to GOE and Poisson distributions, respectively. We find $a\approx0.94$, $0.02$ and $0.26$ for the choices of parameters in Fig.~\ref{int_enspacing_sff}(a-c), respectively. The plots indicate that $P(s)$ approaches GOE distribution in the ergodic phase and Poisson distribution in the MBL phase~\cite{atas2013distribution}. Moreover, for the MBC phase $0.15<a<0.26$, implying distributions intermediate between GOE and Poisson, where maximum value of $a\approx 0.26$ appears along the Widom line ($\lambda \approx 2$) in the MBC phase. 

$P(s)$ captures the short-range correlations of the many-body energy spectrum. There is another quantity, namely the spectral form factor, that includes all possible correlations in the spectrum. One first defines a function $Z(\tau)=\sum_{n=1}^{N_{st}} e^{-iE_n \tau}$ of fictitious time $\tau$ including $N_{st}$ number of eigenstates. Then the spectral form factor can be written as~\cite{haake1991quantum}
\begin{eqnarray}
SFF(\tau) &=& \langle Z^{*}(\tau)Z(\tau)\rangle,\nonumber\\
&=& N_{st} + \langle\sum\limits_{m\neq n} e^{-i (E_m - E_n)\tau}\rangle,
\label{sff_def}
\end{eqnarray}
where $\langle...\rangle$ stands for an average over some ensemble, e.g., realizations of $\phi$ in the EAAH model. 
At $\tau=0$, $SFF=N_{st}^2$. For very small $\tau\ll \tau_{D}$ compared to the Thouless time $\tau_D$, $\langle Z^{*}(\tau)Z(\tau)\rangle=\langle Z^{*}(\tau)\rangle\langle Z(\tau)\rangle$ due to absence of spectral correlation and $SFF$ decreases showing non-universal model-specific spectral features. At $\tau=\tau_H$, the Heisenberg time, $\tau$ becomes comparable to inverse of mean level spacing. As a result, the second term in Eq.~\ref{sff_def} vanishes on average and a plateau at $SFF=N_{st}$ appears. For $\tau_D<\tau<\tau_H$, $SFF$ shows a ramp indicating the development of spectral correlation such that $\langle Z^{*}(\tau)Z(\tau)\rangle \neq \langle Z^{*}(\tau)\rangle\langle Z(\tau)\rangle$. This region shows universal features and does not appear in the absence of any spectral correlation.  
In Fig.~\ref{int_enspacing_sff}(d-f) rescaled $SFF/N_{st}$ is plotted for the three phases corresponding to Fig.~\ref{int_enspacing_sff}(a-c), respectively. The ergodic phase shows a sharper and longer ramp [Fig.~\ref{int_enspacing_sff}(a)], whereas MBC phase shows a weaker and shorter ramp [Fig.~\ref{int_enspacing_sff}(c)], indicating the presence of long-range correlations in the spectrum for both ergodic and MBC phases. In the MBL phase, no ramp regime is found indicating absence of spectral correlation [Fig.~\ref{int_enspacing_sff}(b)]~\cite{prakash2021universal}.

\section{Fock-space self energy and finite-size scaling}\label{appB}
\begin{figure}
\centering
\stackon
{\includegraphics[width=0.7\columnwidth,height=4.2cm]{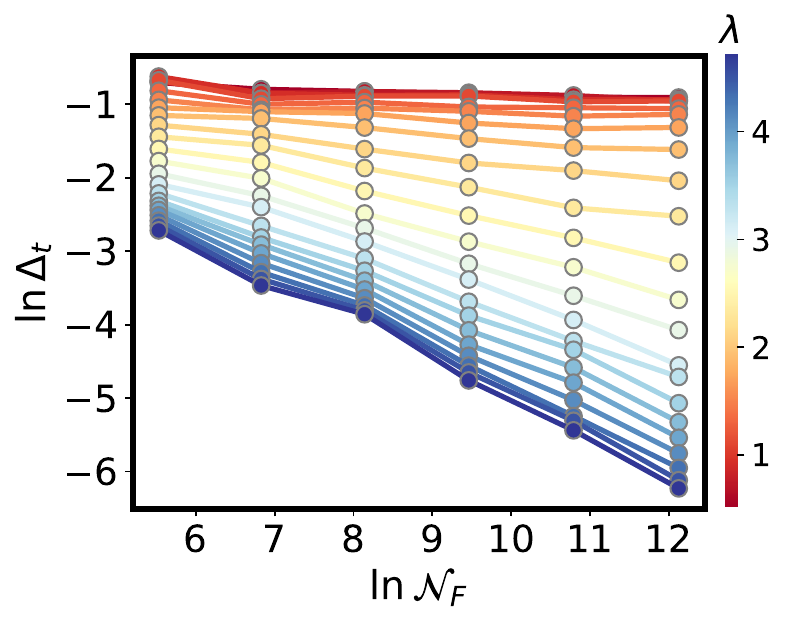}}{(a)}
\stackon{\includegraphics[width=0.85\columnwidth,height=4.3cm]{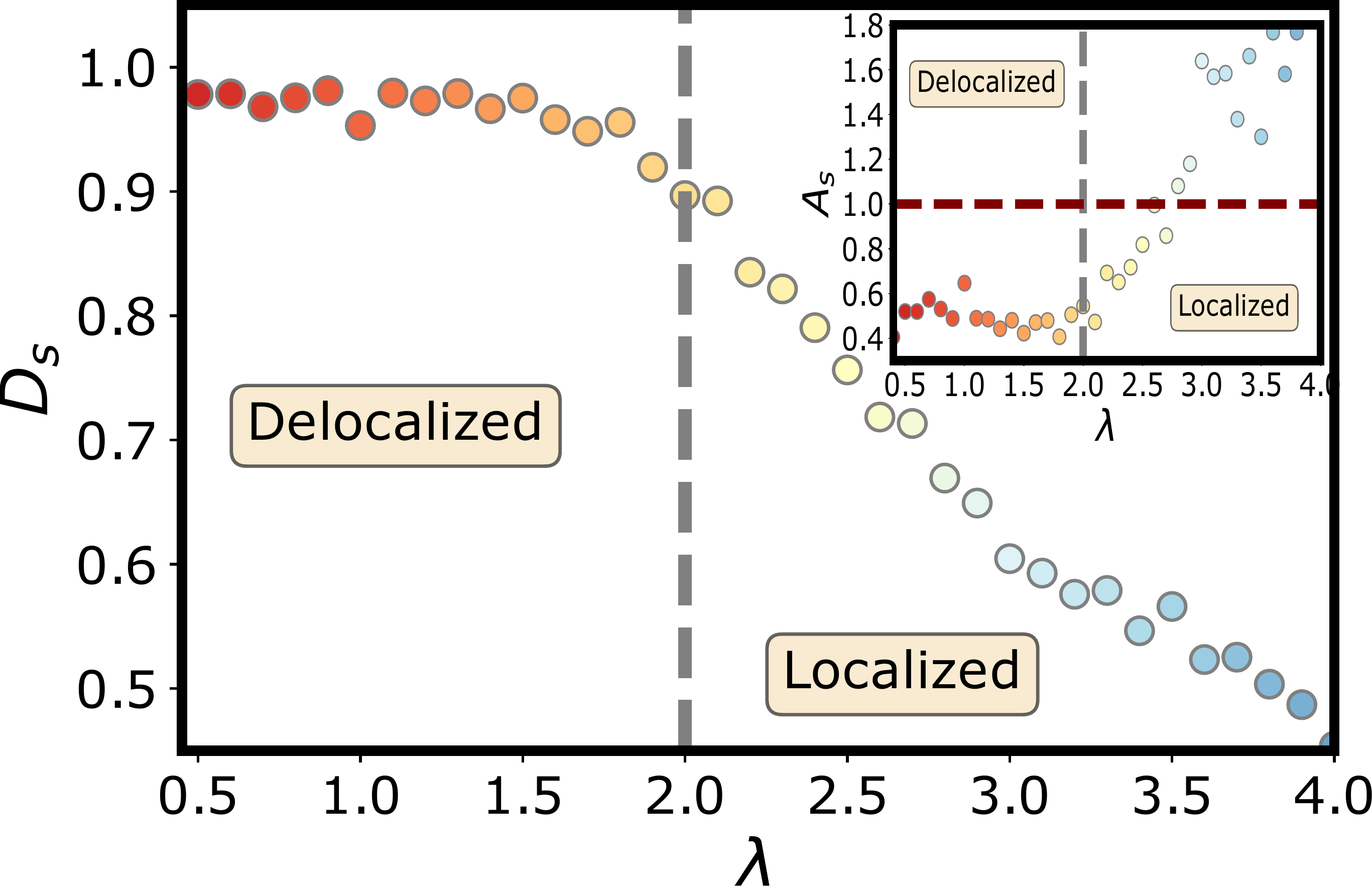}}{(b)}
\caption{{\bf System-size dependence of $\Delta_t$ for non-interacting EAAH model}: 
(a) Log-log plots of $\Delta_t$ as a function of FS dimension $\mathcal{N}_F$ across the delocalization-localization transition with $\lambda$ for $\mu=0.5$.
(b) Spectral dimension $D_s$ and pre-factor $A_s$ (inset), respectively, extracted from $\Delta_t=A_s \mathcal{N}_F^{-1+D_s}$ across the transition in (a). 
The vertical dashed line denotes the phase transition.}
\label{selftyp_Ds_nonint}
\end{figure} 
\begin{figure}
\centering
\stackon{
\includegraphics[width=0.7\columnwidth,height=4.2cm]{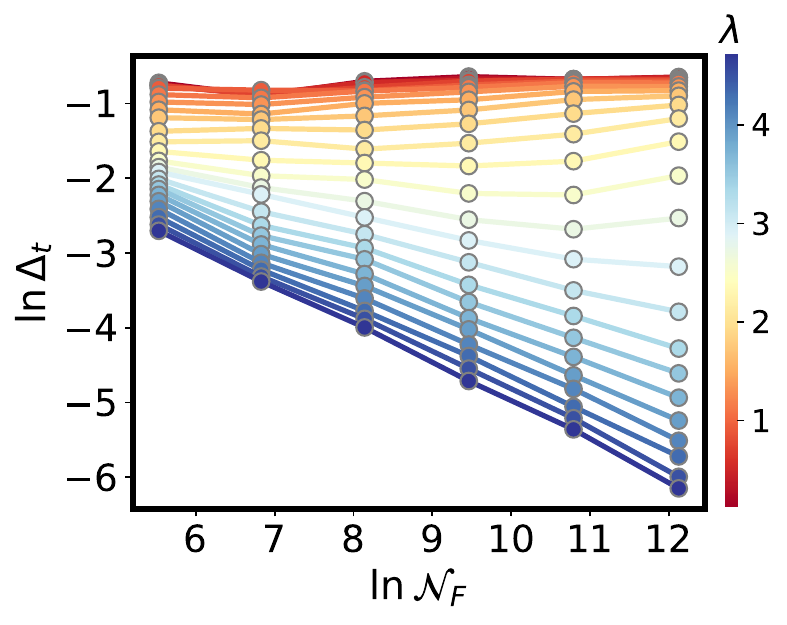}}{(a)}
\stackon{\includegraphics[width=0.8\columnwidth,height=4.2cm]{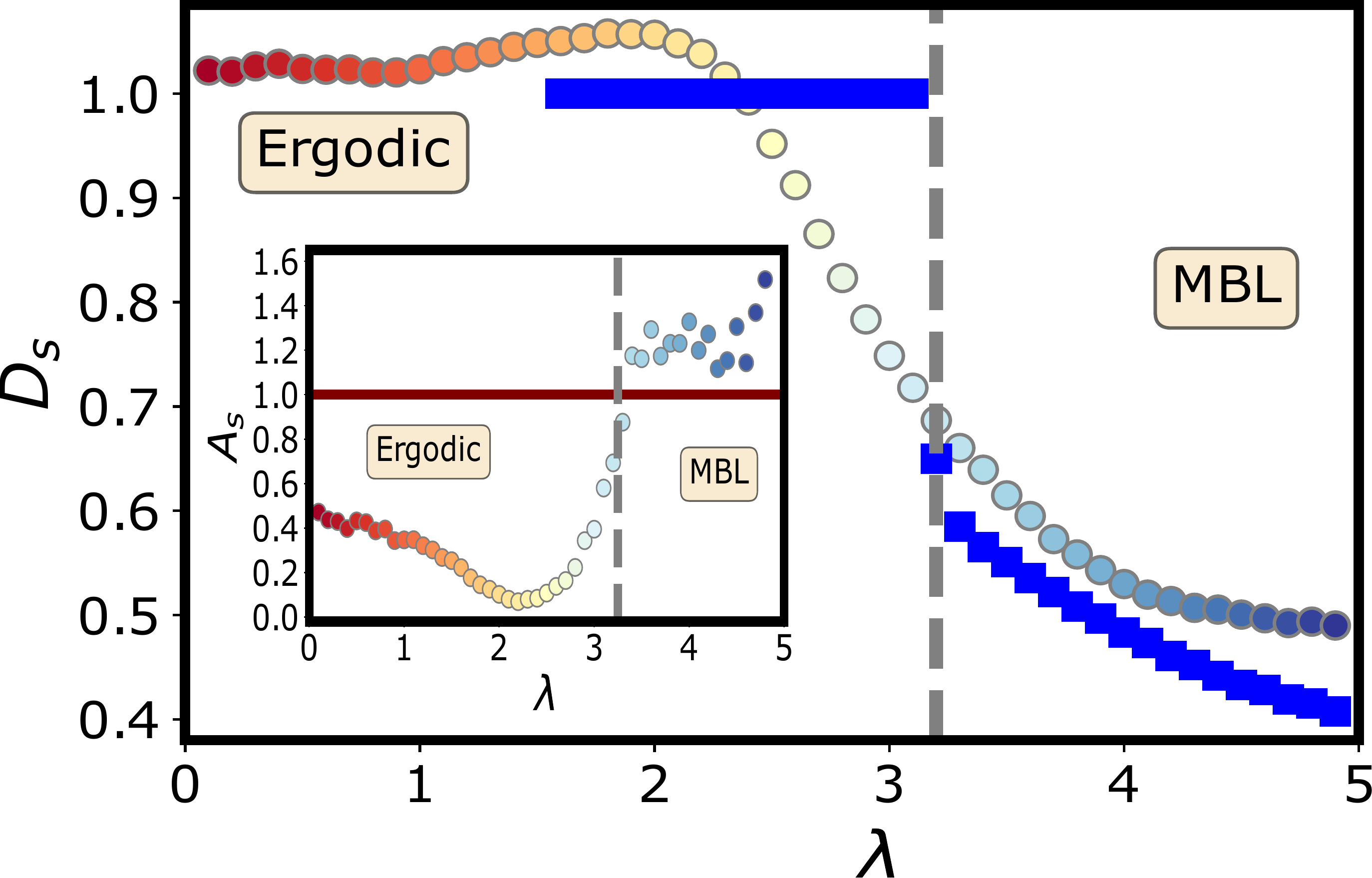}}{(b)}
\stackon{\includegraphics[width=0.8\columnwidth,height=4.2cm]{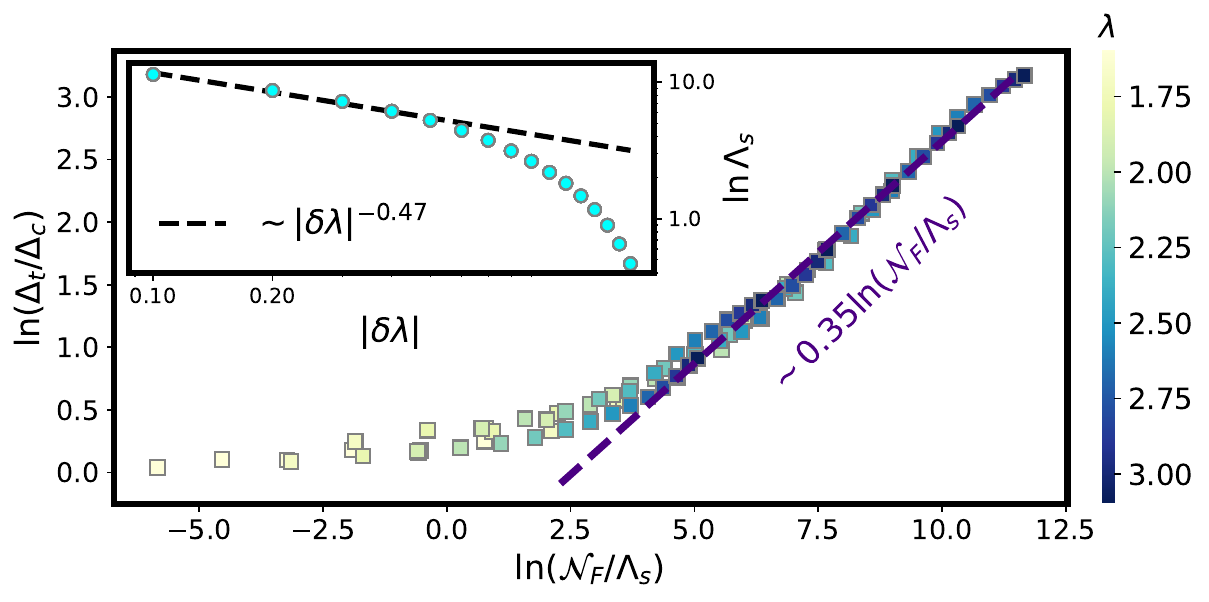}}{(c)}
\stackon{\includegraphics[width=0.8\columnwidth,height=4.2cm]{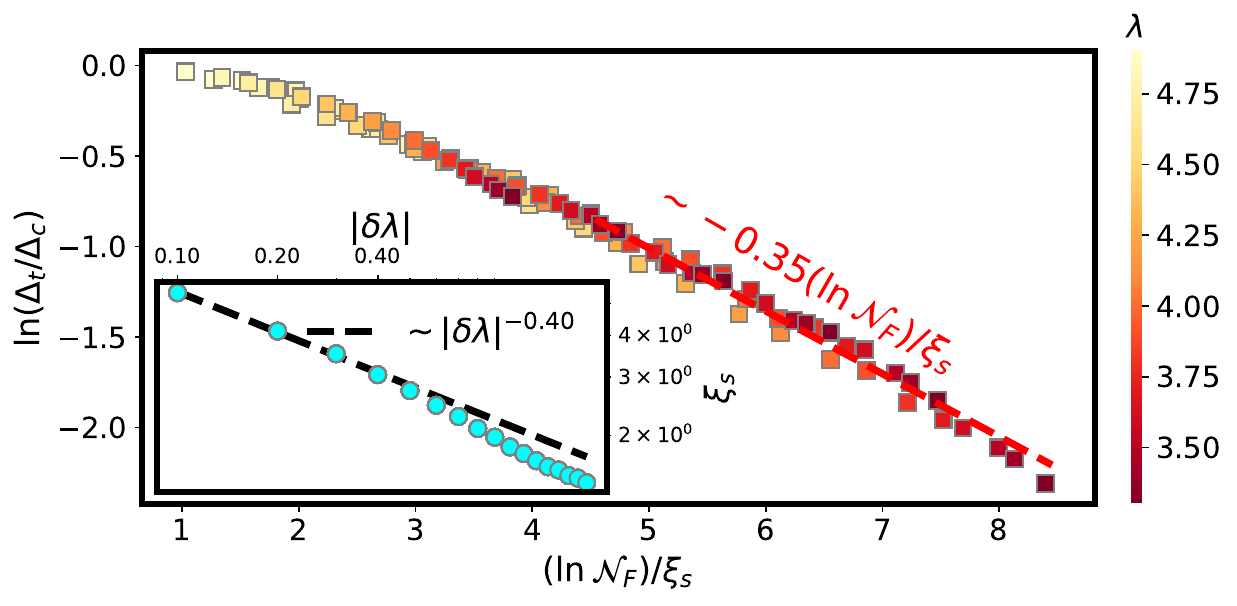}}{(d)}
\caption{{\bf System-size dependence of $\Delta_t$ and finite-size scaling for the interacting EAAH model}: 
(a) Log-log plots of $\Delta_t$ as a function of FS dimension $\mathcal{N}_F$ across the ergodic-MBL transition with $\lambda$ for $\mu=0.5$.
(b) Spectral fractal dimension $D_s$ and the pre-factor $A_s$, respectively, extracted from $\Delta_t=A_s \mathcal{N}_F^{-1+D_s}$ across the transition in (a). The vertical dashed line denotes the transition at $\lambda_c=3.2$.
(c) Volumic scaling of $\Delta_t$ in ergodic phase where nonergodic volume $\Lambda$ shows a KT-like divergence  approaching the ergodic-MBL transition. (d) Linear scaling of $\Delta_t$ in MBL phase where the correlation length $\xi_s$ shows power-law divergence approaching the transition. $D_s$ obtained from finite-size scaling analysis is shown in figure (b) with blue squares.
}
\label{selftyp_Ds_int}
\end{figure}
Here we first make an attempt to extract $A_s$ and $D_s$ from $\Delta_t= A_s \Nf^{-1+D_s}$ for delocalized-localized and ergodic-MBL transitions in non-interacting and interacting systems, respectively.
$A_s$ and $D_s$, extracted through a linear fitting to the plots in Fig.Fig.~\ref{selftyp_Ds_nonint}(a) [also Fig.~\ref{selfenergy_noint}(a)], are shown in Fig.~\ref{selftyp_Ds_nonint}(b) across the delocalized-localized transition as a function of $\lambda$ for $\mu=0.5$ in the the non-interacting EAAH model. We find ($D_s\approx1$, $A_s<1$) and ($0<D_s<1$, $A_s>1$) deep in the delocalized and localized phases, respectively. 

Similarly, we extract $A_s$ and $D_s$ from a linear fitting of our numerical data in Fig.~\ref{selftyp_Ds_int}(a,b) for ergodic-MBL transition shown in Fig.~\ref{selfenergy_int}(a). In the ergodic and MBL phases we find $(D_s\approx1, A_s<1)$ and $(0<D_s<1, A_s>1)$, respectively. We draw the dashed vertical lines to denote the point $A_s=1$, which closely agree with the transition point $\lambda_c=3.2$ obtained from finite-size scaling analysis discussed below. 

Similar to finite-size scaling of IPR discussed in the main text, we use the following scaling forms~\cite{sutradhar2022scaling,roy2023diagnostics,ghosh2024scaling},
\begin{eqnarray}
\ln\frac{\Delta_t}{\Delta_c}=
\begin{cases}
      \mathcal{G}_{vol}\big(\frac{\mathcal{N}_F}{\Lambda_s}\big)  & ~~:~ \text{Ergodic}
      \\
      \mathcal{G}_{lin}\big(\frac{\ln\mathcal{N}_F}{\xi_s}\big)  & ~~:~ \text{MBL}.
    \end{cases} 
\label{scaling_mbl}    
\end{eqnarray}
Here $\Delta_t\sim \Nf^{-1+D_s}$ is the typical value of $\Delta_I$ in the thermodynamic limit, $\mathcal{N}_F\to \infty$; $\Delta_c\sim \Nf^{-1+D_{sc}}$ is its typical value at the critical point. According to the scaling ansatz, the scaling form inside the ergodic phase is volumic controlled by a `nonergodic volume' $\Lambda_s$. In contrast,in the MBL phase, the scaling form is linear controlled by a length scale $\xi_s$. In the asymptotic limit of $\mathcal{N}_F\gg \Lambda_s$, the volumic scaling function in the ergodic phase follows $\mathcal{G}_{vol}\sim (1-D_{sc}) \ln{(\mathcal{N}_F/\Lambda_s)}$, where $\Delta_t\sim \Lambda_s^{(1-D_{sc})}$ \cite{sutradhar2022scaling}. The asymptotic form of the linear scaling function for $\ln{\mathcal{N}_F}\gg \xi_s$ in the MBL phase follows $\mathcal{G}_{lin}\sim -(1-D_{sc}) \ln{(\mathcal{N}_F)}/\xi_s$, where $\xi_s=(1-D_{sc})/(D_{sc}-D_s)$. The latter can be used to extract the spectral fractal dimension $D_s$ in the MBL phase. 
Figs.~\ref{mipr_D2_int}(c,d) show the volumic and linear scaling collpases in the ergodic and MBL phases, respectively, corresponding to phase transition discussed in Figs.~\ref{mipr_D2_int}(a,b). From the scaling analysis in the ergodic phase, we find $\mathcal{G}_{vol}(\mathcal{N}_F\gg \Lambda) \sim 0.35 \ln{(\mathcal{N}_F/\Lambda_s)}$ with $\lambda_c=3.2$, $D_{sc}\approx0.65$ and $\Lambda_s\sim \exp{[B(\lambda_c-\lambda)^{-0.47}]}$ with $B\sim \mathcal{O}(1)$. In the MBL phase, $\mathcal{G}_{lin}(\ln{\mathcal{N}_F}\gg \xi_s)\sim -0.35 \ln{(\mathcal{N}_F)}/\xi_s$ with $\xi_s\sim |\lambda-\lambda_c|^{-0.4}$. These exponents are similar to the ones reported in a recent study on ergodic-MBL transition in quasiperiodic AAH chain~\cite{ghosh2024scaling} and close to the ones obtained from the analysis of IPR in this work. Although, in general, we find $D_{2c}<D_{sc}$, or even $D_{2}<D_{s}$, in the MBC and MBL phases (not shown).

\section{Distributions of the imaginary part of self-energy and inverse participation ratio}\label{appC}
To gain further insights into the differences between MBC, and ergodic and MBL phases, we analyze the distributions of the imaginary part of FS self energy $\Delta_I$ over the FS sites in the middle slice of the FS lattice [Fig.\ref{FSlattice}] and over realizations of disorder in all the phases of the noninteracting and interacting EAAH models. The distribution of $\Delta_I$ has been studied earlier in the ergodic, MBL and even in the NEE phases of other models~\cite{logan2019many,sutradhar2022scaling,roy2023diagnostics}. 
\begin{figure}[h]
\centering
\stackon{\includegraphics[width=0.493\columnwidth,height=3.5cm]{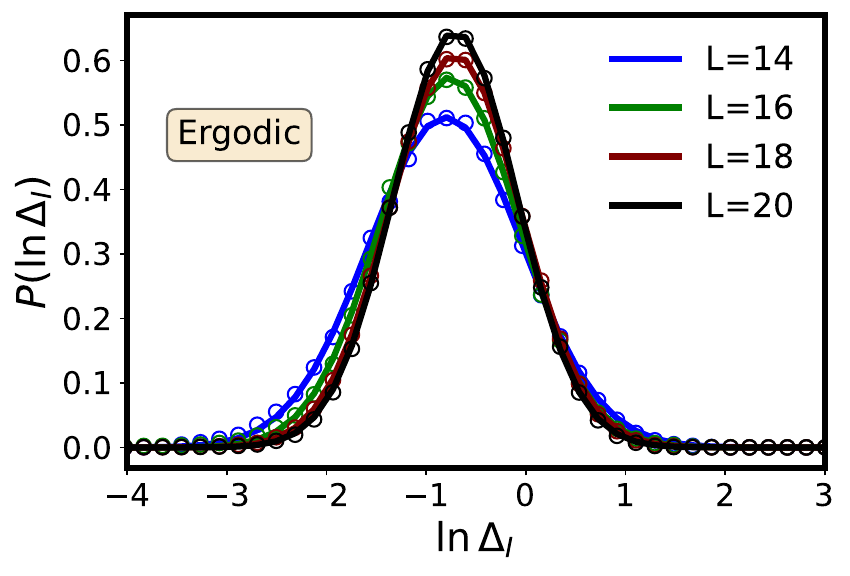}}{(a)}
\stackon{\includegraphics[width=0.493\columnwidth,height=3.5cm]{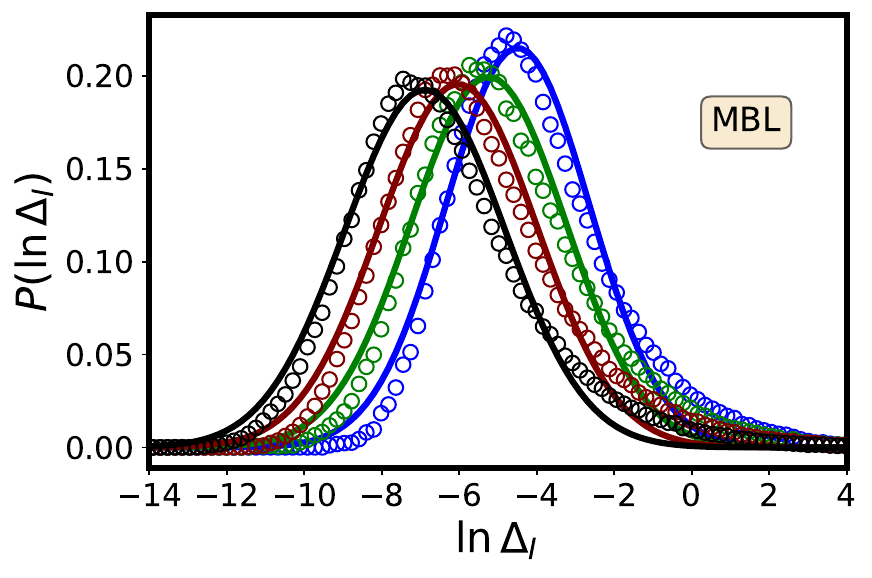}}{(b)}
\stackon{\includegraphics[width=0.493\columnwidth,height=3.5cm]{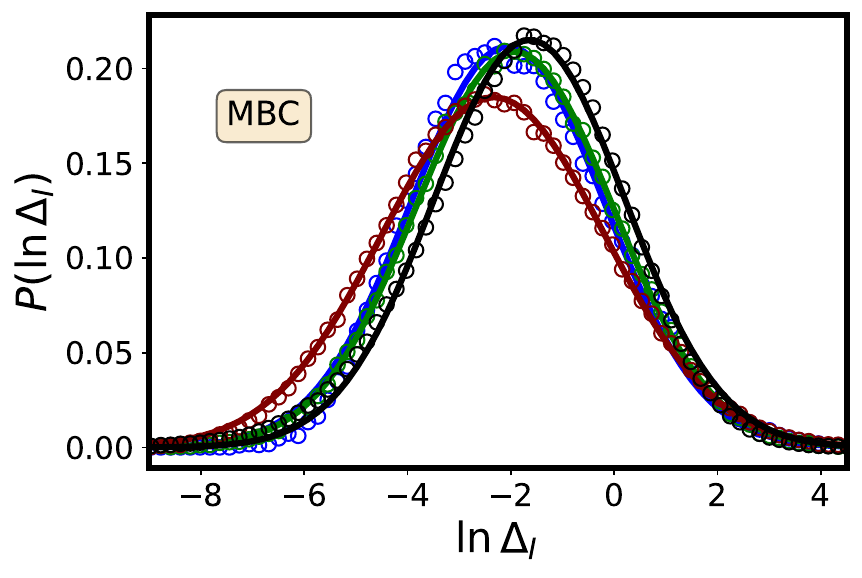}}{(c)}
\stackon{\includegraphics[width=0.493\columnwidth,height=3.5cm]{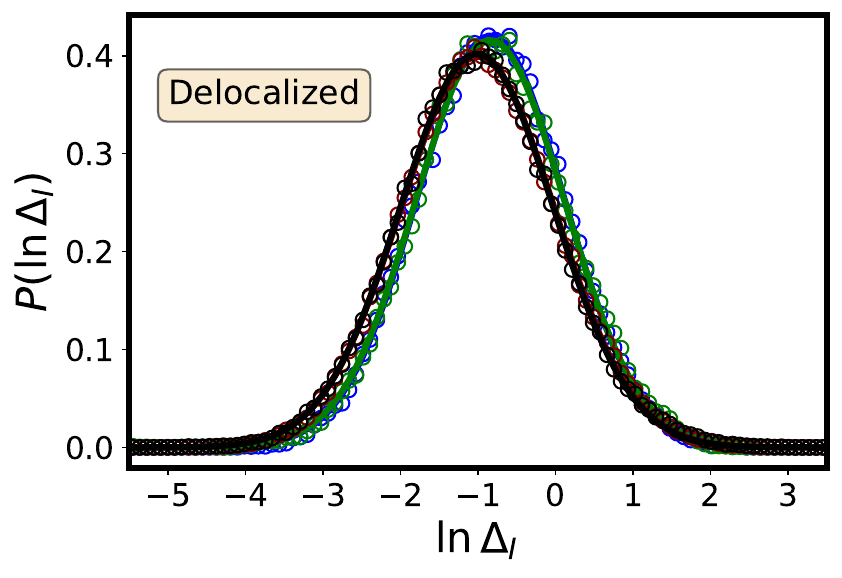}}{(d)}
\caption{ {\bf Distributions of $\ln\Delta_I$}: (a-c) Probability distributions $P(\ln\Delta_I)$ in the ergodic 
$(\mu=0.5,\lambda=1.0)$, MBL  $(\mu=0.5,\lambda=5.0)$ and MBC $(\mu=1.5,\lambda=1.0)$ phases, respectively.
(d) $P(\ln\Delta_I)$ in the delocalized $(\mu=0.5,\lambda=1.0)$ phase of the non-interacting EAAH model. Here solid lines are Gaussian fit to the data.}
\label{logself_dist_int}
\end{figure}   
\begin{figure}[h]
\centering
\stackon{\includegraphics[width=0.493\columnwidth,height=3.5cm]{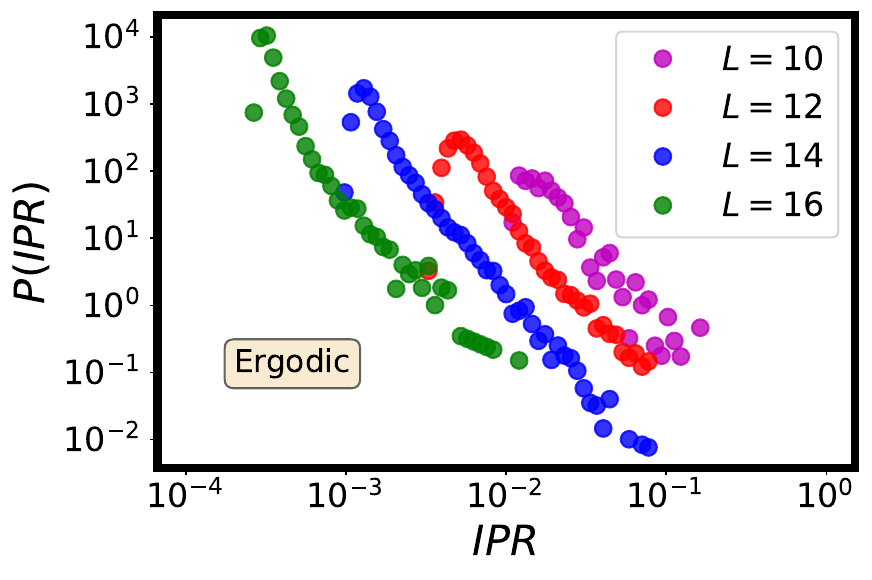}}{(a)}
\stackon{\includegraphics[width=0.493\columnwidth,height=3.5cm]{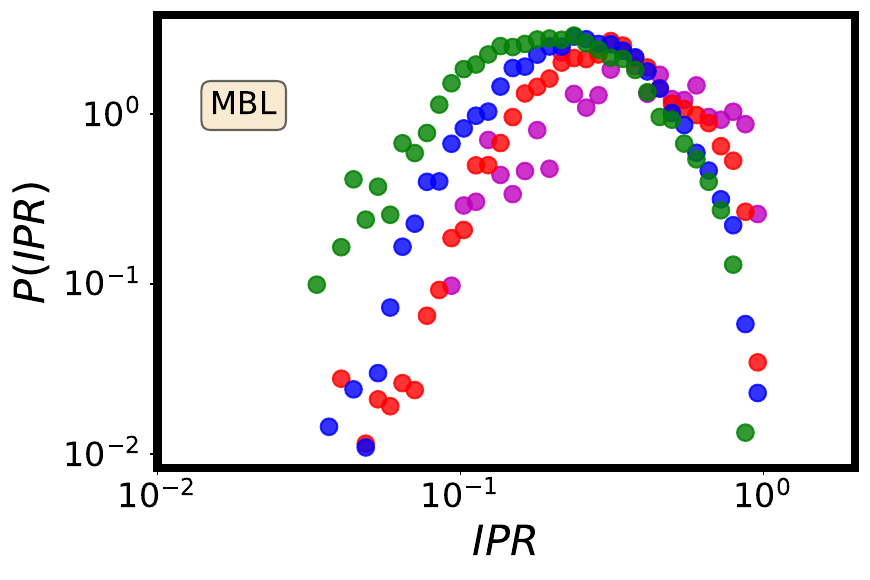}}{(b)}
\stackon{\includegraphics[width=0.493\columnwidth,height=3.5cm]{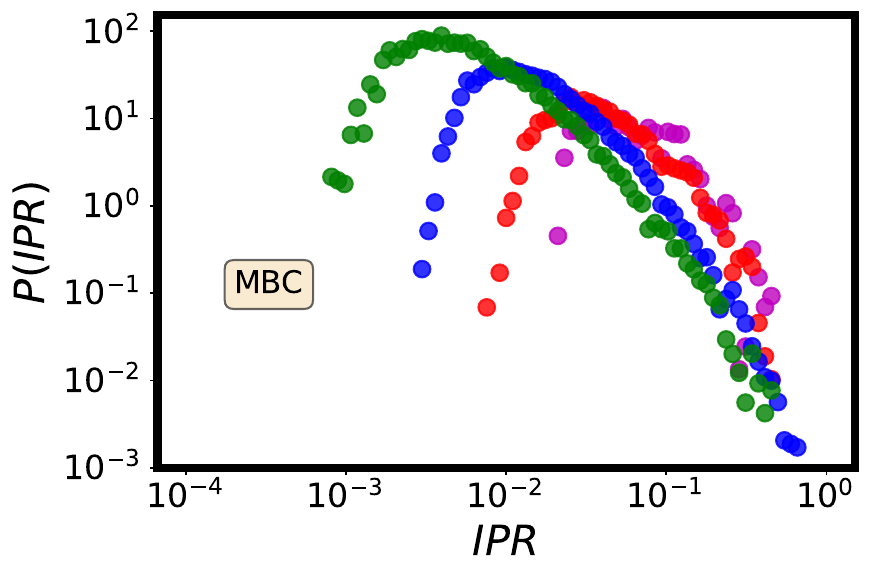}}{(c)}
\stackon{\includegraphics[width=0.493\columnwidth,height=3.5cm]{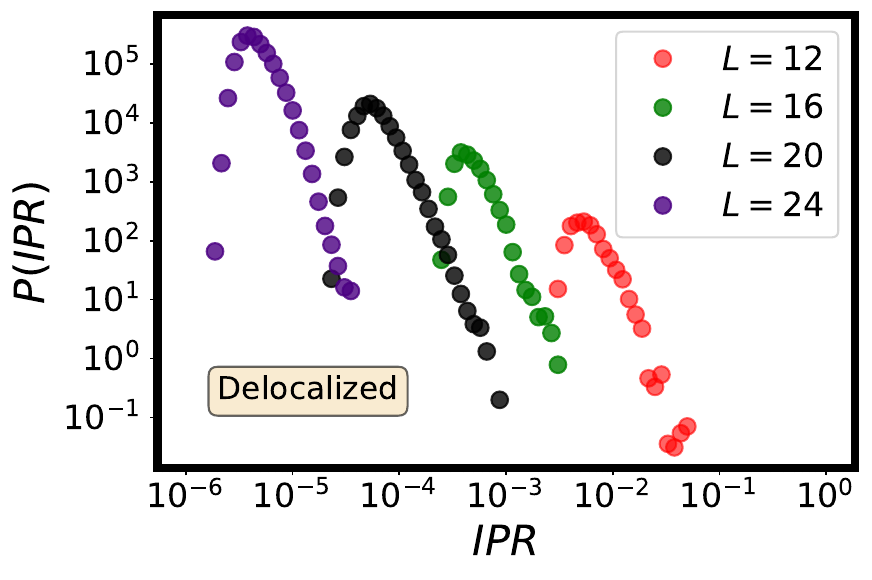}}{(d)}
\caption{ {\bf Distributions of $IPR$}: (a-c) Probability distributions of $IPR$ in the ergodic 
$(\mu=0.5,\lambda=1.0)$, MBL  $(\mu=0.5,\lambda=5.0)$ and MBC $(\mu=1.5,\lambda=1.0)$ phases, respectively.
(d) The same in the delocalized $(\mu=0.5,\lambda=1.0)$ phase of the non-interacting EAAH model. }
\label{ipr_dist_int}
\end{figure}
In the ergodic phase, $\Delta_I$ has been shown to follow an almost log-normal (LN) distribution whereas in the non-ergodic phases it exhibits a broad distribution with a Levy-like power-law tail~\cite{logan2019many,roy2023diagnostics,sutradhar2022scaling,ghosh2024scaling}. In non-interacting systems with uncorrelated disorder on real-space lattices, both the delocalized and localized phases demonstrate LN distributions of real-space local self-energy ~\cite{Schubert2010,Mirlin1996,Mirlin2000}. However, disorder on FS lattice originating from disorder, even if uncorrelated, of an underlying real-space lattice, becomes highly correlated. Thus, deviations from LN distributions have been obtained in the non-ergodic phases in earlier studies~\cite{roy2023diagnostics,ghosh2024scaling}. 

We show distributions of $\ln\Delta_I$ in ergodic, MBL and MBC phases of interacting EAAH model in Figs.~\ref{logself_dist_int}(a-c). For reference, we also show similar distributions of the FS self energy in the delocalized phase of the non-interacting EAAH model in Figs.~\ref{logself_dist_int}(d). $P(\ln\Delta_I)$ is a Gaussian for the LN distribution of $\Delta_I$. In the ergodic phase $P(\ln\Delta_I)$, as shown in Fig.~\ref{logself_dist_int}(a), is close to a Gaussian whose variance decreases with increasing $L$. In the MBL phase, $P(\ln\Delta_I)$ is skewed and clearly deviates from a Gaussian. Here, the distribution broadens as $L$ increases [Fig.~\ref{logself_dist_int}(b)], while developing a long tail. In the MBC phase, $P(\ln\Delta_I)$ is closer to a Gaussian distribution, compared to in MBL. However, the variance of the distribution changes non-monotonically with $L$, as can be noticed in Fig.~\ref{logself_dist_int}(c). 
In the non-interacting delocalized phase, shown in Fig.~\ref{logself_dist_int}(d), $P(\ln\Delta_I)$ is close to a Gaussian, but its variance does not change with $L$, unlike in the ergodic phase of the interacting system. The Anderson localized (AL) phase and critical phase of non-interacting systems show qualitatively similar $P(\ln\Delta_I)$ as the MBL and MBC phases of the interacting EAAH model, respectively (not shown here).
Thus, all the phases of the interacting system show some deviations from the LN distribution of $\Delta_I$, presumably due to the presence of correlated disorder on the FS lattice. 

We also look into the distributions of FS IPR. In ref.~\onlinecite{roy2021fock}, the distributions of $D_2$ have been studied, assuming $A_I$ to be $\delta$-distributed, across the ergodic-MBL transition in a spin chain with random disorder. Here, we directly show the distributions of FS IPR, $P(IPR)$, in Figs.~\ref{ipr_dist_int}(a,b,c) for ergodic, MBL and MBC phases, respectively. The distribution in the ergodic phase [Fig.~\ref{ipr_dist_int}(a)] is quite distinct from those in the MBL [Fig.~\ref{ipr_dist_int}(b)] and MBC [Fig.~\ref{ipr_dist_int}(c)] phases, and even from the distribution in the non-interacting delocalized phase [Fig.~\ref{ipr_dist_int}(d)]. The distribution in the ergodic phase is rather skewed and exhibits longer tails compared to the MBC phase. The tail is absent in the MBL phase. The distribution is significantly less skewed in the delocalized phase of the non-interacting EAAH model. The distributions of IPR in the localized and critical phases (not shown) of the non-interacting EAAH model are similar to those in the MBL and MBC phases, respectively, of the interacting model.

\bibliography{refs}

\begin{thebibliography}{119}%
\makeatletter
\providecommand \@ifxundefined [1]{%
 \@ifx{#1\undefined}
}%
\providecommand \@ifnum [1]{%
 \ifnum #1\expandafter \@firstoftwo
 \else \expandafter \@secondoftwo
 \fi
}%
\providecommand \@ifx [1]{%
 \ifx #1\expandafter \@firstoftwo
 \else \expandafter \@secondoftwo
 \fi
}%
\providecommand \natexlab [1]{#1}%
\providecommand \enquote  [1]{``#1''}%
\providecommand \bibnamefont  [1]{#1}%
\providecommand \bibfnamefont [1]{#1}%
\providecommand \citenamefont [1]{#1}%
\providecommand \href@noop [0]{\@secondoftwo}%
\providecommand \href [0]{\begingroup \@sanitize@url \@href}%
\providecommand \@href[1]{\@@startlink{#1}\@@href}%
\providecommand \@@href[1]{\endgroup#1\@@endlink}%
\providecommand \@sanitize@url [0]{\catcode `\\12\catcode `\$12\catcode
  `\&12\catcode `\#12\catcode `\^12\catcode `\_12\catcode `\%12\relax}%
\providecommand \@@startlink[1]{}%
\providecommand \@@endlink[0]{}%
\providecommand \url  [0]{\begingroup\@sanitize@url \@url }%
\providecommand \@url [1]{\endgroup\@href {#1}{\urlprefix }}%
\providecommand \urlprefix  [0]{URL }%
\providecommand \Eprint [0]{\href }%
\providecommand \doibase [0]{http://dx.doi.org/}%
\providecommand \selectlanguage [0]{\@gobble}%
\providecommand \bibinfo  [0]{\@secondoftwo}%
\providecommand \bibfield  [0]{\@secondoftwo}%
\providecommand \translation [1]{[#1]}%
\providecommand \BibitemOpen [0]{}%
\providecommand \bibitemStop [0]{}%
\providecommand \bibitemNoStop [0]{.\EOS\space}%
\providecommand \EOS [0]{\spacefactor3000\relax}%
\providecommand \BibitemShut  [1]{\csname bibitem#1\endcsname}%
\let\auto@bib@innerbib\@empty
\bibitem [{\citenamefont {Anderson}(1958)}]{anderson1958absence}%
  \BibitemOpen
  \bibfield  {author} {\bibinfo {author} {\bibfnamefont {Philip~W}\
  \bibnamefont {Anderson}},\ }\bibfield  {title} {\enquote {\bibinfo {title}
  {Absence of diffusion in certain random lattices},}\ }\href@noop {}
  {\bibfield  {journal} {\bibinfo  {journal} {Physical review}\ }\textbf
  {\bibinfo {volume} {109}},\ \bibinfo {pages} {1492} (\bibinfo {year}
  {1958})}\BibitemShut {NoStop}%
\bibitem [{\citenamefont {Aubry}\ and\ \citenamefont
  {Andr{\'e}}(1980)}]{aubry1980analyticity}%
  \BibitemOpen
  \bibfield  {author} {\bibinfo {author} {\bibfnamefont {Serge}\ \bibnamefont
  {Aubry}}\ and\ \bibinfo {author} {\bibfnamefont {Gilles}\ \bibnamefont
  {Andr{\'e}}},\ }\bibfield  {title} {\enquote {\bibinfo {title} {Analyticity
  breaking and anderson localization in incommensurate lattices},}\ }\href@noop
  {} {\bibfield  {journal} {\bibinfo  {journal} {Ann. Israel Phys. Soc}\
  }\textbf {\bibinfo {volume} {3}},\ \bibinfo {pages} {18} (\bibinfo {year}
  {1980})}\BibitemShut {NoStop}%
\bibitem [{\citenamefont {Abrahams}\ \emph {et~al.}(1979)\citenamefont
  {Abrahams}, \citenamefont {Anderson}, \citenamefont {Licciardello},\ and\
  \citenamefont {Ramakrishnan}}]{abrahams1979scaling}%
  \BibitemOpen
  \bibfield  {author} {\bibinfo {author} {\bibfnamefont {Elihu}\ \bibnamefont
  {Abrahams}}, \bibinfo {author} {\bibfnamefont {PW}~\bibnamefont {Anderson}},
  \bibinfo {author} {\bibfnamefont {DC}~\bibnamefont {Licciardello}}, \ and\
  \bibinfo {author} {\bibfnamefont {TV}~\bibnamefont {Ramakrishnan}},\
  }\bibfield  {title} {\enquote {\bibinfo {title} {Scaling theory of
  localization: Absence of quantum diffusion in two dimensions},}\ }\href@noop
  {} {\bibfield  {journal} {\bibinfo  {journal} {Physical Review Letters}\
  }\textbf {\bibinfo {volume} {42}},\ \bibinfo {pages} {673} (\bibinfo {year}
  {1979})}\BibitemShut {NoStop}%
\bibitem [{\citenamefont {Evers}\ and\ \citenamefont
  {Mirlin}(2008)}]{evers2008anderson}%
  \BibitemOpen
  \bibfield  {author} {\bibinfo {author} {\bibfnamefont {Ferdinand}\
  \bibnamefont {Evers}}\ and\ \bibinfo {author} {\bibfnamefont {Alexander~D}\
  \bibnamefont {Mirlin}},\ }\bibfield  {title} {\enquote {\bibinfo {title}
  {Anderson transitions},}\ }\href@noop {} {\bibfield  {journal} {\bibinfo
  {journal} {Reviews of Modern Physics}\ }\textbf {\bibinfo {volume} {80}},\
  \bibinfo {pages} {1355--1417} (\bibinfo {year} {2008})}\BibitemShut {NoStop}%
\bibitem [{\citenamefont {Han}\ \emph {et~al.}(1994)\citenamefont {Han},
  \citenamefont {Thouless}, \citenamefont {Hiramoto},\ and\ \citenamefont
  {Kohmoto}}]{han1994critical}%
  \BibitemOpen
  \bibfield  {author} {\bibinfo {author} {\bibfnamefont {JH}~\bibnamefont
  {Han}}, \bibinfo {author} {\bibfnamefont {DJ}~\bibnamefont {Thouless}},
  \bibinfo {author} {\bibfnamefont {H}~\bibnamefont {Hiramoto}}, \ and\
  \bibinfo {author} {\bibfnamefont {M}~\bibnamefont {Kohmoto}},\ }\bibfield
  {title} {\enquote {\bibinfo {title} {Critical and bicritical properties of
  harper’s equation with next-nearest-neighbor coupling},}\ }\href@noop {}
  {\bibfield  {journal} {\bibinfo  {journal} {Physical Review B}\ }\textbf
  {\bibinfo {volume} {50}},\ \bibinfo {pages} {11365} (\bibinfo {year}
  {1994})}\BibitemShut {NoStop}%
\bibitem [{\citenamefont {Chang}\ \emph {et~al.}(1997)\citenamefont {Chang},
  \citenamefont {Ikezawa},\ and\ \citenamefont
  {Kohmoto}}]{chang1997multifractal}%
  \BibitemOpen
  \bibfield  {author} {\bibinfo {author} {\bibfnamefont {Iksoo}\ \bibnamefont
  {Chang}}, \bibinfo {author} {\bibfnamefont {Kazuhiro}\ \bibnamefont
  {Ikezawa}}, \ and\ \bibinfo {author} {\bibfnamefont {Mahito}\ \bibnamefont
  {Kohmoto}},\ }\bibfield  {title} {\enquote {\bibinfo {title} {Multifractal
  properties of the wave functions of the square-lattice tight-binding model
  with next-nearest-neighbor hopping in a magnetic field},}\ }\href@noop {}
  {\bibfield  {journal} {\bibinfo  {journal} {Physical Review B}\ }\textbf
  {\bibinfo {volume} {55}},\ \bibinfo {pages} {12971} (\bibinfo {year}
  {1997})}\BibitemShut {NoStop}%
\bibitem [{\citenamefont {Liu}\ \emph {et~al.}(2016)\citenamefont {Liu},
  \citenamefont {Wang},\ and\ \citenamefont {Xianlong}}]{liu2016phase}%
  \BibitemOpen
  \bibfield  {author} {\bibinfo {author} {\bibfnamefont {Tong}\ \bibnamefont
  {Liu}}, \bibinfo {author} {\bibfnamefont {Pei}\ \bibnamefont {Wang}}, \ and\
  \bibinfo {author} {\bibfnamefont {Gao}\ \bibnamefont {Xianlong}},\ }\bibfield
   {title} {\enquote {\bibinfo {title} {Phase diagram of the off-diagonal
  aubry-andr$\backslash$'e model},}\ }\href@noop {} {\bibfield  {journal}
  {\bibinfo  {journal} {arXiv preprint arXiv:1609.06939}\ } (\bibinfo {year}
  {2016})}\BibitemShut {NoStop}%
\bibitem [{\citenamefont {Kohmoto}\ \emph {et~al.}(1987)\citenamefont
  {Kohmoto}, \citenamefont {Sutherland},\ and\ \citenamefont
  {Tang}}]{kohmoto1987critical}%
  \BibitemOpen
  \bibfield  {author} {\bibinfo {author} {\bibfnamefont {Mahito}\ \bibnamefont
  {Kohmoto}}, \bibinfo {author} {\bibfnamefont {Bill}\ \bibnamefont
  {Sutherland}}, \ and\ \bibinfo {author} {\bibfnamefont {Chao}\ \bibnamefont
  {Tang}},\ }\bibfield  {title} {\enquote {\bibinfo {title} {Critical wave
  functions and a cantor-set spectrum of a one-dimensional quasicrystal
  model},}\ }\href@noop {} {\bibfield  {journal} {\bibinfo  {journal} {Physical
  Review B}\ }\textbf {\bibinfo {volume} {35}},\ \bibinfo {pages} {1020}
  (\bibinfo {year} {1987})}\BibitemShut {NoStop}%
\bibitem [{\citenamefont {Lin}\ \emph {et~al.}(2023)\citenamefont {Lin},
  \citenamefont {Chen}, \citenamefont {Guo},\ and\ \citenamefont
  {Gong}}]{lin2023general}%
  \BibitemOpen
  \bibfield  {author} {\bibinfo {author} {\bibfnamefont {Xiaoshui}\
  \bibnamefont {Lin}}, \bibinfo {author} {\bibfnamefont {Xiaoman}\ \bibnamefont
  {Chen}}, \bibinfo {author} {\bibfnamefont {Guang-Can}\ \bibnamefont {Guo}}, \
  and\ \bibinfo {author} {\bibfnamefont {Ming}\ \bibnamefont {Gong}},\
  }\bibfield  {title} {\enquote {\bibinfo {title} {General approach to the
  critical phase with coupled quasiperiodic chains},}\ }\href@noop {}
  {\bibfield  {journal} {\bibinfo  {journal} {Physical Review B}\ }\textbf
  {\bibinfo {volume} {108}},\ \bibinfo {pages} {174206} (\bibinfo {year}
  {2023})}\BibitemShut {NoStop}%
\bibitem [{\citenamefont {Wang}\ \emph {et~al.}(2022)\citenamefont {Wang},
  \citenamefont {Zhang}, \citenamefont {Sun}, \citenamefont {Poon},\ and\
  \citenamefont {Liu}}]{wang2022quantum}%
  \BibitemOpen
  \bibfield  {author} {\bibinfo {author} {\bibfnamefont {Yucheng}\ \bibnamefont
  {Wang}}, \bibinfo {author} {\bibfnamefont {Long}\ \bibnamefont {Zhang}},
  \bibinfo {author} {\bibfnamefont {Wei}\ \bibnamefont {Sun}}, \bibinfo
  {author} {\bibfnamefont {Ting-Fung~Jeffrey}\ \bibnamefont {Poon}}, \ and\
  \bibinfo {author} {\bibfnamefont {Xiong-Jun}\ \bibnamefont {Liu}},\
  }\bibfield  {title} {\enquote {\bibinfo {title} {Quantum phase with
  coexisting localized, extended, and critical zones},}\ }\href@noop {}
  {\bibfield  {journal} {\bibinfo  {journal} {Physical Review B}\ }\textbf
  {\bibinfo {volume} {106}},\ \bibinfo {pages} {L140203} (\bibinfo {year}
  {2022})}\BibitemShut {NoStop}%
\bibitem [{\citenamefont {Wang}\ \emph
  {et~al.}(2016{\natexlab{a}})\citenamefont {Wang}, \citenamefont {Liu},
  \citenamefont {Xianlong},\ and\ \citenamefont {Hu}}]{wang2016phase}%
  \BibitemOpen
  \bibfield  {author} {\bibinfo {author} {\bibfnamefont {Jun}\ \bibnamefont
  {Wang}}, \bibinfo {author} {\bibfnamefont {Xia-Ji}\ \bibnamefont {Liu}},
  \bibinfo {author} {\bibfnamefont {Gao}\ \bibnamefont {Xianlong}}, \ and\
  \bibinfo {author} {\bibfnamefont {Hui}\ \bibnamefont {Hu}},\ }\bibfield
  {title} {\enquote {\bibinfo {title} {Phase diagram of a non-abelian
  aubry-andr{\'e}-harper model with p-wave superfluidity},}\ }\href@noop {}
  {\bibfield  {journal} {\bibinfo  {journal} {Physical Review B}\ }\textbf
  {\bibinfo {volume} {93}},\ \bibinfo {pages} {104504} (\bibinfo {year}
  {2016}{\natexlab{a}})}\BibitemShut {NoStop}%
\bibitem [{\citenamefont {Wang}\ \emph
  {et~al.}(2016{\natexlab{b}})\citenamefont {Wang}, \citenamefont {Wang},\ and\
  \citenamefont {Chen}}]{wang2016spectral}%
  \BibitemOpen
  \bibfield  {author} {\bibinfo {author} {\bibfnamefont {Yucheng}\ \bibnamefont
  {Wang}}, \bibinfo {author} {\bibfnamefont {Yancheng}\ \bibnamefont {Wang}}, \
  and\ \bibinfo {author} {\bibfnamefont {Shu}\ \bibnamefont {Chen}},\
  }\bibfield  {title} {\enquote {\bibinfo {title} {Spectral statistics,
  finite-size scaling and multifractal analysis of quasiperiodic chain with
  p-wave pairing},}\ }\href@noop {} {\bibfield  {journal} {\bibinfo  {journal}
  {The European Physical Journal B}\ }\textbf {\bibinfo {volume} {89}},\
  \bibinfo {pages} {1--9} (\bibinfo {year} {2016}{\natexlab{b}})}\BibitemShut
  {NoStop}%
\bibitem [{\citenamefont {Li}\ \emph {et~al.}(2023)\citenamefont {Li},
  \citenamefont {Wang}, \citenamefont {Shi}, \citenamefont {Huang},
  \citenamefont {Song}, \citenamefont {Liang}, \citenamefont {Mei},
  \citenamefont {Zhou}, \citenamefont {Zhang}, \citenamefont {Zhang} \emph
  {et~al.}}]{li2023observation}%
  \BibitemOpen
  \bibfield  {author} {\bibinfo {author} {\bibfnamefont {Hao}\ \bibnamefont
  {Li}}, \bibinfo {author} {\bibfnamefont {Yong-Yi}\ \bibnamefont {Wang}},
  \bibinfo {author} {\bibfnamefont {Yun-Hao}\ \bibnamefont {Shi}}, \bibinfo
  {author} {\bibfnamefont {Kaixuan}\ \bibnamefont {Huang}}, \bibinfo {author}
  {\bibfnamefont {Xiaohui}\ \bibnamefont {Song}}, \bibinfo {author}
  {\bibfnamefont {Gui-Han}\ \bibnamefont {Liang}}, \bibinfo {author}
  {\bibfnamefont {Zheng-Yang}\ \bibnamefont {Mei}}, \bibinfo {author}
  {\bibfnamefont {Bozhen}\ \bibnamefont {Zhou}}, \bibinfo {author}
  {\bibfnamefont {He}~\bibnamefont {Zhang}}, \bibinfo {author} {\bibfnamefont
  {Jia-Chi}\ \bibnamefont {Zhang}},  \emph {et~al.},\ }\bibfield  {title}
  {\enquote {\bibinfo {title} {Observation of critical phase transition in a
  generalized aubry-andr{\'e}-harper model with superconducting circuits},}\
  }\href@noop {} {\bibfield  {journal} {\bibinfo  {journal} {npj Quantum
  Information}\ }\textbf {\bibinfo {volume} {9}},\ \bibinfo {pages} {40}
  (\bibinfo {year} {2023})}\BibitemShut {NoStop}%
\bibitem [{\citenamefont {Xiao}\ \emph {et~al.}(2021)\citenamefont {Xiao},
  \citenamefont {Xie}, \citenamefont {Dong}, \citenamefont {Chen},
  \citenamefont {Yi},\ and\ \citenamefont {Yan}}]{xiao2021observation}%
  \BibitemOpen
  \bibfield  {author} {\bibinfo {author} {\bibfnamefont {Teng}\ \bibnamefont
  {Xiao}}, \bibinfo {author} {\bibfnamefont {Dizhou}\ \bibnamefont {Xie}},
  \bibinfo {author} {\bibfnamefont {Zhaoli}\ \bibnamefont {Dong}}, \bibinfo
  {author} {\bibfnamefont {Tao}\ \bibnamefont {Chen}}, \bibinfo {author}
  {\bibfnamefont {Wei}\ \bibnamefont {Yi}}, \ and\ \bibinfo {author}
  {\bibfnamefont {Bo}~\bibnamefont {Yan}},\ }\bibfield  {title} {\enquote
  {\bibinfo {title} {Observation of topological phase with critical
  localization in a quasi-periodic lattice},}\ }\href@noop {} {\bibfield
  {journal} {\bibinfo  {journal} {Science Bulletin}\ }\textbf {\bibinfo
  {volume} {66}},\ \bibinfo {pages} {2175--2180} (\bibinfo {year}
  {2021})}\BibitemShut {NoStop}%
\bibitem [{\citenamefont {Wang}\ \emph {et~al.}(2020)\citenamefont {Wang},
  \citenamefont {Zhang}, \citenamefont {Niu}, \citenamefont {Yu},\ and\
  \citenamefont {Liu}}]{wang2020realization}%
  \BibitemOpen
  \bibfield  {author} {\bibinfo {author} {\bibfnamefont {Yucheng}\ \bibnamefont
  {Wang}}, \bibinfo {author} {\bibfnamefont {Long}\ \bibnamefont {Zhang}},
  \bibinfo {author} {\bibfnamefont {Sen}\ \bibnamefont {Niu}}, \bibinfo
  {author} {\bibfnamefont {Dapeng}\ \bibnamefont {Yu}}, \ and\ \bibinfo
  {author} {\bibfnamefont {Xiong-Jun}\ \bibnamefont {Liu}},\ }\bibfield
  {title} {\enquote {\bibinfo {title} {Realization and detection of nonergodic
  critical phases in an optical raman lattice},}\ }\href@noop {} {\bibfield
  {journal} {\bibinfo  {journal} {Physical Review Letters}\ }\textbf {\bibinfo
  {volume} {125}},\ \bibinfo {pages} {073204} (\bibinfo {year}
  {2020})}\BibitemShut {NoStop}%
\bibitem [{\citenamefont {Zhou}\ \emph {et~al.}(2022)\citenamefont {Zhou},
  \citenamefont {Wang}, \citenamefont {Poon}, \citenamefont {Zhou},\ and\
  \citenamefont {Liu}}]{zhou2022exact}%
  \BibitemOpen
  \bibfield  {author} {\bibinfo {author} {\bibfnamefont {Xin-Chi}\ \bibnamefont
  {Zhou}}, \bibinfo {author} {\bibfnamefont {Yongjian}\ \bibnamefont {Wang}},
  \bibinfo {author} {\bibfnamefont {Ting-Fung~Jeffrey}\ \bibnamefont {Poon}},
  \bibinfo {author} {\bibfnamefont {Qi}~\bibnamefont {Zhou}}, \ and\ \bibinfo
  {author} {\bibfnamefont {Xiong-Jun}\ \bibnamefont {Liu}},\ }\bibfield
  {title} {\enquote {\bibinfo {title} {Exact new mobility edges between
  critical and localized states},}\ }\href@noop {} {\bibfield  {journal}
  {\bibinfo  {journal} {arXiv preprint arXiv:2212.14285}\ } (\bibinfo {year}
  {2022})}\BibitemShut {NoStop}%
\bibitem [{\citenamefont {Deng}\ \emph {et~al.}(2019)\citenamefont {Deng},
  \citenamefont {Ray}, \citenamefont {Sinha}, \citenamefont {Shlyapnikov},\
  and\ \citenamefont {Santos}}]{deng2019one}%
  \BibitemOpen
  \bibfield  {author} {\bibinfo {author} {\bibfnamefont {X}~\bibnamefont
  {Deng}}, \bibinfo {author} {\bibfnamefont {S}~\bibnamefont {Ray}}, \bibinfo
  {author} {\bibfnamefont {S}~\bibnamefont {Sinha}}, \bibinfo {author}
  {\bibfnamefont {GV}~\bibnamefont {Shlyapnikov}}, \ and\ \bibinfo {author}
  {\bibfnamefont {L}~\bibnamefont {Santos}},\ }\bibfield  {title} {\enquote
  {\bibinfo {title} {One-dimensional quasicrystals with power-law hopping},}\
  }\href@noop {} {\bibfield  {journal} {\bibinfo  {journal} {Physical review
  letters}\ }\textbf {\bibinfo {volume} {123}},\ \bibinfo {pages} {025301}
  (\bibinfo {year} {2019})}\BibitemShut {NoStop}%
\bibitem [{\citenamefont {Roy}\ and\ \citenamefont
  {Sharma}(2021)}]{roy2021fraction}%
  \BibitemOpen
  \bibfield  {author} {\bibinfo {author} {\bibfnamefont {Nilanjan}\
  \bibnamefont {Roy}}\ and\ \bibinfo {author} {\bibfnamefont {Auditya}\
  \bibnamefont {Sharma}},\ }\bibfield  {title} {\enquote {\bibinfo {title}
  {Fraction of delocalized eigenstates in the long-range aubry-andr{\'e}-harper
  model},}\ }\href@noop {} {\bibfield  {journal} {\bibinfo  {journal} {Physical
  Review B}\ }\textbf {\bibinfo {volume} {103}},\ \bibinfo {pages} {075124}
  (\bibinfo {year} {2021})}\BibitemShut {NoStop}%
\bibitem [{\citenamefont {Lee}\ \emph {et~al.}(2023)\citenamefont {Lee},
  \citenamefont {Andreanov},\ and\ \citenamefont {Flach}}]{lee2023critical}%
  \BibitemOpen
  \bibfield  {author} {\bibinfo {author} {\bibfnamefont {Sanghoon}\
  \bibnamefont {Lee}}, \bibinfo {author} {\bibfnamefont {Alexei}\ \bibnamefont
  {Andreanov}}, \ and\ \bibinfo {author} {\bibfnamefont {Sergej}\ \bibnamefont
  {Flach}},\ }\bibfield  {title} {\enquote {\bibinfo {title}
  {Critical-to-insulator transitions and fractality edges in perturbed flat
  bands},}\ }\href@noop {} {\bibfield  {journal} {\bibinfo  {journal} {Physical
  Review B}\ }\textbf {\bibinfo {volume} {107}},\ \bibinfo {pages} {014204}
  (\bibinfo {year} {2023})}\BibitemShut {NoStop}%
\bibitem [{\citenamefont {Ahmed}\ \emph {et~al.}(2022)\citenamefont {Ahmed},
  \citenamefont {Ramachandran}, \citenamefont {Khaymovich},\ and\ \citenamefont
  {Sharma}}]{ahmed2022flat}%
  \BibitemOpen
  \bibfield  {author} {\bibinfo {author} {\bibfnamefont {Aamna}\ \bibnamefont
  {Ahmed}}, \bibinfo {author} {\bibfnamefont {Ajith}\ \bibnamefont
  {Ramachandran}}, \bibinfo {author} {\bibfnamefont {Ivan~M}\ \bibnamefont
  {Khaymovich}}, \ and\ \bibinfo {author} {\bibfnamefont {Auditya}\
  \bibnamefont {Sharma}},\ }\bibfield  {title} {\enquote {\bibinfo {title}
  {Flat band based multifractality in the all-band-flat diamond chain},}\
  }\href@noop {} {\bibfield  {journal} {\bibinfo  {journal} {Physical Review
  B}\ }\textbf {\bibinfo {volume} {106}},\ \bibinfo {pages} {205119} (\bibinfo
  {year} {2022})}\BibitemShut {NoStop}%
\bibitem [{\citenamefont {Abanin}\ and\ \citenamefont
  {Papi{\'c}}(2017)}]{abanin2017recent}%
  \BibitemOpen
  \bibfield  {author} {\bibinfo {author} {\bibfnamefont {Dmitry~A}\
  \bibnamefont {Abanin}}\ and\ \bibinfo {author} {\bibfnamefont {Zlatko}\
  \bibnamefont {Papi{\'c}}},\ }\bibfield  {title} {\enquote {\bibinfo {title}
  {Recent progress in many-body localization},}\ }\href@noop {} {\bibfield
  {journal} {\bibinfo  {journal} {Annalen der Physik}\ }\textbf {\bibinfo
  {volume} {529}},\ \bibinfo {pages} {1700169} (\bibinfo {year}
  {2017})}\BibitemShut {NoStop}%
\bibitem [{\citenamefont {Alet}\ and\ \citenamefont
  {Laflorencie}(2018)}]{alet2018many}%
  \BibitemOpen
  \bibfield  {author} {\bibinfo {author} {\bibfnamefont {Fabien}\ \bibnamefont
  {Alet}}\ and\ \bibinfo {author} {\bibfnamefont {Nicolas}\ \bibnamefont
  {Laflorencie}},\ }\bibfield  {title} {\enquote {\bibinfo {title} {Many-body
  localization: An introduction and selected topics},}\ }\href@noop {}
  {\bibfield  {journal} {\bibinfo  {journal} {Comptes Rendus Physique}\
  }\textbf {\bibinfo {volume} {19}},\ \bibinfo {pages} {498--525} (\bibinfo
  {year} {2018})}\BibitemShut {NoStop}%
\bibitem [{\citenamefont {Abanin}\ \emph {et~al.}(2019)\citenamefont {Abanin},
  \citenamefont {Altman}, \citenamefont {Bloch},\ and\ \citenamefont
  {Serbyn}}]{abanin2019colloquium}%
  \BibitemOpen
  \bibfield  {author} {\bibinfo {author} {\bibfnamefont {Dmitry~A}\
  \bibnamefont {Abanin}}, \bibinfo {author} {\bibfnamefont {Ehud}\ \bibnamefont
  {Altman}}, \bibinfo {author} {\bibfnamefont {Immanuel}\ \bibnamefont
  {Bloch}}, \ and\ \bibinfo {author} {\bibfnamefont {Maksym}\ \bibnamefont
  {Serbyn}},\ }\bibfield  {title} {\enquote {\bibinfo {title} {Colloquium:
  Many-body localization, thermalization, and entanglement},}\ }\href@noop {}
  {\bibfield  {journal} {\bibinfo  {journal} {Reviews of Modern Physics}\
  }\textbf {\bibinfo {volume} {91}},\ \bibinfo {pages} {021001} (\bibinfo
  {year} {2019})}\BibitemShut {NoStop}%
\bibitem [{\citenamefont {Basko}\ \emph {et~al.}(2006)\citenamefont {Basko},
  \citenamefont {Aleiner},\ and\ \citenamefont {Altshuler}}]{baa}%
  \BibitemOpen
  \bibfield  {author} {\bibinfo {author} {\bibfnamefont {DM}~\bibnamefont
  {Basko}}, \bibinfo {author} {\bibfnamefont {IL}~\bibnamefont {Aleiner}}, \
  and\ \bibinfo {author} {\bibfnamefont {BL}~\bibnamefont {Altshuler}},\
  }\bibfield  {title} {\enquote {\bibinfo {title} {Metal--insulator transition
  in a weakly interacting many-electron system with localized single-particle
  states},}\ }\href@noop {} {\bibfield  {journal} {\bibinfo  {journal} {Annals
  of physics}\ }\textbf {\bibinfo {volume} {321}},\ \bibinfo {pages}
  {1126--1205} (\bibinfo {year} {2006})}\BibitemShut {NoStop}%
\bibitem [{\citenamefont {Gornyi}\ \emph {et~al.}(2005)\citenamefont {Gornyi},
  \citenamefont {Mirlin},\ and\ \citenamefont
  {Polyakov}}]{gornyi2005interacting}%
  \BibitemOpen
  \bibfield  {author} {\bibinfo {author} {\bibfnamefont {Igor~V}\ \bibnamefont
  {Gornyi}}, \bibinfo {author} {\bibfnamefont {Alexander~D}\ \bibnamefont
  {Mirlin}}, \ and\ \bibinfo {author} {\bibfnamefont {Dmitry~G}\ \bibnamefont
  {Polyakov}},\ }\bibfield  {title} {\enquote {\bibinfo {title} {Interacting
  electrons in disordered wires: Anderson localization and low-t transport},}\
  }\href@noop {} {\bibfield  {journal} {\bibinfo  {journal} {Physical review
  letters}\ }\textbf {\bibinfo {volume} {95}},\ \bibinfo {pages} {206603}
  (\bibinfo {year} {2005})}\BibitemShut {NoStop}%
\bibitem [{\citenamefont {Oganesyan}\ and\ \citenamefont
  {Huse}(2007)}]{oganesyan2007localization}%
  \BibitemOpen
  \bibfield  {author} {\bibinfo {author} {\bibfnamefont {Vadim}\ \bibnamefont
  {Oganesyan}}\ and\ \bibinfo {author} {\bibfnamefont {David~A}\ \bibnamefont
  {Huse}},\ }\bibfield  {title} {\enquote {\bibinfo {title} {Localization of
  interacting fermions at high temperature},}\ }\href@noop {} {\bibfield
  {journal} {\bibinfo  {journal} {Physical review b}\ }\textbf {\bibinfo
  {volume} {75}},\ \bibinfo {pages} {155111} (\bibinfo {year}
  {2007})}\BibitemShut {NoStop}%
\bibitem [{\citenamefont {Pal}\ and\ \citenamefont {Huse}(2010)}]{pal2010many}%
  \BibitemOpen
  \bibfield  {author} {\bibinfo {author} {\bibfnamefont {Arijeet}\ \bibnamefont
  {Pal}}\ and\ \bibinfo {author} {\bibfnamefont {David~A}\ \bibnamefont
  {Huse}},\ }\bibfield  {title} {\enquote {\bibinfo {title} {Many-body
  localization phase transition},}\ }\href@noop {} {\bibfield  {journal}
  {\bibinfo  {journal} {Physical review b}\ }\textbf {\bibinfo {volume} {82}},\
  \bibinfo {pages} {174411} (\bibinfo {year} {2010})}\BibitemShut {NoStop}%
\bibitem [{\citenamefont {Luitz}\ \emph {et~al.}(2015)\citenamefont {Luitz},
  \citenamefont {Laflorencie},\ and\ \citenamefont {Alet}}]{luitz2015many}%
  \BibitemOpen
  \bibfield  {author} {\bibinfo {author} {\bibfnamefont {David~J}\ \bibnamefont
  {Luitz}}, \bibinfo {author} {\bibfnamefont {Nicolas}\ \bibnamefont
  {Laflorencie}}, \ and\ \bibinfo {author} {\bibfnamefont {Fabien}\
  \bibnamefont {Alet}},\ }\bibfield  {title} {\enquote {\bibinfo {title}
  {Many-body localization edge in the random-field heisenberg chain},}\
  }\href@noop {} {\bibfield  {journal} {\bibinfo  {journal} {Physical Review
  B}\ }\textbf {\bibinfo {volume} {91}},\ \bibinfo {pages} {081103} (\bibinfo
  {year} {2015})}\BibitemShut {NoStop}%
\bibitem [{\citenamefont {Imbrie}(2016)}]{imbrie2016many}%
  \BibitemOpen
  \bibfield  {author} {\bibinfo {author} {\bibfnamefont {John~Z}\ \bibnamefont
  {Imbrie}},\ }\bibfield  {title} {\enquote {\bibinfo {title} {On many-body
  localization for quantum spin chains},}\ }\href@noop {} {\bibfield  {journal}
  {\bibinfo  {journal} {Journal of Statistical Physics}\ }\textbf {\bibinfo
  {volume} {163}},\ \bibinfo {pages} {998--1048} (\bibinfo {year}
  {2016})}\BibitemShut {NoStop}%
\bibitem [{\citenamefont {De~Roeck}\ and\ \citenamefont
  {Huveneers}(2017)}]{de2017stability}%
  \BibitemOpen
  \bibfield  {author} {\bibinfo {author} {\bibfnamefont {Wojciech}\
  \bibnamefont {De~Roeck}}\ and\ \bibinfo {author} {\bibfnamefont
  {Fran{\c{c}}ois}\ \bibnamefont {Huveneers}},\ }\bibfield  {title} {\enquote
  {\bibinfo {title} {Stability and instability towards delocalization in
  many-body localization systems},}\ }\href@noop {} {\bibfield  {journal}
  {\bibinfo  {journal} {Physical Review B}\ }\textbf {\bibinfo {volume} {95}},\
  \bibinfo {pages} {155129} (\bibinfo {year} {2017})}\BibitemShut {NoStop}%
\bibitem [{\citenamefont {Schreiber}\ \emph {et~al.}(2015)\citenamefont
  {Schreiber}, \citenamefont {Hodgman}, \citenamefont {Bordia}, \citenamefont
  {L{\"u}schen}, \citenamefont {Fischer}, \citenamefont {Vosk}, \citenamefont
  {Altman}, \citenamefont {Schneider},\ and\ \citenamefont
  {Bloch}}]{schreiber2015observation}%
  \BibitemOpen
  \bibfield  {author} {\bibinfo {author} {\bibfnamefont {Michael}\ \bibnamefont
  {Schreiber}}, \bibinfo {author} {\bibfnamefont {Sean~S}\ \bibnamefont
  {Hodgman}}, \bibinfo {author} {\bibfnamefont {Pranjal}\ \bibnamefont
  {Bordia}}, \bibinfo {author} {\bibfnamefont {Henrik~P}\ \bibnamefont
  {L{\"u}schen}}, \bibinfo {author} {\bibfnamefont {Mark~H}\ \bibnamefont
  {Fischer}}, \bibinfo {author} {\bibfnamefont {Ronen}\ \bibnamefont {Vosk}},
  \bibinfo {author} {\bibfnamefont {Ehud}\ \bibnamefont {Altman}}, \bibinfo
  {author} {\bibfnamefont {Ulrich}\ \bibnamefont {Schneider}}, \ and\ \bibinfo
  {author} {\bibfnamefont {Immanuel}\ \bibnamefont {Bloch}},\ }\bibfield
  {title} {\enquote {\bibinfo {title} {Observation of many-body localization of
  interacting fermions in a quasirandom optical lattice},}\ }\href@noop {}
  {\bibfield  {journal} {\bibinfo  {journal} {Science}\ }\textbf {\bibinfo
  {volume} {349}},\ \bibinfo {pages} {842--845} (\bibinfo {year}
  {2015})}\BibitemShut {NoStop}%
\bibitem [{\citenamefont {Rispoli}\ \emph {et~al.}(2019)\citenamefont
  {Rispoli}, \citenamefont {Lukin}, \citenamefont {Schittko}, \citenamefont
  {Kim}, \citenamefont {Tai}, \citenamefont {L{\'e}onard},\ and\ \citenamefont
  {Greiner}}]{rispoli2019quantum}%
  \BibitemOpen
  \bibfield  {author} {\bibinfo {author} {\bibfnamefont {Matthew}\ \bibnamefont
  {Rispoli}}, \bibinfo {author} {\bibfnamefont {Alexander}\ \bibnamefont
  {Lukin}}, \bibinfo {author} {\bibfnamefont {Robert}\ \bibnamefont
  {Schittko}}, \bibinfo {author} {\bibfnamefont {Sooshin}\ \bibnamefont {Kim}},
  \bibinfo {author} {\bibfnamefont {M~Eric}\ \bibnamefont {Tai}}, \bibinfo
  {author} {\bibfnamefont {Julian}\ \bibnamefont {L{\'e}onard}}, \ and\
  \bibinfo {author} {\bibfnamefont {Markus}\ \bibnamefont {Greiner}},\
  }\bibfield  {title} {\enquote {\bibinfo {title} {Quantum critical behaviour
  at the many-body localization transition},}\ }\href@noop {} {\bibfield
  {journal} {\bibinfo  {journal} {Nature}\ }\textbf {\bibinfo {volume} {573}},\
  \bibinfo {pages} {385--389} (\bibinfo {year} {2019})}\BibitemShut {NoStop}%
\bibitem [{\citenamefont {Lukin}\ \emph {et~al.}(2019)\citenamefont {Lukin},
  \citenamefont {Rispoli}, \citenamefont {Schittko}, \citenamefont {Tai},
  \citenamefont {Kaufman}, \citenamefont {Choi}, \citenamefont {Khemani},
  \citenamefont {L{\'e}onard},\ and\ \citenamefont
  {Greiner}}]{lukin2019probing}%
  \BibitemOpen
  \bibfield  {author} {\bibinfo {author} {\bibfnamefont {Alexander}\
  \bibnamefont {Lukin}}, \bibinfo {author} {\bibfnamefont {Matthew}\
  \bibnamefont {Rispoli}}, \bibinfo {author} {\bibfnamefont {Robert}\
  \bibnamefont {Schittko}}, \bibinfo {author} {\bibfnamefont {M~Eric}\
  \bibnamefont {Tai}}, \bibinfo {author} {\bibfnamefont {Adam~M}\ \bibnamefont
  {Kaufman}}, \bibinfo {author} {\bibfnamefont {Soonwon}\ \bibnamefont {Choi}},
  \bibinfo {author} {\bibfnamefont {Vedika}\ \bibnamefont {Khemani}}, \bibinfo
  {author} {\bibfnamefont {Julian}\ \bibnamefont {L{\'e}onard}}, \ and\
  \bibinfo {author} {\bibfnamefont {Markus}\ \bibnamefont {Greiner}},\
  }\bibfield  {title} {\enquote {\bibinfo {title} {Probing entanglement in a
  many-body--localized system},}\ }\href@noop {} {\bibfield  {journal}
  {\bibinfo  {journal} {Science}\ }\textbf {\bibinfo {volume} {364}},\ \bibinfo
  {pages} {256--260} (\bibinfo {year} {2019})}\BibitemShut {NoStop}%
\bibitem [{\citenamefont {Sels}\ and\ \citenamefont
  {Polkovnikov}(2021)}]{sels2021dynamical}%
  \BibitemOpen
  \bibfield  {author} {\bibinfo {author} {\bibfnamefont {Dries}\ \bibnamefont
  {Sels}}\ and\ \bibinfo {author} {\bibfnamefont {Anatoli}\ \bibnamefont
  {Polkovnikov}},\ }\bibfield  {title} {\enquote {\bibinfo {title} {Dynamical
  obstruction to localization in a disordered spin chain},}\ }\href@noop {}
  {\bibfield  {journal} {\bibinfo  {journal} {Physical Review E}\ }\textbf
  {\bibinfo {volume} {104}},\ \bibinfo {pages} {054105} (\bibinfo {year}
  {2021})}\BibitemShut {NoStop}%
\bibitem [{\citenamefont {Morningstar}\ \emph {et~al.}(2022)\citenamefont
  {Morningstar}, \citenamefont {Colmenarez}, \citenamefont {Khemani},
  \citenamefont {Luitz},\ and\ \citenamefont
  {Huse}}]{morningstar2022avalanches}%
  \BibitemOpen
  \bibfield  {author} {\bibinfo {author} {\bibfnamefont {Alan}\ \bibnamefont
  {Morningstar}}, \bibinfo {author} {\bibfnamefont {Luis}\ \bibnamefont
  {Colmenarez}}, \bibinfo {author} {\bibfnamefont {Vedika}\ \bibnamefont
  {Khemani}}, \bibinfo {author} {\bibfnamefont {David~J}\ \bibnamefont
  {Luitz}}, \ and\ \bibinfo {author} {\bibfnamefont {David~A}\ \bibnamefont
  {Huse}},\ }\bibfield  {title} {\enquote {\bibinfo {title} {Avalanches and
  many-body resonances in many-body localized systems},}\ }\href@noop {}
  {\bibfield  {journal} {\bibinfo  {journal} {Physical Review B}\ }\textbf
  {\bibinfo {volume} {105}},\ \bibinfo {pages} {174205} (\bibinfo {year}
  {2022})}\BibitemShut {NoStop}%
\bibitem [{\citenamefont {Crowley}\ and\ \citenamefont
  {Chandran}(2022)}]{crowley2022constructive}%
  \BibitemOpen
  \bibfield  {author} {\bibinfo {author} {\bibfnamefont {Philip}\ \bibnamefont
  {Crowley}}\ and\ \bibinfo {author} {\bibfnamefont {Anushya}\ \bibnamefont
  {Chandran}},\ }\bibfield  {title} {\enquote {\bibinfo {title} {A constructive
  theory of the numerically accessible many-body localized to thermal
  crossover},}\ }\href@noop {} {\bibfield  {journal} {\bibinfo  {journal}
  {SciPost Physics}\ }\textbf {\bibinfo {volume} {12}},\ \bibinfo {pages} {201}
  (\bibinfo {year} {2022})}\BibitemShut {NoStop}%
\bibitem [{\citenamefont {Long}\ \emph {et~al.}(2023)\citenamefont {Long},
  \citenamefont {Crowley}, \citenamefont {Khemani},\ and\ \citenamefont
  {Chandran}}]{long2022phenomenology}%
  \BibitemOpen
  \bibfield  {author} {\bibinfo {author} {\bibfnamefont {David~M}\ \bibnamefont
  {Long}}, \bibinfo {author} {\bibfnamefont {Philip~JD}\ \bibnamefont
  {Crowley}}, \bibinfo {author} {\bibfnamefont {Vedika}\ \bibnamefont
  {Khemani}}, \ and\ \bibinfo {author} {\bibfnamefont {Anushya}\ \bibnamefont
  {Chandran}},\ }\bibfield  {title} {\enquote {\bibinfo {title} {Phenomenology
  of the prethermal many-body localized regime},}\ }\href@noop {} {\bibfield
  {journal} {\bibinfo  {journal} {Physical Review Letters}\ }\textbf {\bibinfo
  {volume} {131}},\ \bibinfo {pages} {106301} (\bibinfo {year}
  {2023})}\BibitemShut {NoStop}%
\bibitem [{\citenamefont {Sels}(2022)}]{sels2022bath}%
  \BibitemOpen
  \bibfield  {author} {\bibinfo {author} {\bibfnamefont {Dries}\ \bibnamefont
  {Sels}},\ }\bibfield  {title} {\enquote {\bibinfo {title} {Bath-induced
  delocalization in interacting disordered spin chains},}\ }\href@noop {}
  {\bibfield  {journal} {\bibinfo  {journal} {Physical Review B}\ }\textbf
  {\bibinfo {volume} {106}},\ \bibinfo {pages} {L020202} (\bibinfo {year}
  {2022})}\BibitemShut {NoStop}%
\bibitem [{\citenamefont {{\v{S}}untajs}\ \emph {et~al.}(2020)\citenamefont
  {{\v{S}}untajs}, \citenamefont {Bon{\v{c}}a}, \citenamefont {Prosen},\ and\
  \citenamefont {Vidmar}}]{Suntajs2020quantum}%
  \BibitemOpen
  \bibfield  {author} {\bibinfo {author} {\bibfnamefont {Jan}\ \bibnamefont
  {{\v{S}}untajs}}, \bibinfo {author} {\bibfnamefont {Janez}\ \bibnamefont
  {Bon{\v{c}}a}}, \bibinfo {author} {\bibfnamefont {Toma{\v{z}}}\ \bibnamefont
  {Prosen}}, \ and\ \bibinfo {author} {\bibfnamefont {Lev}\ \bibnamefont
  {Vidmar}},\ }\bibfield  {title} {\enquote {\bibinfo {title} {Quantum chaos
  challenges many-body localization},}\ }\href@noop {} {\bibfield  {journal}
  {\bibinfo  {journal} {Physical Review E}\ }\textbf {\bibinfo {volume}
  {102}},\ \bibinfo {pages} {062144} (\bibinfo {year} {2020})}\BibitemShut
  {NoStop}%
\bibitem [{\citenamefont {Sierant}\ and\ \citenamefont
  {Zakrzewski}(2022)}]{sierant2022challenges}%
  \BibitemOpen
  \bibfield  {author} {\bibinfo {author} {\bibfnamefont {Piotr}\ \bibnamefont
  {Sierant}}\ and\ \bibinfo {author} {\bibfnamefont {Jakub}\ \bibnamefont
  {Zakrzewski}},\ }\bibfield  {title} {\enquote {\bibinfo {title} {Challenges
  to observation of many-body localization},}\ }\href@noop {} {\bibfield
  {journal} {\bibinfo  {journal} {Physical Review B}\ }\textbf {\bibinfo
  {volume} {105}},\ \bibinfo {pages} {224203} (\bibinfo {year}
  {2022})}\BibitemShut {NoStop}%
\bibitem [{\citenamefont {Sierant}\ \emph {et~al.}()\citenamefont {Sierant},
  \citenamefont {Lewenstein}, \citenamefont {Scardicchio}, \citenamefont
  {Vidmar},\ and\ \citenamefont {Zakrzewski}}]{sierant2403many}%
  \BibitemOpen
  \bibfield  {author} {\bibinfo {author} {\bibfnamefont {P}~\bibnamefont
  {Sierant}}, \bibinfo {author} {\bibfnamefont {M}~\bibnamefont {Lewenstein}},
  \bibinfo {author} {\bibfnamefont {A}~\bibnamefont {Scardicchio}}, \bibinfo
  {author} {\bibfnamefont {L}~\bibnamefont {Vidmar}}, \ and\ \bibinfo {author}
  {\bibfnamefont {J}~\bibnamefont {Zakrzewski}},\ }\bibfield  {title} {\enquote
  {\bibinfo {title} {Many-body localization in the age of classical computing
  (2024)},}\ }\href@noop {} {\bibinfo  {journal} {arXiv preprint
  arXiv:2403.07111}\ }\BibitemShut {NoStop}%
\bibitem [{\citenamefont {Deutsch}(1991)}]{deutsch1991quantum}%
  \BibitemOpen
\bibfield  {journal} {  }\bibfield  {author} {\bibinfo {author} {\bibfnamefont
  {Josh~M}\ \bibnamefont {Deutsch}},\ }\bibfield  {title} {\enquote {\bibinfo
  {title} {Quantum statistical mechanics in a closed system},}\ }\href@noop {}
  {\bibfield  {journal} {\bibinfo  {journal} {Physical Review A}\ }\textbf
  {\bibinfo {volume} {43}},\ \bibinfo {pages} {2046} (\bibinfo {year}
  {1991})}\BibitemShut {NoStop}%
\bibitem [{\citenamefont {Srednicki}(1994)}]{srednicki1994chaos}%
  \BibitemOpen
  \bibfield  {author} {\bibinfo {author} {\bibfnamefont {Mark}\ \bibnamefont
  {Srednicki}},\ }\bibfield  {title} {\enquote {\bibinfo {title} {Chaos and
  quantum thermalization},}\ }\href@noop {} {\bibfield  {journal} {\bibinfo
  {journal} {Physical Review E}\ }\textbf {\bibinfo {volume} {50}},\ \bibinfo
  {pages} {888} (\bibinfo {year} {1994})}\BibitemShut {NoStop}%
\bibitem [{\citenamefont {Zhang}\ and\ \citenamefont
  {Yao}(2018)}]{zhang2018universal}%
  \BibitemOpen
  \bibfield  {author} {\bibinfo {author} {\bibfnamefont {Shi-Xin}\ \bibnamefont
  {Zhang}}\ and\ \bibinfo {author} {\bibfnamefont {Hong}\ \bibnamefont {Yao}},\
  }\bibfield  {title} {\enquote {\bibinfo {title} {Universal properties of
  many-body localization transitions in quasiperiodic systems},}\ }\href@noop
  {} {\bibfield  {journal} {\bibinfo  {journal} {Physical review letters}\
  }\textbf {\bibinfo {volume} {121}},\ \bibinfo {pages} {206601} (\bibinfo
  {year} {2018})}\BibitemShut {NoStop}%
\bibitem [{\citenamefont {Dumitrescu}\ \emph {et~al.}(2019)\citenamefont
  {Dumitrescu}, \citenamefont {Goremykina}, \citenamefont {Parameswaran},
  \citenamefont {Serbyn},\ and\ \citenamefont
  {Vasseur}}]{dumitrescu2019kosterlitz}%
  \BibitemOpen
  \bibfield  {author} {\bibinfo {author} {\bibfnamefont {Philipp~T}\
  \bibnamefont {Dumitrescu}}, \bibinfo {author} {\bibfnamefont {Anna}\
  \bibnamefont {Goremykina}}, \bibinfo {author} {\bibfnamefont {Siddharth~A}\
  \bibnamefont {Parameswaran}}, \bibinfo {author} {\bibfnamefont {Maksym}\
  \bibnamefont {Serbyn}}, \ and\ \bibinfo {author} {\bibfnamefont {Romain}\
  \bibnamefont {Vasseur}},\ }\bibfield  {title} {\enquote {\bibinfo {title}
  {Kosterlitz-thouless scaling at many-body localization phase transitions},}\
  }\href@noop {} {\bibfield  {journal} {\bibinfo  {journal} {Physical Review
  B}\ }\textbf {\bibinfo {volume} {99}},\ \bibinfo {pages} {094205} (\bibinfo
  {year} {2019})}\BibitemShut {NoStop}%
\bibitem [{\citenamefont {Kiefer-Emmanouilidis}\ \emph
  {et~al.}(2020)\citenamefont {Kiefer-Emmanouilidis}, \citenamefont {Unanyan},
  \citenamefont {Fleischhauer},\ and\ \citenamefont
  {Sirker}}]{kiefer2020evidence}%
  \BibitemOpen
  \bibfield  {author} {\bibinfo {author} {\bibfnamefont {Maximilian}\
  \bibnamefont {Kiefer-Emmanouilidis}}, \bibinfo {author} {\bibfnamefont
  {Razmik}\ \bibnamefont {Unanyan}}, \bibinfo {author} {\bibfnamefont
  {Michael}\ \bibnamefont {Fleischhauer}}, \ and\ \bibinfo {author}
  {\bibfnamefont {Jesko}\ \bibnamefont {Sirker}},\ }\bibfield  {title}
  {\enquote {\bibinfo {title} {Evidence for unbounded growth of the number
  entropy in many-body localized phases},}\ }\href@noop {} {\bibfield
  {journal} {\bibinfo  {journal} {Physical Review Letters}\ }\textbf {\bibinfo
  {volume} {124}},\ \bibinfo {pages} {243601} (\bibinfo {year}
  {2020})}\BibitemShut {NoStop}%
\bibitem [{\citenamefont {Abanin}\ \emph {et~al.}(2021)\citenamefont {Abanin},
  \citenamefont {Bardarson}, \citenamefont {De~Tomasi}, \citenamefont
  {Gopalakrishnan}, \citenamefont {Khemani}, \citenamefont {Parameswaran},
  \citenamefont {Pollmann}, \citenamefont {Potter}, \citenamefont {Serbyn},\
  and\ \citenamefont {Vasseur}}]{abanin2021distinguishing}%
  \BibitemOpen
  \bibfield  {author} {\bibinfo {author} {\bibfnamefont {DA}~\bibnamefont
  {Abanin}}, \bibinfo {author} {\bibfnamefont {Jens~H}\ \bibnamefont
  {Bardarson}}, \bibinfo {author} {\bibfnamefont {G}~\bibnamefont {De~Tomasi}},
  \bibinfo {author} {\bibfnamefont {S}~\bibnamefont {Gopalakrishnan}}, \bibinfo
  {author} {\bibfnamefont {V}~\bibnamefont {Khemani}}, \bibinfo {author}
  {\bibfnamefont {SA}~\bibnamefont {Parameswaran}}, \bibinfo {author}
  {\bibfnamefont {F}~\bibnamefont {Pollmann}}, \bibinfo {author} {\bibfnamefont
  {AC}~\bibnamefont {Potter}}, \bibinfo {author} {\bibfnamefont {Maksym}\
  \bibnamefont {Serbyn}}, \ and\ \bibinfo {author} {\bibfnamefont
  {R}~\bibnamefont {Vasseur}},\ }\bibfield  {title} {\enquote {\bibinfo {title}
  {Distinguishing localization from chaos: Challenges in finite-size
  systems},}\ }\href@noop {} {\bibfield  {journal} {\bibinfo  {journal} {Annals
  of Physics}\ }\textbf {\bibinfo {volume} {427}},\ \bibinfo {pages} {168415}
  (\bibinfo {year} {2021})}\BibitemShut {NoStop}%
\bibitem [{\citenamefont {Sierant}\ \emph {et~al.}(2020)\citenamefont
  {Sierant}, \citenamefont {Lewenstein},\ and\ \citenamefont
  {Zakrzewski}}]{Sierant2020}%
  \BibitemOpen
  \bibfield  {author} {\bibinfo {author} {\bibfnamefont {Piotr}\ \bibnamefont
  {Sierant}}, \bibinfo {author} {\bibfnamefont {Maciej}\ \bibnamefont
  {Lewenstein}}, \ and\ \bibinfo {author} {\bibfnamefont {Jakub}\ \bibnamefont
  {Zakrzewski}},\ }\bibfield  {title} {\enquote {\bibinfo {title} {Polynomially
  filtered exact diagonalization approach to many-body localization},}\
  }\href@noop {} {\bibfield  {journal} {\bibinfo  {journal} {Physical Review
  Letters}\ }\textbf {\bibinfo {volume} {125}},\ \bibinfo {pages} {156601}
  (\bibinfo {year} {2020})}\BibitemShut {NoStop}%
\bibitem [{\citenamefont {Laflorencie}\ \emph {et~al.}(2020)\citenamefont
  {Laflorencie}, \citenamefont {Lemari{\'e}},\ and\ \citenamefont
  {Mac{\'e}}}]{Laflorencie2020}%
  \BibitemOpen
  \bibfield  {author} {\bibinfo {author} {\bibfnamefont {Nicolas}\ \bibnamefont
  {Laflorencie}}, \bibinfo {author} {\bibfnamefont {Gabriel}\ \bibnamefont
  {Lemari{\'e}}}, \ and\ \bibinfo {author} {\bibfnamefont {Nicolas}\
  \bibnamefont {Mac{\'e}}},\ }\bibfield  {title} {\enquote {\bibinfo {title}
  {Chain breaking and kosterlitz-thouless scaling at the many-body localization
  transition in the random-field heisenberg spin chain},}\ }\href@noop {}
  {\bibfield  {journal} {\bibinfo  {journal} {Physical Review Research}\
  }\textbf {\bibinfo {volume} {2}},\ \bibinfo {pages} {042033} (\bibinfo {year}
  {2020})}\BibitemShut {NoStop}%
\bibitem [{\citenamefont {Kudo}\ and\ \citenamefont
  {Deguchi}(2018)}]{kudo2018finite}%
  \BibitemOpen
  \bibfield  {author} {\bibinfo {author} {\bibfnamefont {Kazue}\ \bibnamefont
  {Kudo}}\ and\ \bibinfo {author} {\bibfnamefont {Tetsuo}\ \bibnamefont
  {Deguchi}},\ }\bibfield  {title} {\enquote {\bibinfo {title} {Finite-size
  scaling with respect to interaction and disorder strength at the many-body
  localization transition},}\ }\href@noop {} {\bibfield  {journal} {\bibinfo
  {journal} {Physical Review B}\ }\textbf {\bibinfo {volume} {97}},\ \bibinfo
  {pages} {220201} (\bibinfo {year} {2018})}\BibitemShut {NoStop}%
\bibitem [{\citenamefont {Bertrand}\ and\ \citenamefont
  {Garc{\'\i}a-Garc{\'\i}a}(2016)}]{bertrand2016anomalous}%
  \BibitemOpen
  \bibfield  {author} {\bibinfo {author} {\bibfnamefont {Corentin~L}\
  \bibnamefont {Bertrand}}\ and\ \bibinfo {author} {\bibfnamefont {Antonio~M}\
  \bibnamefont {Garc{\'\i}a-Garc{\'\i}a}},\ }\bibfield  {title} {\enquote
  {\bibinfo {title} {Anomalous thouless energy and critical statistics on the
  metallic side of the many-body localization transition},}\ }\href@noop {}
  {\bibfield  {journal} {\bibinfo  {journal} {Physical Review B}\ }\textbf
  {\bibinfo {volume} {94}},\ \bibinfo {pages} {144201} (\bibinfo {year}
  {2016})}\BibitemShut {NoStop}%
\bibitem [{\citenamefont {Setiawan}\ \emph {et~al.}(2017)\citenamefont
  {Setiawan}, \citenamefont {Deng},\ and\ \citenamefont
  {Pixley}}]{setiawan2017transport}%
  \BibitemOpen
  \bibfield  {author} {\bibinfo {author} {\bibfnamefont {F}~\bibnamefont
  {Setiawan}}, \bibinfo {author} {\bibfnamefont {Dong-Ling}\ \bibnamefont
  {Deng}}, \ and\ \bibinfo {author} {\bibfnamefont {JH}~\bibnamefont
  {Pixley}},\ }\bibfield  {title} {\enquote {\bibinfo {title} {Transport
  properties across the many-body localization transition in quasiperiodic and
  random systems},}\ }\href@noop {} {\bibfield  {journal} {\bibinfo  {journal}
  {Physical Review B}\ }\textbf {\bibinfo {volume} {96}},\ \bibinfo {pages}
  {104205} (\bibinfo {year} {2017})}\BibitemShut {NoStop}%
\bibitem [{\citenamefont {Sutradhar}\ \emph {et~al.}(2022)\citenamefont
  {Sutradhar}, \citenamefont {Ghosh}, \citenamefont {Roy}, \citenamefont
  {Logan}, \citenamefont {Mukerjee},\ and\ \citenamefont
  {Banerjee}}]{sutradhar2022scaling}%
  \BibitemOpen
  \bibfield  {author} {\bibinfo {author} {\bibfnamefont {Jagannath}\
  \bibnamefont {Sutradhar}}, \bibinfo {author} {\bibfnamefont {Soumi}\
  \bibnamefont {Ghosh}}, \bibinfo {author} {\bibfnamefont {Sthitadhi}\
  \bibnamefont {Roy}}, \bibinfo {author} {\bibfnamefont {David~E}\ \bibnamefont
  {Logan}}, \bibinfo {author} {\bibfnamefont {Subroto}\ \bibnamefont
  {Mukerjee}}, \ and\ \bibinfo {author} {\bibfnamefont {Sumilan}\ \bibnamefont
  {Banerjee}},\ }\bibfield  {title} {\enquote {\bibinfo {title} {Scaling of the
  fock-space propagator and multifractality across the many-body localization
  transition},}\ }\href@noop {} {\bibfield  {journal} {\bibinfo  {journal}
  {Physical Review B}\ }\textbf {\bibinfo {volume} {106}},\ \bibinfo {pages}
  {054203} (\bibinfo {year} {2022})}\BibitemShut {NoStop}%
\bibitem [{\citenamefont {Vosk}\ and\ \citenamefont
  {Altman}(2013)}]{vosk2013many}%
  \BibitemOpen
  \bibfield  {author} {\bibinfo {author} {\bibfnamefont {Ronen}\ \bibnamefont
  {Vosk}}\ and\ \bibinfo {author} {\bibfnamefont {Ehud}\ \bibnamefont
  {Altman}},\ }\bibfield  {title} {\enquote {\bibinfo {title} {Many-body
  localization in one dimension as a dynamical<? format?> renormalization group
  fixed point},}\ }\href@noop {} {\bibfield  {journal} {\bibinfo  {journal}
  {Physical review letters}\ }\textbf {\bibinfo {volume} {110}},\ \bibinfo
  {pages} {067204} (\bibinfo {year} {2013})}\BibitemShut {NoStop}%
\bibitem [{\citenamefont {Goremykina}\ \emph {et~al.}(2019)\citenamefont
  {Goremykina}, \citenamefont {Vasseur},\ and\ \citenamefont
  {Serbyn}}]{goremykina2019analytically}%
  \BibitemOpen
  \bibfield  {author} {\bibinfo {author} {\bibfnamefont {Anna}\ \bibnamefont
  {Goremykina}}, \bibinfo {author} {\bibfnamefont {Romain}\ \bibnamefont
  {Vasseur}}, \ and\ \bibinfo {author} {\bibfnamefont {Maksym}\ \bibnamefont
  {Serbyn}},\ }\bibfield  {title} {\enquote {\bibinfo {title} {Analytically
  solvable renormalization group for the many-body localization transition},}\
  }\href@noop {} {\bibfield  {journal} {\bibinfo  {journal} {Physical review
  letters}\ }\textbf {\bibinfo {volume} {122}},\ \bibinfo {pages} {040601}
  (\bibinfo {year} {2019})}\BibitemShut {NoStop}%
\bibitem [{\citenamefont {Morningstar}\ \emph {et~al.}(2020)\citenamefont
  {Morningstar}, \citenamefont {Huse},\ and\ \citenamefont
  {Imbrie}}]{morningstar2020many}%
  \BibitemOpen
  \bibfield  {author} {\bibinfo {author} {\bibfnamefont {Alan}\ \bibnamefont
  {Morningstar}}, \bibinfo {author} {\bibfnamefont {David~A}\ \bibnamefont
  {Huse}}, \ and\ \bibinfo {author} {\bibfnamefont {John~Z}\ \bibnamefont
  {Imbrie}},\ }\bibfield  {title} {\enquote {\bibinfo {title} {Many-body
  localization near the critical point},}\ }\href@noop {} {\bibfield  {journal}
  {\bibinfo  {journal} {Physical Review B}\ }\textbf {\bibinfo {volume}
  {102}},\ \bibinfo {pages} {125134} (\bibinfo {year} {2020})}\BibitemShut
  {NoStop}%
\bibitem [{\citenamefont {Wang}\ \emph {et~al.}(2021)\citenamefont {Wang},
  \citenamefont {Cheng}, \citenamefont {Liu},\ and\ \citenamefont
  {Yu}}]{wang2021many}%
  \BibitemOpen
  \bibfield  {author} {\bibinfo {author} {\bibfnamefont {Yucheng}\ \bibnamefont
  {Wang}}, \bibinfo {author} {\bibfnamefont {Chen}\ \bibnamefont {Cheng}},
  \bibinfo {author} {\bibfnamefont {Xiong-Jun}\ \bibnamefont {Liu}}, \ and\
  \bibinfo {author} {\bibfnamefont {Dapeng}\ \bibnamefont {Yu}},\ }\bibfield
  {title} {\enquote {\bibinfo {title} {Many-body critical phase: extended and
  nonthermal},}\ }\href@noop {} {\bibfield  {journal} {\bibinfo  {journal}
  {Physical Review Letters}\ }\textbf {\bibinfo {volume} {126}},\ \bibinfo
  {pages} {080602} (\bibinfo {year} {2021})}\BibitemShut {NoStop}%
\bibitem [{\citenamefont {Welsh}\ and\ \citenamefont
  {Logan}(2018)}]{welsh2018simple}%
  \BibitemOpen
  \bibfield  {author} {\bibinfo {author} {\bibfnamefont {Staszek}\ \bibnamefont
  {Welsh}}\ and\ \bibinfo {author} {\bibfnamefont {David~E}\ \bibnamefont
  {Logan}},\ }\bibfield  {title} {\enquote {\bibinfo {title} {Simple
  probability distributions on a fock-space lattice},}\ }\href@noop {}
  {\bibfield  {journal} {\bibinfo  {journal} {Journal of Physics: Condensed
  Matter}\ }\textbf {\bibinfo {volume} {30}},\ \bibinfo {pages} {405601}
  (\bibinfo {year} {2018})}\BibitemShut {NoStop}%
\bibitem [{\citenamefont {Logan}\ and\ \citenamefont
  {Welsh}(2019)}]{logan2019many}%
  \BibitemOpen
  \bibfield  {author} {\bibinfo {author} {\bibfnamefont {David~E}\ \bibnamefont
  {Logan}}\ and\ \bibinfo {author} {\bibfnamefont {Staszek}\ \bibnamefont
  {Welsh}},\ }\bibfield  {title} {\enquote {\bibinfo {title} {Many-body
  localization in fock space: A local perspective},}\ }\href@noop {} {\bibfield
   {journal} {\bibinfo  {journal} {Physical Review B}\ }\textbf {\bibinfo
  {volume} {99}},\ \bibinfo {pages} {045131} (\bibinfo {year}
  {2019})}\BibitemShut {NoStop}%
\bibitem [{\citenamefont {Roy}\ and\ \citenamefont
  {Logan}(2020{\natexlab{a}})}]{roy2020fock}%
  \BibitemOpen
  \bibfield  {author} {\bibinfo {author} {\bibfnamefont {Sthitadhi}\
  \bibnamefont {Roy}}\ and\ \bibinfo {author} {\bibfnamefont {David~E}\
  \bibnamefont {Logan}},\ }\bibfield  {title} {\enquote {\bibinfo {title}
  {Fock-space correlations and the origins of many-body localization},}\
  }\href@noop {} {\bibfield  {journal} {\bibinfo  {journal} {Physical Review
  B}\ }\textbf {\bibinfo {volume} {101}},\ \bibinfo {pages} {134202} (\bibinfo
  {year} {2020}{\natexlab{a}})}\BibitemShut {NoStop}%
\bibitem [{\citenamefont {Roy}\ and\ \citenamefont
  {Logan}(2021)}]{roy2021fock}%
  \BibitemOpen
  \bibfield  {author} {\bibinfo {author} {\bibfnamefont {Sthitadhi}\
  \bibnamefont {Roy}}\ and\ \bibinfo {author} {\bibfnamefont {David~E}\
  \bibnamefont {Logan}},\ }\bibfield  {title} {\enquote {\bibinfo {title}
  {Fock-space anatomy of eigenstates across the many-body localization
  transition},}\ }\href@noop {} {\bibfield  {journal} {\bibinfo  {journal}
  {Physical Review B}\ }\textbf {\bibinfo {volume} {104}},\ \bibinfo {pages}
  {174201} (\bibinfo {year} {2021})}\BibitemShut {NoStop}%
\bibitem [{\citenamefont {Roy}(2022)}]{roy2022hilbert}%
  \BibitemOpen
  \bibfield  {author} {\bibinfo {author} {\bibfnamefont {Sthitadhi}\
  \bibnamefont {Roy}},\ }\bibfield  {title} {\enquote {\bibinfo {title}
  {Hilbert-space correlations beyond multifractality and bipartite entanglement
  in many-body localized systems},}\ }\href@noop {} {\bibfield  {journal}
  {\bibinfo  {journal} {Physical Review B}\ }\textbf {\bibinfo {volume}
  {106}},\ \bibinfo {pages} {L140204} (\bibinfo {year} {2022})}\BibitemShut
  {NoStop}%
\bibitem [{\citenamefont {Roy}\ \emph {et~al.}(2023)\citenamefont {Roy},
  \citenamefont {Sutradhar},\ and\ \citenamefont
  {Banerjee}}]{roy2023diagnostics}%
  \BibitemOpen
  \bibfield  {author} {\bibinfo {author} {\bibfnamefont {Nilanjan}\
  \bibnamefont {Roy}}, \bibinfo {author} {\bibfnamefont {Jagannath}\
  \bibnamefont {Sutradhar}}, \ and\ \bibinfo {author} {\bibfnamefont {Sumilan}\
  \bibnamefont {Banerjee}},\ }\bibfield  {title} {\enquote {\bibinfo {title}
  {Diagnostics of nonergodic extended states and many body localization
  proximity effect through real-space and fock-space excitations},}\
  }\href@noop {} {\bibfield  {journal} {\bibinfo  {journal} {Physical Review
  B}\ }\textbf {\bibinfo {volume} {107}},\ \bibinfo {pages} {115155} (\bibinfo
  {year} {2023})}\BibitemShut {NoStop}%
\bibitem [{\citenamefont {Ghosh}\ \emph {et~al.}(2019)\citenamefont {Ghosh},
  \citenamefont {Acharya}, \citenamefont {Sahu},\ and\ \citenamefont
  {Mukerjee}}]{ghosh2019many}%
  \BibitemOpen
  \bibfield  {author} {\bibinfo {author} {\bibfnamefont {Soumi}\ \bibnamefont
  {Ghosh}}, \bibinfo {author} {\bibfnamefont {Atithi}\ \bibnamefont {Acharya}},
  \bibinfo {author} {\bibfnamefont {Subhayan}\ \bibnamefont {Sahu}}, \ and\
  \bibinfo {author} {\bibfnamefont {Subroto}\ \bibnamefont {Mukerjee}},\
  }\bibfield  {title} {\enquote {\bibinfo {title} {Many-body localization due
  to correlated disorder in fock space},}\ }\href@noop {} {\bibfield  {journal}
  {\bibinfo  {journal} {Physical Review B}\ }\textbf {\bibinfo {volume} {99}},\
  \bibinfo {pages} {165131} (\bibinfo {year} {2019})}\BibitemShut {NoStop}%
\bibitem [{\citenamefont {Ghosh}\ \emph {et~al.}(2025)\citenamefont {Ghosh},
  \citenamefont {Sutradhar}, \citenamefont {Mukerjee},\ and\ \citenamefont
  {Banerjee}}]{ghosh2024scaling}%
  \BibitemOpen
  \bibfield  {author} {\bibinfo {author} {\bibfnamefont {Soumi}\ \bibnamefont
  {Ghosh}}, \bibinfo {author} {\bibfnamefont {Jagannath}\ \bibnamefont
  {Sutradhar}}, \bibinfo {author} {\bibfnamefont {Subroto}\ \bibnamefont
  {Mukerjee}}, \ and\ \bibinfo {author} {\bibfnamefont {Sumilan}\ \bibnamefont
  {Banerjee}},\ }\bibfield  {title} {\enquote {\bibinfo {title} {Scaling of
  fock space propagator in quasiperiodic many-body localizing systems},}\
  }\href@noop {} {\bibfield  {journal} {\bibinfo  {journal} {Annals of
  Physics}\ ,\ \bibinfo {pages} {170001}} (\bibinfo {year} {2025})}\BibitemShut
  {NoStop}%
\bibitem [{\citenamefont {Altshuler}\ \emph {et~al.}(2016)\citenamefont
  {Altshuler}, \citenamefont {Cuevas}, \citenamefont {Ioffe},\ and\
  \citenamefont {Kravtsov}}]{altshuler2016nonergodic}%
  \BibitemOpen
  \bibfield  {author} {\bibinfo {author} {\bibfnamefont {BL}~\bibnamefont
  {Altshuler}}, \bibinfo {author} {\bibfnamefont {E}~\bibnamefont {Cuevas}},
  \bibinfo {author} {\bibfnamefont {LB}~\bibnamefont {Ioffe}}, \ and\ \bibinfo
  {author} {\bibfnamefont {VE}~\bibnamefont {Kravtsov}},\ }\bibfield  {title}
  {\enquote {\bibinfo {title} {Nonergodic phases in strongly disordered random
  regular graphs},}\ }\href@noop {} {\bibfield  {journal} {\bibinfo  {journal}
  {Physical review letters}\ }\textbf {\bibinfo {volume} {117}},\ \bibinfo
  {pages} {156601} (\bibinfo {year} {2016})}\BibitemShut {NoStop}%
\bibitem [{\citenamefont {Tikhonov}\ \emph {et~al.}(2016)\citenamefont
  {Tikhonov}, \citenamefont {Mirlin},\ and\ \citenamefont
  {Skvortsov}}]{tikhonov2016anderson}%
  \BibitemOpen
  \bibfield  {author} {\bibinfo {author} {\bibfnamefont {Konstantin~S}\
  \bibnamefont {Tikhonov}}, \bibinfo {author} {\bibfnamefont {Alexander~D}\
  \bibnamefont {Mirlin}}, \ and\ \bibinfo {author} {\bibfnamefont {Mikhail~A}\
  \bibnamefont {Skvortsov}},\ }\bibfield  {title} {\enquote {\bibinfo {title}
  {Anderson localization and ergodicity on random regular graphs},}\
  }\href@noop {} {\bibfield  {journal} {\bibinfo  {journal} {Physical Review
  B}\ }\textbf {\bibinfo {volume} {94}},\ \bibinfo {pages} {220203} (\bibinfo
  {year} {2016})}\BibitemShut {NoStop}%
\bibitem [{\citenamefont {Kravtsov}\ \emph {et~al.}(2018)\citenamefont
  {Kravtsov}, \citenamefont {Altshuler},\ and\ \citenamefont
  {Ioffe}}]{kravtsov2018non}%
  \BibitemOpen
  \bibfield  {author} {\bibinfo {author} {\bibfnamefont {VE}~\bibnamefont
  {Kravtsov}}, \bibinfo {author} {\bibfnamefont {BL}~\bibnamefont {Altshuler}},
  \ and\ \bibinfo {author} {\bibfnamefont {LB3762015}\ \bibnamefont {Ioffe}},\
  }\bibfield  {title} {\enquote {\bibinfo {title} {Non-ergodic delocalized
  phase in anderson model on bethe lattice and regular graph},}\ }\href@noop {}
  {\bibfield  {journal} {\bibinfo  {journal} {Annals of Physics}\ }\textbf
  {\bibinfo {volume} {389}},\ \bibinfo {pages} {148--191} (\bibinfo {year}
  {2018})}\BibitemShut {NoStop}%
\bibitem [{\citenamefont {Herre}\ \emph {et~al.}(2023)\citenamefont {Herre},
  \citenamefont {Karcher}, \citenamefont {Tikhonov},\ and\ \citenamefont
  {Mirlin}}]{herre2023ergodicity}%
  \BibitemOpen
  \bibfield  {author} {\bibinfo {author} {\bibfnamefont {Jan-Niklas}\
  \bibnamefont {Herre}}, \bibinfo {author} {\bibfnamefont {Jonas~F}\
  \bibnamefont {Karcher}}, \bibinfo {author} {\bibfnamefont {Konstantin~S}\
  \bibnamefont {Tikhonov}}, \ and\ \bibinfo {author} {\bibfnamefont
  {Alexander~D}\ \bibnamefont {Mirlin}},\ }\bibfield  {title} {\enquote
  {\bibinfo {title} {Ergodicity-to-localization transition on random regular
  graphs with large connectivity and in many-body quantum dots},}\ }\href@noop
  {} {\bibfield  {journal} {\bibinfo  {journal} {Physical Review B}\ }\textbf
  {\bibinfo {volume} {108}},\ \bibinfo {pages} {014203} (\bibinfo {year}
  {2023})}\BibitemShut {NoStop}%
\bibitem [{\citenamefont {Vanoni}\ \emph {et~al.}(2024)\citenamefont {Vanoni},
  \citenamefont {Altshuler}, \citenamefont {Kravtsov},\ and\ \citenamefont
  {Scardicchio}}]{vanoni2024renormalization}%
  \BibitemOpen
  \bibfield  {author} {\bibinfo {author} {\bibfnamefont {Carlo}\ \bibnamefont
  {Vanoni}}, \bibinfo {author} {\bibfnamefont {Boris~L}\ \bibnamefont
  {Altshuler}}, \bibinfo {author} {\bibfnamefont {Vladimir~E}\ \bibnamefont
  {Kravtsov}}, \ and\ \bibinfo {author} {\bibfnamefont {Antonello}\
  \bibnamefont {Scardicchio}},\ }\bibfield  {title} {\enquote {\bibinfo {title}
  {Renormalization group analysis of the anderson model on random regular
  graphs},}\ }\href@noop {} {\bibfield  {journal} {\bibinfo  {journal}
  {Proceedings of the National Academy of Sciences}\ }\textbf {\bibinfo
  {volume} {121}},\ \bibinfo {pages} {e2401955121} (\bibinfo {year}
  {2024})}\BibitemShut {NoStop}%
\bibitem [{\citenamefont {Li}\ \emph {et~al.}(2015)\citenamefont {Li},
  \citenamefont {Ganeshan}, \citenamefont {Pixley},\ and\ \citenamefont
  {Sarma}}]{li2015many}%
  \BibitemOpen
  \bibfield  {author} {\bibinfo {author} {\bibfnamefont {Xiaopeng}\
  \bibnamefont {Li}}, \bibinfo {author} {\bibfnamefont {Sriram}\ \bibnamefont
  {Ganeshan}}, \bibinfo {author} {\bibfnamefont {JH}~\bibnamefont {Pixley}}, \
  and\ \bibinfo {author} {\bibfnamefont {S~Das}\ \bibnamefont {Sarma}},\
  }\bibfield  {title} {\enquote {\bibinfo {title} {Many-body localization and
  quantum nonergodicity in a model with a single-particle mobility edge},}\
  }\href@noop {} {\bibfield  {journal} {\bibinfo  {journal} {Physical review
  letters}\ }\textbf {\bibinfo {volume} {115}},\ \bibinfo {pages} {186601}
  (\bibinfo {year} {2015})}\BibitemShut {NoStop}%
\bibitem [{\citenamefont {Modak}\ and\ \citenamefont
  {Mukerjee}(2015)}]{modak2015many}%
  \BibitemOpen
  \bibfield  {author} {\bibinfo {author} {\bibfnamefont {Ranjan}\ \bibnamefont
  {Modak}}\ and\ \bibinfo {author} {\bibfnamefont {Subroto}\ \bibnamefont
  {Mukerjee}},\ }\bibfield  {title} {\enquote {\bibinfo {title} {Many-body
  localization in the presence of a single-particle mobility edge},}\
  }\href@noop {} {\bibfield  {journal} {\bibinfo  {journal} {Physical review
  letters}\ }\textbf {\bibinfo {volume} {115}},\ \bibinfo {pages} {230401}
  (\bibinfo {year} {2015})}\BibitemShut {NoStop}%
\bibitem [{\citenamefont {Deng}\ \emph {et~al.}(2017)\citenamefont {Deng},
  \citenamefont {Ganeshan}, \citenamefont {Li}, \citenamefont {Modak},
  \citenamefont {Mukerjee},\ and\ \citenamefont {Pixley}}]{deng2017many}%
  \BibitemOpen
  \bibfield  {author} {\bibinfo {author} {\bibfnamefont {Dong-Ling}\
  \bibnamefont {Deng}}, \bibinfo {author} {\bibfnamefont {Sriram}\ \bibnamefont
  {Ganeshan}}, \bibinfo {author} {\bibfnamefont {Xiaopeng}\ \bibnamefont {Li}},
  \bibinfo {author} {\bibfnamefont {Ranjan}\ \bibnamefont {Modak}}, \bibinfo
  {author} {\bibfnamefont {Subroto}\ \bibnamefont {Mukerjee}}, \ and\ \bibinfo
  {author} {\bibfnamefont {JH}~\bibnamefont {Pixley}},\ }\bibfield  {title}
  {\enquote {\bibinfo {title} {Many-body localization in incommensurate models
  with a mobility edge},}\ }\href@noop {} {\bibfield  {journal} {\bibinfo
  {journal} {Annalen der Physik}\ }\textbf {\bibinfo {volume} {529}},\ \bibinfo
  {pages} {1600399} (\bibinfo {year} {2017})}\BibitemShut {NoStop}%
\bibitem [{\citenamefont {Modak}\ \emph {et~al.}(2018)\citenamefont {Modak},
  \citenamefont {Ghosh},\ and\ \citenamefont {Mukerjee}}]{Modak2018}%
  \BibitemOpen
  \bibfield  {author} {\bibinfo {author} {\bibfnamefont {Ranjan}\ \bibnamefont
  {Modak}}, \bibinfo {author} {\bibfnamefont {Soumi}\ \bibnamefont {Ghosh}}, \
  and\ \bibinfo {author} {\bibfnamefont {Subroto}\ \bibnamefont {Mukerjee}},\
  }\bibfield  {title} {\enquote {\bibinfo {title} {Criterion for the occurrence
  of many-body localization in the presence of a single-particle mobility
  edge},}\ }\href@noop {} {\bibfield  {journal} {\bibinfo  {journal} {Physical
  Review B}\ }\textbf {\bibinfo {volume} {97}},\ \bibinfo {pages} {104204}
  (\bibinfo {year} {2018})}\BibitemShut {NoStop}%
\bibitem [{\citenamefont {Ghosh}\ \emph {et~al.}(2020)\citenamefont {Ghosh},
  \citenamefont {Gidugu},\ and\ \citenamefont {Mukerjee}}]{ghosh2020transport}%
  \BibitemOpen
  \bibfield  {author} {\bibinfo {author} {\bibfnamefont {Soumi}\ \bibnamefont
  {Ghosh}}, \bibinfo {author} {\bibfnamefont {Jyotsna}\ \bibnamefont {Gidugu}},
  \ and\ \bibinfo {author} {\bibfnamefont {Subroto}\ \bibnamefont {Mukerjee}},\
  }\bibfield  {title} {\enquote {\bibinfo {title} {Transport in the nonergodic
  extended phase of interacting quasiperiodic systems},}\ }\href@noop {}
  {\bibfield  {journal} {\bibinfo  {journal} {Physical Review B}\ }\textbf
  {\bibinfo {volume} {102}},\ \bibinfo {pages} {224203} (\bibinfo {year}
  {2020})}\BibitemShut {NoStop}%
\bibitem [{\citenamefont {Ganeshan}\ \emph {et~al.}(2015)\citenamefont
  {Ganeshan}, \citenamefont {Pixley},\ and\ \citenamefont
  {Sarma}}]{ganeshan2015nearest}%
  \BibitemOpen
  \bibfield  {author} {\bibinfo {author} {\bibfnamefont {Sriram}\ \bibnamefont
  {Ganeshan}}, \bibinfo {author} {\bibfnamefont {JH}~\bibnamefont {Pixley}}, \
  and\ \bibinfo {author} {\bibfnamefont {S~Das}\ \bibnamefont {Sarma}},\
  }\bibfield  {title} {\enquote {\bibinfo {title} {Nearest neighbor tight
  binding models with an exact mobility edge in one dimension},}\ }\href@noop
  {} {\bibfield  {journal} {\bibinfo  {journal} {Physical review letters}\
  }\textbf {\bibinfo {volume} {114}},\ \bibinfo {pages} {146601} (\bibinfo
  {year} {2015})}\BibitemShut {NoStop}%
\bibitem [{\citenamefont {{Xu}}\ \emph {et~al.}(2005)\citenamefont {{Xu}},
  \citenamefont {{Kumar}}, \citenamefont {{Buldyrev}}, \citenamefont {{Chen}},
  \citenamefont {{Poole}}, \citenamefont {{Sciortino}},\ and\ \citenamefont
  {{Stanley}}}]{Xu2005}%
  \BibitemOpen
  \bibfield  {author} {\bibinfo {author} {\bibfnamefont {Limei}\ \bibnamefont
  {{Xu}}}, \bibinfo {author} {\bibfnamefont {Pradeep}\ \bibnamefont {{Kumar}}},
  \bibinfo {author} {\bibfnamefont {S.~V.}\ \bibnamefont {{Buldyrev}}},
  \bibinfo {author} {\bibfnamefont {S.~H.}\ \bibnamefont {{Chen}}}, \bibinfo
  {author} {\bibfnamefont {P.~H.}\ \bibnamefont {{Poole}}}, \bibinfo {author}
  {\bibfnamefont {F.}~\bibnamefont {{Sciortino}}}, \ and\ \bibinfo {author}
  {\bibfnamefont {H.~E.}\ \bibnamefont {{Stanley}}},\ }\bibfield  {title}
  {\enquote {\bibinfo {title} {{Relation between the Widom line and the dynamic
  crossover in systems with a liquid-liquid phase transition}},}\ }\href
  {\doibase 10.1073/pnas.0507870102} {\bibfield  {journal} {\bibinfo  {journal}
  {Proceedings of the National Academy of Science}\ }\textbf {\bibinfo {volume}
  {102}},\ \bibinfo {pages} {16558--16562} (\bibinfo {year} {2005})},\ \Eprint
  {http://arxiv.org/abs/cond-mat/0509616} {arXiv:cond-mat/0509616
  [cond-mat.stat-mech]} \BibitemShut {NoStop}%
\bibitem [{\citenamefont {{Franzese}}\ and\ \citenamefont
  {{Stanley}}(2007)}]{Franzese2007}%
  \BibitemOpen
  \bibfield  {author} {\bibinfo {author} {\bibfnamefont {Giancarlo}\
  \bibnamefont {{Franzese}}}\ and\ \bibinfo {author} {\bibfnamefont
  {H.~Eugene}\ \bibnamefont {{Stanley}}},\ }\bibfield  {title} {\enquote
  {\bibinfo {title} {{The Widom line of supercooled water}},}\ }\href {\doibase
  10.1088/0953-8984/19/20/205126} {\bibfield  {journal} {\bibinfo  {journal}
  {Journal of Physics Condensed Matter}\ }\textbf {\bibinfo {volume} {19}},\
  \bibinfo {eid} {205126} (\bibinfo {year} {2007})}\BibitemShut {NoStop}%
\bibitem [{\citenamefont {{Simeoni}}\ \emph {et~al.}(2010)\citenamefont
  {{Simeoni}}, \citenamefont {{Bryk}}, \citenamefont {{Gorelli}}, \citenamefont
  {{Krisch}}, \citenamefont {{Ruocco}}, \citenamefont {{Santoro}},\ and\
  \citenamefont {{Scopigno}}}]{Simeoni2010}%
  \BibitemOpen
  \bibfield  {author} {\bibinfo {author} {\bibfnamefont {G.~G.}\ \bibnamefont
  {{Simeoni}}}, \bibinfo {author} {\bibfnamefont {T.}~\bibnamefont {{Bryk}}},
  \bibinfo {author} {\bibfnamefont {F.~A.}\ \bibnamefont {{Gorelli}}}, \bibinfo
  {author} {\bibfnamefont {M.}~\bibnamefont {{Krisch}}}, \bibinfo {author}
  {\bibfnamefont {G.}~\bibnamefont {{Ruocco}}}, \bibinfo {author}
  {\bibfnamefont {M.}~\bibnamefont {{Santoro}}}, \ and\ \bibinfo {author}
  {\bibfnamefont {T.}~\bibnamefont {{Scopigno}}},\ }\bibfield  {title}
  {\enquote {\bibinfo {title} {{The Widom line as the crossover between
  liquid-like and gas-like behaviour in supercritical fluids}},}\ }\href
  {\doibase 10.1038/nphys1683} {\bibfield  {journal} {\bibinfo  {journal}
  {Nature Physics}\ }\textbf {\bibinfo {volume} {6}},\ \bibinfo {pages}
  {503--507} (\bibinfo {year} {2010})}\BibitemShut {NoStop}%
\bibitem [{\citenamefont {Luo}\ \emph {et~al.}(2014)\citenamefont {Luo},
  \citenamefont {Xu}, \citenamefont {Lascaris}, \citenamefont {Stanley},\ and\
  \citenamefont {Buldyrev}}]{Luo2014}%
  \BibitemOpen
  \bibfield  {author} {\bibinfo {author} {\bibfnamefont {Jiayuan}\ \bibnamefont
  {Luo}}, \bibinfo {author} {\bibfnamefont {Limei}\ \bibnamefont {Xu}},
  \bibinfo {author} {\bibfnamefont {Erik}\ \bibnamefont {Lascaris}}, \bibinfo
  {author} {\bibfnamefont {H.~Eugene}\ \bibnamefont {Stanley}}, \ and\ \bibinfo
  {author} {\bibfnamefont {Sergey~V.}\ \bibnamefont {Buldyrev}},\ }\bibfield
  {title} {\enquote {\bibinfo {title} {Behavior of the widom line in critical
  phenomena},}\ }\href {\doibase 10.1103/PhysRevLett.112.135701} {\bibfield
  {journal} {\bibinfo  {journal} {Phys. Rev. Lett.}\ }\textbf {\bibinfo
  {volume} {112}},\ \bibinfo {pages} {135701} (\bibinfo {year}
  {2014})}\BibitemShut {NoStop}%
\bibitem [{\citenamefont {Sordi}\ and\ \citenamefont
  {Tremblay}(2024)}]{Sordi2024}%
  \BibitemOpen
  \bibfield  {author} {\bibinfo {author} {\bibfnamefont {G.}~\bibnamefont
  {Sordi}}\ and\ \bibinfo {author} {\bibfnamefont {A.-M.~S.}\ \bibnamefont
  {Tremblay}},\ }\bibfield  {title} {\enquote {\bibinfo {title} {Introducing
  the concept of the widom line in the qcd phase diagram},}\ }\href {\doibase
  10.1103/PhysRevD.109.114020} {\bibfield  {journal} {\bibinfo  {journal}
  {Phys. Rev. D}\ }\textbf {\bibinfo {volume} {109}},\ \bibinfo {pages}
  {114020} (\bibinfo {year} {2024})}\BibitemShut {NoStop}%
\bibitem [{\citenamefont {Altland}\ and\ \citenamefont
  {Micklitz}(2017)}]{altland2017field}%
  \BibitemOpen
  \bibfield  {author} {\bibinfo {author} {\bibfnamefont {Alexander}\
  \bibnamefont {Altland}}\ and\ \bibinfo {author} {\bibfnamefont {Tobias}\
  \bibnamefont {Micklitz}},\ }\bibfield  {title} {\enquote {\bibinfo {title}
  {Field theory approach to many-body localization},}\ }\href@noop {}
  {\bibfield  {journal} {\bibinfo  {journal} {Physical Review Letters}\
  }\textbf {\bibinfo {volume} {118}},\ \bibinfo {pages} {127202} (\bibinfo
  {year} {2017})}\BibitemShut {NoStop}%
\bibitem [{\citenamefont {Pietracaprina}\ \emph {et~al.}(2017)\citenamefont
  {Pietracaprina}, \citenamefont {Gogolin},\ and\ \citenamefont
  {Goold}}]{pietracaprina2017total}%
  \BibitemOpen
  \bibfield  {author} {\bibinfo {author} {\bibfnamefont {Francesca}\
  \bibnamefont {Pietracaprina}}, \bibinfo {author} {\bibfnamefont {Christian}\
  \bibnamefont {Gogolin}}, \ and\ \bibinfo {author} {\bibfnamefont {John}\
  \bibnamefont {Goold}},\ }\bibfield  {title} {\enquote {\bibinfo {title}
  {Total correlations of the diagonal ensemble as a generic indicator for
  ergodicity breaking in quantum systems},}\ }\href@noop {} {\bibfield
  {journal} {\bibinfo  {journal} {Physical Review B}\ }\textbf {\bibinfo
  {volume} {95}},\ \bibinfo {pages} {125118} (\bibinfo {year}
  {2017})}\BibitemShut {NoStop}%
\bibitem [{\citenamefont {Scoquart}\ \emph {et~al.}(2024)\citenamefont
  {Scoquart}, \citenamefont {Gornyi},\ and\ \citenamefont
  {Mirlin}}]{scoquart2024role}%
  \BibitemOpen
  \bibfield  {author} {\bibinfo {author} {\bibfnamefont {Thibault}\
  \bibnamefont {Scoquart}}, \bibinfo {author} {\bibfnamefont {Igor~V}\
  \bibnamefont {Gornyi}}, \ and\ \bibinfo {author} {\bibfnamefont
  {Alexander~D}\ \bibnamefont {Mirlin}},\ }\bibfield  {title} {\enquote
  {\bibinfo {title} {Role of fock-space correlations in many-body
  localization},}\ }\href@noop {} {\bibfield  {journal} {\bibinfo  {journal}
  {Physical Review B}\ }\textbf {\bibinfo {volume} {109}},\ \bibinfo {pages}
  {214203} (\bibinfo {year} {2024})}\BibitemShut {NoStop}%
\bibitem [{\citenamefont {De~Luca}\ and\ \citenamefont
  {Scardicchio}(2013)}]{de2013ergodicity}%
  \BibitemOpen
  \bibfield  {author} {\bibinfo {author} {\bibfnamefont {Andrea}\ \bibnamefont
  {De~Luca}}\ and\ \bibinfo {author} {\bibfnamefont {Antonello}\ \bibnamefont
  {Scardicchio}},\ }\bibfield  {title} {\enquote {\bibinfo {title} {Ergodicity
  breaking in a model showing many-body localization},}\ }\href@noop {}
  {\bibfield  {journal} {\bibinfo  {journal} {EPL (Europhysics Letters)}\
  }\textbf {\bibinfo {volume} {101}},\ \bibinfo {pages} {37003} (\bibinfo
  {year} {2013})}\BibitemShut {NoStop}%
\bibitem [{\citenamefont {Mac{\'e}}\ \emph {et~al.}(2019)\citenamefont
  {Mac{\'e}}, \citenamefont {Alet},\ and\ \citenamefont
  {Laflorencie}}]{mace2019multifractal}%
  \BibitemOpen
  \bibfield  {author} {\bibinfo {author} {\bibfnamefont {Nicolas}\ \bibnamefont
  {Mac{\'e}}}, \bibinfo {author} {\bibfnamefont {Fabien}\ \bibnamefont {Alet}},
  \ and\ \bibinfo {author} {\bibfnamefont {Nicolas}\ \bibnamefont
  {Laflorencie}},\ }\bibfield  {title} {\enquote {\bibinfo {title}
  {Multifractal scalings across the many-body localization transition},}\
  }\href@noop {} {\bibfield  {journal} {\bibinfo  {journal} {Physical review
  letters}\ }\textbf {\bibinfo {volume} {123}},\ \bibinfo {pages} {180601}
  (\bibinfo {year} {2019})}\BibitemShut {NoStop}%
\bibitem [{\citenamefont {De~Tomasi}\ \emph {et~al.}(2021)\citenamefont
  {De~Tomasi}, \citenamefont {Khaymovich}, \citenamefont {Pollmann},\ and\
  \citenamefont {Warzel}}]{de2021rare}%
  \BibitemOpen
  \bibfield  {author} {\bibinfo {author} {\bibfnamefont {Giuseppe}\
  \bibnamefont {De~Tomasi}}, \bibinfo {author} {\bibfnamefont {Ivan~M}\
  \bibnamefont {Khaymovich}}, \bibinfo {author} {\bibfnamefont {Frank}\
  \bibnamefont {Pollmann}}, \ and\ \bibinfo {author} {\bibfnamefont {Simone}\
  \bibnamefont {Warzel}},\ }\bibfield  {title} {\enquote {\bibinfo {title}
  {Rare thermal bubbles at the many-body localization transition from the fock
  space point of view},}\ }\href@noop {} {\bibfield  {journal} {\bibinfo
  {journal} {Physical Review B}\ }\textbf {\bibinfo {volume} {104}},\ \bibinfo
  {pages} {024202} (\bibinfo {year} {2021})}\BibitemShut {NoStop}%
\bibitem [{\citenamefont {Garcia-Mata}\ \emph {et~al.}(2017)\citenamefont
  {Garcia-Mata}, \citenamefont {Giraud}, \citenamefont {Georgeot},
  \citenamefont {Martin}, \citenamefont {Dubertrand},\ and\ \citenamefont
  {Lemari{\'e}}}]{garcia2017scaling}%
  \BibitemOpen
  \bibfield  {author} {\bibinfo {author} {\bibfnamefont {Ignacio}\ \bibnamefont
  {Garcia-Mata}}, \bibinfo {author} {\bibfnamefont {Olivier}\ \bibnamefont
  {Giraud}}, \bibinfo {author} {\bibfnamefont {Bertrand}\ \bibnamefont
  {Georgeot}}, \bibinfo {author} {\bibfnamefont {John}\ \bibnamefont {Martin}},
  \bibinfo {author} {\bibfnamefont {R{\'e}my}\ \bibnamefont {Dubertrand}}, \
  and\ \bibinfo {author} {\bibfnamefont {Gabriel}\ \bibnamefont
  {Lemari{\'e}}},\ }\bibfield  {title} {\enquote {\bibinfo {title} {Scaling
  theory of the anderson transition in random graphs: ergodicity and
  universality},}\ }\href@noop {} {\bibfield  {journal} {\bibinfo  {journal}
  {Physical review letters}\ }\textbf {\bibinfo {volume} {118}},\ \bibinfo
  {pages} {166801} (\bibinfo {year} {2017})}\BibitemShut {NoStop}%
\bibitem [{\citenamefont {Biddle}\ and\ \citenamefont
  {Das~Sarma}(2010)}]{biddle2010predicted}%
  \BibitemOpen
  \bibfield  {author} {\bibinfo {author} {\bibfnamefont {J}~\bibnamefont
  {Biddle}}\ and\ \bibinfo {author} {\bibfnamefont {S}~\bibnamefont
  {Das~Sarma}},\ }\bibfield  {title} {\enquote {\bibinfo {title} {Predicted
  mobility edges in one-dimensional incommensurate optical lattices: An exactly
  solvable model of anderson localization},}\ }\href@noop {} {\bibfield
  {journal} {\bibinfo  {journal} {Physical review letters}\ }\textbf {\bibinfo
  {volume} {104}},\ \bibinfo {pages} {070601} (\bibinfo {year}
  {2010})}\BibitemShut {NoStop}%
\bibitem [{\citenamefont {Takada}\ \emph {et~al.}(2004)\citenamefont {Takada},
  \citenamefont {Ino},\ and\ \citenamefont {Yamanaka}}]{takada2004statistics}%
  \BibitemOpen
  \bibfield  {author} {\bibinfo {author} {\bibfnamefont {Yoshihiro}\
  \bibnamefont {Takada}}, \bibinfo {author} {\bibfnamefont {Kazusumi}\
  \bibnamefont {Ino}}, \ and\ \bibinfo {author} {\bibfnamefont {Masanori}\
  \bibnamefont {Yamanaka}},\ }\bibfield  {title} {\enquote {\bibinfo {title}
  {Statistics of spectra for critical quantum chaos in one-dimensional
  quasiperiodic systems},}\ }\href@noop {} {\bibfield  {journal} {\bibinfo
  {journal} {Physical Review E}\ }\textbf {\bibinfo {volume} {70}},\ \bibinfo
  {pages} {066203} (\bibinfo {year} {2004})}\BibitemShut {NoStop}%
\bibitem [{\citenamefont {Aditya}\ and\ \citenamefont
  {Roy}(2024)}]{aditya2024family}%
  \BibitemOpen
  \bibfield  {author} {\bibinfo {author} {\bibfnamefont {Sreemayee}\
  \bibnamefont {Aditya}}\ and\ \bibinfo {author} {\bibfnamefont {Nilanjan}\
  \bibnamefont {Roy}},\ }\bibfield  {title} {\enquote {\bibinfo {title}
  {Family-vicsek dynamical scaling and kardar-parisi-zhang-like superdiffusive
  growth of surface roughness in a driven one-dimensional quasiperiodic
  model},}\ }\href@noop {} {\bibfield  {journal} {\bibinfo  {journal} {Physical
  Review B}\ }\textbf {\bibinfo {volume} {109}},\ \bibinfo {pages} {035164}
  (\bibinfo {year} {2024})}\BibitemShut {NoStop}%
\bibitem [{\citenamefont {Dobrosavljevi{\'c}}\ \emph
  {et~al.}(2003)\citenamefont {Dobrosavljevi{\'c}}, \citenamefont {Pastor},\
  and\ \citenamefont {Nikoli{\'c}}}]{dobrosavljevic2003typical}%
  \BibitemOpen
  \bibfield  {author} {\bibinfo {author} {\bibfnamefont {V}~\bibnamefont
  {Dobrosavljevi{\'c}}}, \bibinfo {author} {\bibfnamefont {AA}~\bibnamefont
  {Pastor}}, \ and\ \bibinfo {author} {\bibfnamefont {BK}~\bibnamefont
  {Nikoli{\'c}}},\ }\bibfield  {title} {\enquote {\bibinfo {title} {Typical
  medium theory of anderson localization: A local order parameter approach to
  strong-disorder effects},}\ }\href@noop {} {\bibfield  {journal} {\bibinfo
  {journal} {Europhysics Letters}\ }\textbf {\bibinfo {volume} {62}},\ \bibinfo
  {pages} {76} (\bibinfo {year} {2003})}\BibitemShut {NoStop}%
\bibitem [{\citenamefont {Dobrosavljevi{\'c}}\ and\ \citenamefont
  {Kotliar}(1997)}]{dobrosavljevic1997mean}%
  \BibitemOpen
  \bibfield  {author} {\bibinfo {author} {\bibfnamefont {Vladimir}\
  \bibnamefont {Dobrosavljevi{\'c}}}\ and\ \bibinfo {author} {\bibfnamefont
  {Gabriel}\ \bibnamefont {Kotliar}},\ }\bibfield  {title} {\enquote {\bibinfo
  {title} {Mean field theory of the mott-anderson transition},}\ }\href@noop {}
  {\bibfield  {journal} {\bibinfo  {journal} {Physical review letters}\
  }\textbf {\bibinfo {volume} {78}},\ \bibinfo {pages} {3943} (\bibinfo {year}
  {1997})}\BibitemShut {NoStop}%
\bibitem [{\citenamefont {Jana}\ \emph {et~al.}(2021)\citenamefont {Jana},
  \citenamefont {Chandra},\ and\ \citenamefont {Garg}}]{jana2021local}%
  \BibitemOpen
  \bibfield  {author} {\bibinfo {author} {\bibfnamefont {Atanu}\ \bibnamefont
  {Jana}}, \bibinfo {author} {\bibfnamefont {V~Ravi}\ \bibnamefont {Chandra}},
  \ and\ \bibinfo {author} {\bibfnamefont {Arti}\ \bibnamefont {Garg}},\
  }\bibfield  {title} {\enquote {\bibinfo {title} {Local density of states and
  scattering rates across the many-body localization transition},}\ }\href@noop
  {} {\bibfield  {journal} {\bibinfo  {journal} {Physical Review B}\ }\textbf
  {\bibinfo {volume} {104}},\ \bibinfo {pages} {L140201} (\bibinfo {year}
  {2021})}\BibitemShut {NoStop}%
\bibitem [{\citenamefont {Nandkishore}(2015)}]{nandkishore2015many}%
  \BibitemOpen
  \bibfield  {author} {\bibinfo {author} {\bibfnamefont {Rahul}\ \bibnamefont
  {Nandkishore}},\ }\bibfield  {title} {\enquote {\bibinfo {title} {Many-body
  localization proximity effect},}\ }\href@noop {} {\bibfield  {journal}
  {\bibinfo  {journal} {Physical Review B}\ }\textbf {\bibinfo {volume} {92}},\
  \bibinfo {pages} {245141} (\bibinfo {year} {2015})}\BibitemShut {NoStop}%
\bibitem [{\citenamefont {Roy}\ and\ \citenamefont
  {Logan}(2020{\natexlab{b}})}]{roy2020localization}%
  \BibitemOpen
  \bibfield  {author} {\bibinfo {author} {\bibfnamefont {Sthitadhi}\
  \bibnamefont {Roy}}\ and\ \bibinfo {author} {\bibfnamefont {David~E}\
  \bibnamefont {Logan}},\ }\bibfield  {title} {\enquote {\bibinfo {title}
  {Localization on certain graphs with strongly correlated disorder},}\
  }\href@noop {} {\bibfield  {journal} {\bibinfo  {journal} {Physical Review
  Letters}\ }\textbf {\bibinfo {volume} {125}},\ \bibinfo {pages} {250402}
  (\bibinfo {year} {2020}{\natexlab{b}})}\BibitemShut {NoStop}%
\bibitem [{\citenamefont {MacKinnon}(1980)}]{mackinnon1980conductivity}%
  \BibitemOpen
  \bibfield  {author} {\bibinfo {author} {\bibfnamefont {A}~\bibnamefont
  {MacKinnon}},\ }\bibfield  {title} {\enquote {\bibinfo {title} {The
  conductivity of the one-dimensional disordered anderson model: a new
  numerical method},}\ }\href@noop {} {\bibfield  {journal} {\bibinfo
  {journal} {Journal of Physics C: Solid State Physics}\ }\textbf {\bibinfo
  {volume} {13}},\ \bibinfo {pages} {L1031} (\bibinfo {year}
  {1980})}\BibitemShut {NoStop}%
\bibitem [{\citenamefont {MacKinnon}\ and\ \citenamefont
  {Kramer}(1983)}]{mackinnon1983scaling}%
  \BibitemOpen
  \bibfield  {author} {\bibinfo {author} {\bibfnamefont {A}~\bibnamefont
  {MacKinnon}}\ and\ \bibinfo {author} {\bibfnamefont {B}~\bibnamefont
  {Kramer}},\ }\bibfield  {title} {\enquote {\bibinfo {title} {The scaling
  theory of electrons in disordered solids: Additional numerical results},}\
  }\href@noop {} {\bibfield  {journal} {\bibinfo  {journal} {Zeitschrift
  f{\"u}r Physik B Condensed Matter}\ }\textbf {\bibinfo {volume} {53}},\
  \bibinfo {pages} {1--13} (\bibinfo {year} {1983})}\BibitemShut {NoStop}%
\bibitem [{\citenamefont {Lee}\ and\ \citenamefont
  {Fisher}(1981)}]{lee1981anderson}%
  \BibitemOpen
  \bibfield  {author} {\bibinfo {author} {\bibfnamefont {Patrick~A}\
  \bibnamefont {Lee}}\ and\ \bibinfo {author} {\bibfnamefont {Daniel~S}\
  \bibnamefont {Fisher}},\ }\bibfield  {title} {\enquote {\bibinfo {title}
  {Anderson localization in two dimensions},}\ }\href@noop {} {\bibfield
  {journal} {\bibinfo  {journal} {Physical Review Letters}\ }\textbf {\bibinfo
  {volume} {47}},\ \bibinfo {pages} {882} (\bibinfo {year} {1981})}\BibitemShut
  {NoStop}%
\bibitem [{\citenamefont {Verg{\'e}s}(1999)}]{verges1999computational}%
  \BibitemOpen
  \bibfield  {author} {\bibinfo {author} {\bibfnamefont {JA}~\bibnamefont
  {Verg{\'e}s}},\ }\bibfield  {title} {\enquote {\bibinfo {title}
  {Computational implementation of the kubo formula for the static conductance:
  application to two-dimensional quantum dots},}\ }\href@noop {} {\bibfield
  {journal} {\bibinfo  {journal} {Computer physics communications}\ }\textbf
  {\bibinfo {volume} {118}},\ \bibinfo {pages} {71--80} (\bibinfo {year}
  {1999})}\BibitemShut {NoStop}%
\bibitem [{\citenamefont {{Sordi}}\ \emph {et~al.}(2012)\citenamefont
  {{Sordi}}, \citenamefont {{S{\'e}mon}}, \citenamefont {{Haule}},\ and\
  \citenamefont {{Tremblay}}}]{Sordi2012}%
  \BibitemOpen
  \bibfield  {author} {\bibinfo {author} {\bibfnamefont {G.}~\bibnamefont
  {{Sordi}}}, \bibinfo {author} {\bibfnamefont {P.}~\bibnamefont
  {{S{\'e}mon}}}, \bibinfo {author} {\bibfnamefont {K.}~\bibnamefont
  {{Haule}}}, \ and\ \bibinfo {author} {\bibfnamefont {A.~M.~S.}\ \bibnamefont
  {{Tremblay}}},\ }\bibfield  {title} {\enquote {\bibinfo {title} {{Pseudogap
  temperature as a Widom line in doped Mott insulators}},}\ }\href {\doibase
  10.1038/srep00547} {\bibfield  {journal} {\bibinfo  {journal} {Scientific
  Reports}\ }\textbf {\bibinfo {volume} {2}},\ \bibinfo {eid} {547} (\bibinfo
  {year} {2012})},\ \Eprint {http://arxiv.org/abs/1110.1392} {arXiv:1110.1392
  [cond-mat.str-el]} \BibitemShut {NoStop}%
\bibitem [{\citenamefont {Sordi}\ \emph {et~al.}(2013)\citenamefont {Sordi},
  \citenamefont {S\'emon}, \citenamefont {Haule},\ and\ \citenamefont
  {Tremblay}}]{Sordi2013}%
  \BibitemOpen
  \bibfield  {author} {\bibinfo {author} {\bibfnamefont {G.}~\bibnamefont
  {Sordi}}, \bibinfo {author} {\bibfnamefont {P.}~\bibnamefont {S\'emon}},
  \bibinfo {author} {\bibfnamefont {K.}~\bibnamefont {Haule}}, \ and\ \bibinfo
  {author} {\bibfnamefont {A.-M.~S.}\ \bibnamefont {Tremblay}},\ }\bibfield
  {title} {\enquote {\bibinfo {title} {$c$-axis resistivity, pseudogap,
  superconductivity, and widom line in doped mott insulators},}\ }\href
  {\doibase 10.1103/PhysRevB.87.041101} {\bibfield  {journal} {\bibinfo
  {journal} {Phys. Rev. B}\ }\textbf {\bibinfo {volume} {87}},\ \bibinfo
  {pages} {041101} (\bibinfo {year} {2013})}\BibitemShut {NoStop}%
\bibitem [{\citenamefont {Vu\ifmmode \check{c}\else \v{c}\fi{}i\ifmmode
  \check{c}\else \v{c}\fi{}evi\ifmmode~\acute{c}\else \'{c}\fi{}}\ \emph
  {et~al.}(2013)\citenamefont {Vu\ifmmode \check{c}\else \v{c}\fi{}i\ifmmode
  \check{c}\else \v{c}\fi{}evi\ifmmode~\acute{c}\else \'{c}\fi{}},
  \citenamefont {Terletska}, \citenamefont {Tanaskovi\ifmmode~\acute{c}\else
  \'{c}\fi{}},\ and\ \citenamefont {Dobrosavljevi\ifmmode~\acute{c}\else
  \'{c}\fi{}}}]{Vucicevic2013}%
  \BibitemOpen
  \bibfield  {author} {\bibinfo {author} {\bibfnamefont {J.}~\bibnamefont
  {Vu\ifmmode \check{c}\else \v{c}\fi{}i\ifmmode \check{c}\else
  \v{c}\fi{}evi\ifmmode~\acute{c}\else \'{c}\fi{}}}, \bibinfo {author}
  {\bibfnamefont {H.}~\bibnamefont {Terletska}}, \bibinfo {author}
  {\bibfnamefont {D.}~\bibnamefont {Tanaskovi\ifmmode~\acute{c}\else
  \'{c}\fi{}}}, \ and\ \bibinfo {author} {\bibfnamefont {V.}~\bibnamefont
  {Dobrosavljevi\ifmmode~\acute{c}\else \'{c}\fi{}}},\ }\bibfield  {title}
  {\enquote {\bibinfo {title} {Finite-temperature crossover and the quantum
  widom line near the mott transition},}\ }\href {\doibase
  10.1103/PhysRevB.88.075143} {\bibfield  {journal} {\bibinfo  {journal} {Phys.
  Rev. B}\ }\textbf {\bibinfo {volume} {88}},\ \bibinfo {pages} {075143}
  (\bibinfo {year} {2013})}\BibitemShut {NoStop}%
\bibitem [{\citenamefont {Turkeshi}\ and\ \citenamefont
  {Sierant}(2024)}]{turkeshi2024hilbert}%
  \BibitemOpen
  \bibfield  {author} {\bibinfo {author} {\bibfnamefont {Xhek}\ \bibnamefont
  {Turkeshi}}\ and\ \bibinfo {author} {\bibfnamefont {Piotr}\ \bibnamefont
  {Sierant}},\ }\bibfield  {title} {\enquote {\bibinfo {title} {Hilbert space
  delocalization under random unitary circuits},}\ }\href@noop {} {\bibfield
  {journal} {\bibinfo  {journal} {Entropy}\ }\textbf {\bibinfo {volume} {26}},\
  \bibinfo {pages} {471} (\bibinfo {year} {2024})}\BibitemShut {NoStop}%
\bibitem [{\citenamefont {Ahmed}\ and\ \citenamefont
  {Roy}(2025)}]{ahmed2025supervised}%
  \BibitemOpen
  \bibfield  {author} {\bibinfo {author} {\bibfnamefont {Aamna}\ \bibnamefont
  {Ahmed}}\ and\ \bibinfo {author} {\bibfnamefont {Nilanjan}\ \bibnamefont
  {Roy}},\ }\bibfield  {title} {\enquote {\bibinfo {title} {Supervised and
  unsupervised learning the many-body critical phase, phase transitions and
  critical exponents in disordered quantum systems},}\ }\href@noop {}
  {\bibfield  {journal} {\bibinfo  {journal} {arXiv preprint arXiv:2501.03981}\
  } (\bibinfo {year} {2025})}\BibitemShut {NoStop}%
\bibitem [{\citenamefont {Agarwal}\ \emph {et~al.}(2017)\citenamefont
  {Agarwal}, \citenamefont {Altman}, \citenamefont {Demler}, \citenamefont
  {Gopalakrishnan}, \citenamefont {Huse},\ and\ \citenamefont
  {Knap}}]{agarwal2017rare}%
  \BibitemOpen
  \bibfield  {author} {\bibinfo {author} {\bibfnamefont {Kartiek}\ \bibnamefont
  {Agarwal}}, \bibinfo {author} {\bibfnamefont {Ehud}\ \bibnamefont {Altman}},
  \bibinfo {author} {\bibfnamefont {Eugene}\ \bibnamefont {Demler}}, \bibinfo
  {author} {\bibfnamefont {Sarang}\ \bibnamefont {Gopalakrishnan}}, \bibinfo
  {author} {\bibfnamefont {David~A}\ \bibnamefont {Huse}}, \ and\ \bibinfo
  {author} {\bibfnamefont {Michael}\ \bibnamefont {Knap}},\ }\bibfield  {title}
  {\enquote {\bibinfo {title} {Rare-region effects and dynamics near the
  many-body localization transition},}\ }\href@noop {} {\bibfield  {journal}
  {\bibinfo  {journal} {Annalen der Physik}\ }\textbf {\bibinfo {volume}
  {529}},\ \bibinfo {pages} {1600326} (\bibinfo {year} {2017})}\BibitemShut
  {NoStop}%
\bibitem [{\citenamefont {Pomata}\ \emph {et~al.}(2023)\citenamefont {Pomata},
  \citenamefont {Ganeshan},\ and\ \citenamefont {Wei}}]{Pomata2020}%
  \BibitemOpen
  \bibfield  {author} {\bibinfo {author} {\bibfnamefont {Nicholas}\
  \bibnamefont {Pomata}}, \bibinfo {author} {\bibfnamefont {Sriram}\
  \bibnamefont {Ganeshan}}, \ and\ \bibinfo {author} {\bibfnamefont
  {Tzu-Chieh}\ \bibnamefont {Wei}},\ }\bibfield  {title} {\enquote {\bibinfo
  {title} {Seeking a many-body mobility edge with matrix product states in a
  quasiperiodic model},}\ }\href@noop {} {\bibfield  {journal} {\bibinfo
  {journal} {Physical Review B}\ }\textbf {\bibinfo {volume} {108}},\ \bibinfo
  {pages} {094201} (\bibinfo {year} {2023})}\BibitemShut {NoStop}%
\bibitem [{\citenamefont {Kollath}\ \emph {et~al.}(2007)\citenamefont
  {Kollath}, \citenamefont {K{\"o}hl},\ and\ \citenamefont
  {Giamarchi}}]{kollath2007scanning}%
  \BibitemOpen
  \bibfield  {author} {\bibinfo {author} {\bibfnamefont {Corinna}\ \bibnamefont
  {Kollath}}, \bibinfo {author} {\bibfnamefont {Michael}\ \bibnamefont
  {K{\"o}hl}}, \ and\ \bibinfo {author} {\bibfnamefont {Thierry}\ \bibnamefont
  {Giamarchi}},\ }\bibfield  {title} {\enquote {\bibinfo {title} {Scanning
  tunneling microscopy for ultracold atoms},}\ }\href@noop {} {\bibfield
  {journal} {\bibinfo  {journal} {Physical Review A}\ }\textbf {\bibinfo
  {volume} {76}},\ \bibinfo {pages} {063602} (\bibinfo {year}
  {2007})}\BibitemShut {NoStop}%
\bibitem [{\citenamefont {Stewart}\ \emph {et~al.}(2008)\citenamefont
  {Stewart}, \citenamefont {Gaebler},\ and\ \citenamefont
  {Jin}}]{stewart2008using}%
  \BibitemOpen
  \bibfield  {author} {\bibinfo {author} {\bibfnamefont {JT}~\bibnamefont
  {Stewart}}, \bibinfo {author} {\bibfnamefont {JP}~\bibnamefont {Gaebler}}, \
  and\ \bibinfo {author} {\bibfnamefont {DS}~\bibnamefont {Jin}},\ }\bibfield
  {title} {\enquote {\bibinfo {title} {Using photoemission spectroscopy to
  probe a strongly interacting fermi gas},}\ }\href@noop {} {\bibfield
  {journal} {\bibinfo  {journal} {Nature}\ }\textbf {\bibinfo {volume} {454}},\
  \bibinfo {pages} {744--747} (\bibinfo {year} {2008})}\BibitemShut {NoStop}%
\bibitem [{\citenamefont {Perali}\ \emph {et~al.}(2011)\citenamefont {Perali},
  \citenamefont {Palestini}, \citenamefont {Pieri}, \citenamefont {Strinati},
  \citenamefont {Stewart}, \citenamefont {Gaebler}, \citenamefont {Drake},\
  and\ \citenamefont {Jin}}]{perali2011evolution}%
  \BibitemOpen
  \bibfield  {author} {\bibinfo {author} {\bibfnamefont {Andrea}\ \bibnamefont
  {Perali}}, \bibinfo {author} {\bibfnamefont {Fabrizio}\ \bibnamefont
  {Palestini}}, \bibinfo {author} {\bibfnamefont {Pierbiagio}\ \bibnamefont
  {Pieri}}, \bibinfo {author} {\bibfnamefont {G~Calvanese}\ \bibnamefont
  {Strinati}}, \bibinfo {author} {\bibfnamefont {JT}~\bibnamefont {Stewart}},
  \bibinfo {author} {\bibfnamefont {JP}~\bibnamefont {Gaebler}}, \bibinfo
  {author} {\bibfnamefont {TE}~\bibnamefont {Drake}}, \ and\ \bibinfo {author}
  {\bibfnamefont {DS}~\bibnamefont {Jin}},\ }\bibfield  {title} {\enquote
  {\bibinfo {title} {Evolution of the normal state of a strongly interacting
  fermi gas from a pseudogap phase to a molecular bose gas},}\ }\href@noop {}
  {\bibfield  {journal} {\bibinfo  {journal} {Physical review letters}\
  }\textbf {\bibinfo {volume} {106}},\ \bibinfo {pages} {060402} (\bibinfo
  {year} {2011})}\BibitemShut {NoStop}%
\bibitem [{\citenamefont {Jiang}\ \emph {et~al.}(2011)\citenamefont {Jiang},
  \citenamefont {Baksmaty}, \citenamefont {Hu}, \citenamefont {Chen},\ and\
  \citenamefont {Pu}}]{jiang2011single}%
  \BibitemOpen
  \bibfield  {author} {\bibinfo {author} {\bibfnamefont {Lei}\ \bibnamefont
  {Jiang}}, \bibinfo {author} {\bibfnamefont {Leslie~O}\ \bibnamefont
  {Baksmaty}}, \bibinfo {author} {\bibfnamefont {Hui}\ \bibnamefont {Hu}},
  \bibinfo {author} {\bibfnamefont {Yan}\ \bibnamefont {Chen}}, \ and\ \bibinfo
  {author} {\bibfnamefont {Han}\ \bibnamefont {Pu}},\ }\bibfield  {title}
  {\enquote {\bibinfo {title} {Single impurity in ultracold fermi
  superfluids},}\ }\href@noop {} {\bibfield  {journal} {\bibinfo  {journal}
  {Physical Review A}\ }\textbf {\bibinfo {volume} {83}},\ \bibinfo {pages}
  {061604} (\bibinfo {year} {2011})}\BibitemShut {NoStop}%
\bibitem [{\citenamefont {Garratt}\ and\ \citenamefont
  {Roy}(2022)}]{garratt2022resonant}%
  \BibitemOpen
  \bibfield  {author} {\bibinfo {author} {\bibfnamefont {Samuel~J}\
  \bibnamefont {Garratt}}\ and\ \bibinfo {author} {\bibfnamefont {Sthitadhi}\
  \bibnamefont {Roy}},\ }\bibfield  {title} {\enquote {\bibinfo {title}
  {Resonant energy scales and local observables in the many-body localized
  phase},}\ }\href@noop {} {\bibfield  {journal} {\bibinfo  {journal} {Physical
  Review B}\ }\textbf {\bibinfo {volume} {106}},\ \bibinfo {pages} {054309}
  (\bibinfo {year} {2022})}\BibitemShut {NoStop}%
\bibitem [{\citenamefont {Garratt}\ \emph {et~al.}(2021)\citenamefont
  {Garratt}, \citenamefont {Roy},\ and\ \citenamefont
  {Chalker}}]{garratt2021local}%
  \BibitemOpen
  \bibfield  {author} {\bibinfo {author} {\bibfnamefont {SJ}~\bibnamefont
  {Garratt}}, \bibinfo {author} {\bibfnamefont {Sthitadhi}\ \bibnamefont
  {Roy}}, \ and\ \bibinfo {author} {\bibfnamefont {JT}~\bibnamefont
  {Chalker}},\ }\bibfield  {title} {\enquote {\bibinfo {title} {Local
  resonances and parametric level dynamics in the many-body localized phase},}\
  }\href@noop {} {\bibfield  {journal} {\bibinfo  {journal} {Physical Review
  B}\ }\textbf {\bibinfo {volume} {104}},\ \bibinfo {pages} {184203} (\bibinfo
  {year} {2021})}\BibitemShut {NoStop}%
\bibitem [{\citenamefont {Atas}\ \emph {et~al.}(2013)\citenamefont {Atas},
  \citenamefont {Bogomolny}, \citenamefont {Giraud},\ and\ \citenamefont
  {Roux}}]{atas2013distribution}%
  \BibitemOpen
  \bibfield  {author} {\bibinfo {author} {\bibfnamefont {YY}~\bibnamefont
  {Atas}}, \bibinfo {author} {\bibfnamefont {Eugene}\ \bibnamefont
  {Bogomolny}}, \bibinfo {author} {\bibfnamefont {O}~\bibnamefont {Giraud}}, \
  and\ \bibinfo {author} {\bibfnamefont {G}~\bibnamefont {Roux}},\ }\bibfield
  {title} {\enquote {\bibinfo {title} {Distribution of the ratio of consecutive
  level spacings in random matrix ensembles},}\ }\href@noop {} {\bibfield
  {journal} {\bibinfo  {journal} {Physical review letters}\ }\textbf {\bibinfo
  {volume} {110}},\ \bibinfo {pages} {084101} (\bibinfo {year}
  {2013})}\BibitemShut {NoStop}%
\bibitem [{\citenamefont {Haake}(1991)}]{haake1991quantum}%
  \BibitemOpen
  \bibfield  {author} {\bibinfo {author} {\bibfnamefont {Fritz}\ \bibnamefont
  {Haake}},\ }\href@noop {} {\emph {\bibinfo {title} {Quantum signatures of
  chaos}}}\ (\bibinfo  {publisher} {Springer},\ \bibinfo {year}
  {1991})\BibitemShut {NoStop}%
\bibitem [{\citenamefont {Prakash}\ \emph {et~al.}(2021)\citenamefont
  {Prakash}, \citenamefont {Pixley},\ and\ \citenamefont
  {Kulkarni}}]{prakash2021universal}%
  \BibitemOpen
  \bibfield  {author} {\bibinfo {author} {\bibfnamefont {Abhishodh}\
  \bibnamefont {Prakash}}, \bibinfo {author} {\bibfnamefont {JH}~\bibnamefont
  {Pixley}}, \ and\ \bibinfo {author} {\bibfnamefont {Manas}\ \bibnamefont
  {Kulkarni}},\ }\bibfield  {title} {\enquote {\bibinfo {title} {Universal
  spectral form factor for many-body localization},}\ }\href@noop {} {\bibfield
   {journal} {\bibinfo  {journal} {Physical Review Research}\ }\textbf
  {\bibinfo {volume} {3}},\ \bibinfo {pages} {L012019} (\bibinfo {year}
  {2021})}\BibitemShut {NoStop}%
\bibitem [{\citenamefont {Schubert}\ \emph {et~al.}(2010)\citenamefont
  {Schubert}, \citenamefont {Schleede}, \citenamefont {Byczuk}, \citenamefont
  {Fehske},\ and\ \citenamefont {Vollhardt}}]{Schubert2010}%
  \BibitemOpen
  \bibfield  {author} {\bibinfo {author} {\bibfnamefont {Gerald}\ \bibnamefont
  {Schubert}}, \bibinfo {author} {\bibfnamefont {Jens}\ \bibnamefont
  {Schleede}}, \bibinfo {author} {\bibfnamefont {Krzysztof}\ \bibnamefont
  {Byczuk}}, \bibinfo {author} {\bibfnamefont {Holger}\ \bibnamefont {Fehske}},
  \ and\ \bibinfo {author} {\bibfnamefont {Dieter}\ \bibnamefont {Vollhardt}},\
  }\bibfield  {title} {\enquote {\bibinfo {title} {Distribution of the local
  density of states as a criterion for anderson localization: Numerically exact
  results for various lattices in two and three dimensions},}\ }\href@noop {}
  {\bibfield  {journal} {\bibinfo  {journal} {Physical Review B}\ }\textbf
  {\bibinfo {volume} {81}},\ \bibinfo {pages} {155106} (\bibinfo {year}
  {2010})}\BibitemShut {NoStop}%
\bibitem [{\citenamefont {Mirlin}\ \emph {et~al.}(1996)\citenamefont {Mirlin},
  \citenamefont {Fyodorov}, \citenamefont {Dittes}, \citenamefont {Quezada},\
  and\ \citenamefont {Seligman}}]{Mirlin1996}%
  \BibitemOpen
  \bibfield  {author} {\bibinfo {author} {\bibfnamefont {Alexander~D}\
  \bibnamefont {Mirlin}}, \bibinfo {author} {\bibfnamefont {Yan~V}\
  \bibnamefont {Fyodorov}}, \bibinfo {author} {\bibfnamefont {Frank-Michael}\
  \bibnamefont {Dittes}}, \bibinfo {author} {\bibfnamefont {Javier}\
  \bibnamefont {Quezada}}, \ and\ \bibinfo {author} {\bibfnamefont {Thomas~H}\
  \bibnamefont {Seligman}},\ }\bibfield  {title} {\enquote {\bibinfo {title}
  {Transition from localized to extended eigenstates in the ensemble of
  power-law random banded matrices},}\ }\href@noop {} {\bibfield  {journal}
  {\bibinfo  {journal} {Physical Review E}\ }\textbf {\bibinfo {volume} {54}},\
  \bibinfo {pages} {3221} (\bibinfo {year} {1996})}\BibitemShut {NoStop}%
\bibitem [{\citenamefont {Mirlin}(2000)}]{Mirlin2000}%
  \BibitemOpen
  \bibfield  {author} {\bibinfo {author} {\bibfnamefont {Alexander~D}\
  \bibnamefont {Mirlin}},\ }\bibfield  {title} {\enquote {\bibinfo {title}
  {Statistics of energy levels and eigenfunctions in disordered systems},}\
  }\href@noop {} {\bibfield  {journal} {\bibinfo  {journal} {Physics Reports}\
  }\textbf {\bibinfo {volume} {326}},\ \bibinfo {pages} {259--382} (\bibinfo
  {year} {2000})}\BibitemShut {NoStop}%
\end{thebibliography}%

\end{document}